\documentclass[twocolumn,floatfix,superscriptaddress, aps, prb, dvipsnames,longbibliography]{revtex4-2}
\usepackage[utf8]{inputenc}
\usepackage[english]{babel}
\usepackage[T1]{fontenc}
\usepackage{tikz}
\usepackage{epstopdf}
\usepackage{pgfplots}
\pgfplotsset{compat=1.13}
\usetikzlibrary{arrows.meta, positioning, bending, matrix}
\usepackage{braids}
\usepackage[colorlinks, hypertexnames=false]{hyperref}
\usepackage[normalem]{ulem}
\usepackage{amsmath, mathtools, amssymb, amsthm, amsfonts}
\usepackage{physics}
\usepackage{bm}
\usepackage{bbold}
\usepackage{xcolor}
\usepackage{ragged2e}
\usepackage{float}
\usepackage[export]{adjustbox}
\definecolor{bluish}{RGB}{50,100,190}
\hypersetup{citecolor={bluish},
linkcolor={bluish},
urlcolor={bluish}}
\newcommand{\arrowIn}{\tikz \draw[-{Latex[length=5mm, width=2mm]}] (3,0) --++ (1,0); }

\newcommand{\eref}[1]{\hyperref[#1]{{Eq.~\ref{#1}}}}  
\newcommand{\eqsref}[1]{\hyperref[#1]{{Eqs.~\ref{#1}}}}  
\newcommand{\fref}[1]{\hyperref[#1]{{Fig.~\ref{#1}}}}
\newcommand{\frefadd}[2]{\hyperref[#1]{{Fig.~\ref*{#1}#2}}}
\newcommand{\sref}[1]{\hyperref[#1]{{Sec.~\ref{#1}}}}
\newcommand{\aref}[1]{\hyperref[#1]{{App.~\ref{#1}}}}

\begin{document}

\title{Topological transitions in quantum jump dynamics: 
Hidden exceptional points
}

\author{Andrei I. Pavlov}
\email{andrei.pavlov@kit.edu}
\affiliation{IQMT, Karlsruhe Institute of Technology, 76131 Karlsruhe,
Germany}

\author{Yuval Gefen}
\affiliation{Department of Condensed Matter Physics, Weizmann Institute of Science, 7610001 Rehovot, Israel}

\author{Alexander Shnirman}
\affiliation{IQMT, Karlsruhe Institute of Technology, 76131 Karlsruhe,
Germany}
\affiliation{TKM, Karlsruhe Institute of Technology, 76131 Karlsruhe, Germany}

\begin{abstract}
    Complex spectra of dissipative quantum systems may exhibit degeneracies known as exceptional points (EPs). At these points the systems' dynamics may undergo drastic changes. 
    Phenomena associated with EPs and their applications have been extensively studied in relation to various experimental platforms, including, i.e., the superconducting circuits. While most of the studies focus on EPs appearing due to the variation of the system's physical parameters, we focus on EPs emerging in the full counting statistics of the system. We consider a monitored three-level system and find multiple EPs in the Lindbladian eigenvalues considered as functions of a counting field. These "hidden" EPs are not accessible without the insertion of the counting field into the Linbladian, i.e., if only the density matrix of the system is studied. Nevertheless, we show that the hidden EPs are accessible experimentally.
    We demonstrate that these EPs signify transitions between different topological classes which are rigorously characterized in terms of the braid theory. Furthermore, we identify dynamical observables affected by these transitions and demonstrate how experimentally measured quantum jump distributions can be used to spot transitions between different topological regimes. Additionally, we establish a duality between the conventional Lindbladian EPs (zero counting field) and some of the ``hidden'' ones. 
    Our findings allow for easier experimental observations of the EP transitions, normally concealed by the Lindbladian steady state, without applying postselection schemes. These results can be directly generalized to any monitored open system as long as it is described within the Lindbladian formalism.
\end{abstract}

\maketitle

\section{Introduction}
Complex spectra of non-Hermitian operators may possess exceptional points (EPs), degeneracies in eigenvalues with simultaneous coalescence of the corresponding eigenvectors \cite{Moiseyev2011, Uzdin2011, Heiss2012, Ashida2020}. The presence of EPs can have drastic consequences for system's properties in their vicinity, such as non-adiabatic switching \cite{Berry2011, ElGanainy2018, Guria2024}, dynamical phase transitions \cite{Alvarez2006, Pastawski2007, Heyl2013, Eleuch2013, Fernandez2015, Eleuch2016, Deng2023}, quantum Zeno effect \cite{Kumar2020, Froml2020, Mouloudakis2022, Dubey2023}, spontaneous symmetry breaking \cite{Makris2008, Bittner2012, Minganti2018}, topological phase transitions \cite{Gong2018, Lieu2018, Martinez2018, Martinez2018b, FoaTorres2019, Kawabata2019}. These sudden changes in the system's behavior at the EP can be understood in the framework of the catastrophe theory \cite{Seyranian2003, Znojil2022, Hu2023} and originate from non-analyticities of the Riemann surfaces for complex eigenvalues as functions of the system's parameters. Among the phenomena associated with EP transitions, the topological phase transitions, studied within the topological band theory \cite{Bansil2016, Wojcik2020}, stand apart. In addition to the dependence on the external parameters, which can be explicitly controlled, the bands' EPs emerge there at specific values of the momentum. The whole spectrum of such a system is periodic over the Brillouin zone, though particular bands may have different periodicity, and is characterized by a set of integer topological invariants; a change of these invariants results in a topological phase transition which necessarily involves an EP \cite{Lieu2018}.

In recent years, EPs have been extensively studied in the fields of quantum optics and quantum superconducting hardware, as they find multiple applications for sensing \cite{Chen2017, Hodaei2017}, amplification \cite{Metelmann2015, Zhong2020}, optimal steering \cite{Kumar2022}, and entanglement generation \cite{Li2023}. The description of such dissipative systems usually requires going beyond the Hamiltonian dynamics and employing the Lindbladian formalism. In this case, the experimental analysis of phenomena associated with EPs is challenging due to the fact that the steady state of the full Lindbladian cannot be involved in an EP \cite{Minganti2019}. Thus, only the decaying states can host EPs. It means that EP-related effects can be observed only in the transient dynamics of the system \cite{Chen2022}. Consequently, the experimental studies of EPs in dissipative systems have to deal with this limitation, focusing on the Hamiltonian EPs. The latter emerge, e.g., in the postselection schemes, in which the effects of quantum jumps are suppressed. In particular such schemes allow accessing the systems' dynamics conditioned on the temporal absence of quantum jumps \cite{Naghiloo2019, Minev2019, Chen2021}.
Complementary, the dynamics of the system can be characterized in terms of the full counting statistics (FCS) of quantum jumps \cite{Levitov1993, Levitov1996, Taddei2002, Nazarov2002, NazarovB2003, Nazarov2003, Flindt2008}, an approach well known within the field of mesoscopic transport. It allows identifying the quantum jumps probability distribution and its evolution in time. Within this approach, the spectrum of the system becomes a periodic function of the counting field. Interestingly, this spectrum can acquire nontrivial topological properties reminiscent of the topological band theory, as reported in \cite{Li2014} for a two-level system interacting with a detector. 

In the context of FCS, one usually resorts to studying the set of cumulants of the observable at hand.
The work of Li et al. \cite{Li2014}, followed by  \cite{Riwar2019, Kleinherbers2023}, presents the challenge of studying and classifying the topologies that can occur in FCS. Such a classification should explicitly employ the emergent topological invariants.    
What we argue here is that to achieve a broad classification of the emergent topologies, knowledge of cumulants does not suffice. 
One needs to consider ``hidden exceptional points'', that do appear at finite values of the counting fields. Our approach to  topological classification  presented here, which fuses between emergent topological invariants  and the physics of  hidden exceptional points, is followed by a detailed discussion, pointing out the possibility to obtain the necessary information from realistic (accuracy-limited) experimental sets of data.

In this study, we focus on the behavior of a three-level dissipative system with monitored quantum jumps and address the open problems mentioned above. This choice is inspired by the recent experiment \cite{Minev2019} focused on detection and reversal of quantum jumps in a three-level artificial atom realized using superconducting qubits. However, the developed approach and the presented results are platform-agnostic and can be equally applied for Lindbladians of arbitrary size and with arbitrary structures of dissipators.

The paper is structured in the following way. We summarize our main results in Sec.\ref{sec:results}. In Sec. \ref{sec:Model} we introduce the model under consideration and discuss its relation to the existing experimental setups of superconducting qubits, Rydberg atoms and trapped ions. We introduce the topological classification of the system, calculate the corresponding invariants and discuss their relation to the non-Abelian braid theory in Sec. \ref{sec:invariants}. In Sec. \ref{sec:observable}
we discuss how the drastic changes in the vicinity of the EPs 
are related to the observables of the full counting statistics of quantum jumps.
In Sec. \ref{sec:Exp}, we discuss how these observables can be constructed from the typical experimental quantum jump histograms and illustrate how one can recover the underlying topology from these data. Finally, in Sec. \ref{sec:SD}, we propose a universal duality connection between the Lindbladian EPs at the zero and nonzero 
values of the counting field.
In Sec. \ref{sec:conslusion} we discuss the implications of this duality for observing Lindbladian EP dynamics at arbitrary times without using postselection as well as possible consequences for the error correction protocols. In addition, we argue that our approach allows for gaining additional insights about errors in Pauli-Lindblad noise models \cite{vandenBerg2023}.

\section{Main results}
\label{sec:results}
Here we briefly summarize the main findings of our research. 

$\bullet$ We reveal a deep connection between the dynamics of quantum jumps and the topological band theory. Namely, we identify qualitatively different classes of time-dependent quantum jump distributions which are defined by topological structures of the Lindbladian eigenvalues considered as functions of the counting field for the quantum jumps. We establish the rigorous topological characterization of these dynamical classes in terms of the non-Abelian braid theory and find an unexpectedly rich variety of nontrivial topological classes even in small systems. These classes are topologically protected, i.e. they are resilient against external perturbations and noise as long as these perturbations are not too strong.

$\bullet$ We demonstrate that transitions between different topological classes of quantum jumps occur through exceptional points of the system's Lindbladian equipped with the jump-counting field.

$\bullet$ We identify a dynamical observable that is affected by these topological transitions and demonstrate how the underlying topological structures can be restored from quantum jump distributions collected in state-of-the-art experiments.

$\bullet$ We show that the analysis of the Lindbladian eigenvalues at the nonzero values of the counting field allows bypassing the no-go theorem that prohibits formation of EPs in the longest-living Lindbladian eigenstate \cite{Minganti2019, Minganti2020}. Moreover, there is a duality between EPs of the Lindbladian eigenvalues at zero and half-period values of the counting field, so that these EPs appear in pairs. These findings provide tools for studying phenomena associated with EPs in the full Lindbladian picture without employing any postselection protocols.

Our results show the role of the EPs in formation of the non-Poissonian quantum jump statistics and establish a connection between non-equilibrium dynamics and the underlying topological characterization of dissipative systems, demonstrating that dynamical phase transitions are, in fact, topological transitions. They provide tools for studying phenomena associated with EPs in the full Lindbladian picture without dependence on postselection protocols.\\

\section{Model}
\label{sec:Model}
\begin{figure}
\center
\includegraphics[width=\linewidth]{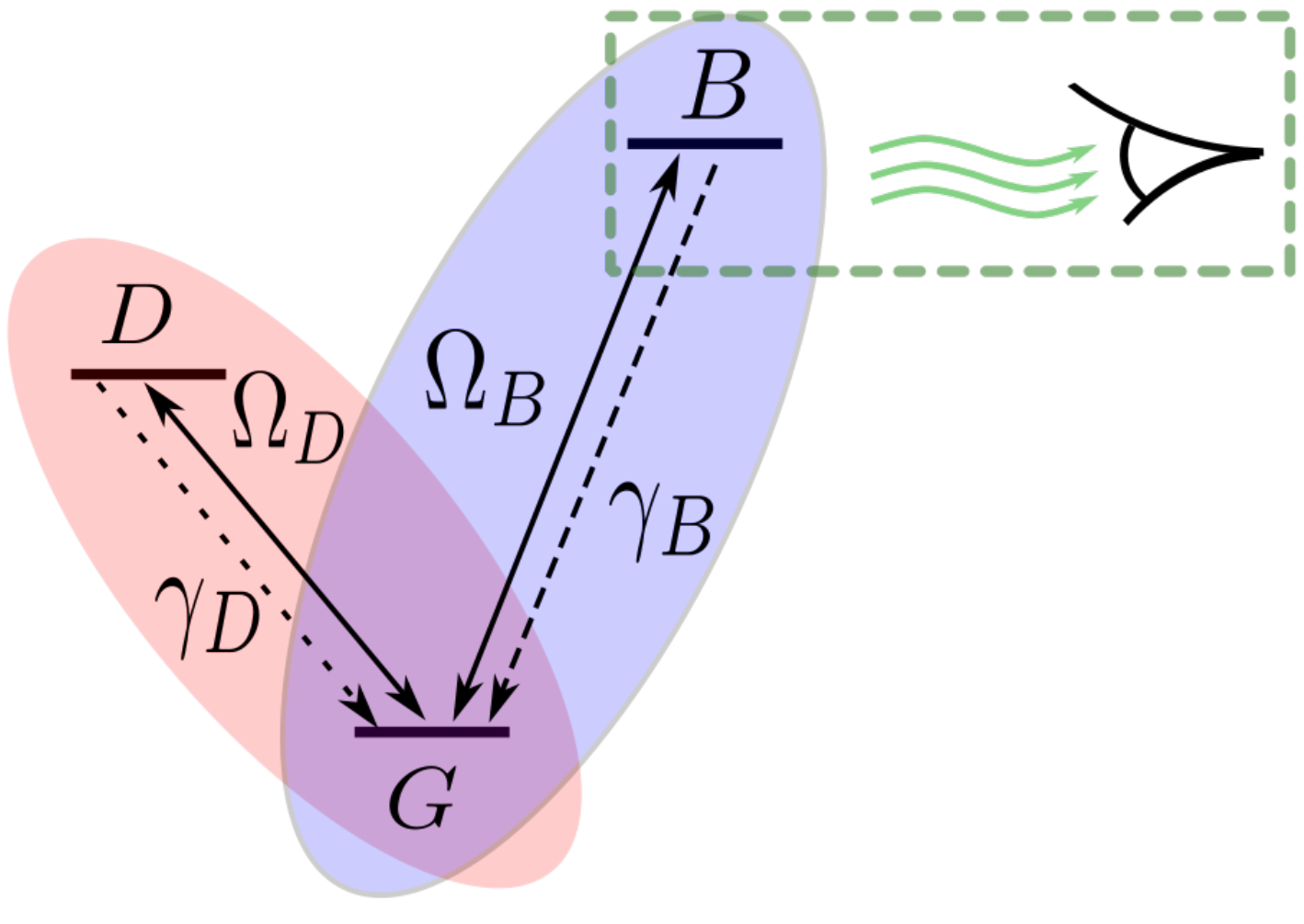}
\caption{A cartoon picture of the three-level system under consideration. The bright state (B) is continuously monitored and quickly decays into the ground state (G). The dark state (D) is not monitored. The Rabi drives $\Omega_B$ and $\Omega_D$ induce transitions between $G\rightleftharpoons B$ and $G\rightleftharpoons D$ states, respectively. $\gamma_B\gg\gamma_D$ are decay rates.}
\label{fig:ThLS}
\end{figure}
Superconducting qubits have become one of paradigmatic platforms for experimental studies of quantum jumps. Over recent years, quantum jumps occurring in such systems have been successfully observed and controlled in various experiments \cite{Vijay2011, Vool2014, Slichter2016, Minev2019, Chen2021, Spiecker2023, Spiecker2024}. We consider a setup that closely matches the dissipative three-level system studied in \cite{Minev2019}. In particular, we focus on the fluorescent quantum jump measurements \cite{Campagn2014, Naghiloo2016, Naghiloo2017} model described there. This setup is reminiscent of earlier observations of quantum jumps in trapped ions with fluorescence detection \cite{Cook1984, Nagourney1986, Sauter1986, Bergquist1986}. Namely, we consider a three-level $V$-shaped system comprised of the ground (G), bright (B), and dark (D) states. Its schematic representation is shown in Fig. \ref{fig:ThLS}. This system is known to exhibit dynamical phase transitions between a rich variety of dynamical phases \cite{Garrahan2010, Lesanovsky2013, Garrahan2018, Perfetto2022}. The bright state relaxes spontaneously at a very high rate into the ground state, this relaxation process is constantly monitored, while the dark state is not monitored at all. In addition, there are two periodic real Rabi drives $\Omega_D$ and $\Omega_B$ which enable transitions between $G\rightleftharpoons D$ and $G\rightleftharpoons B$ states, respectively. This $V$-shaped system also closely resembles typical setups for experiments with Rydberg atoms \cite{Mattioli2015, Begoc2023}, so our results can be directly applied to such experiments.

The detailed derivation of the effective three-level Hamiltonian from two coupled qubits under resonant driving is given in Appendix \ref{sec:RWA}.
The effective three-level description in the interaction picture with the rotating wave approximation is given by 
\begin{align}
 \label{Hint}   H_{I}=\Omega_DS^x +\Omega_BT^x,
\end{align}
$S^x=\frac{1}{2}(S^++S^-), \, T^x=\frac{1}{2}(T^++T^-)$, where $S^{\pm}$ and $T^{\pm}$ are $3\times 3$ matrices that form raising and lowering operators in the basis $\{\ket{G},\ket{D}, \ket{B} \}$. They account for transitions between $G\rightleftharpoons D$ and $G\rightleftharpoons B$ states correspondingly:
\begin{align}
 \label{STp}  &S^{+}=\begin{pmatrix}
        0 & 0 & 0 \\
        1 & 0 & 0 \\
        0 & 0 & 0
    \end{pmatrix}, \,\,\, T^{+}=\begin{pmatrix}
        0 & 0 & 0 \\
        0 & 0 & 0 \\
        1 & 0 & 0
    \end{pmatrix},&\\
 \label{STm}  &S^{-}=\,\left(S^{+}\right)^{\dagger}, \,\,\,\,\,\,\,\,\,\,\,\,\,\, T^{-}=\,\left(T^{+}\right)^{\dagger}.&
\end{align}
The decay from exited states $B$ and $D$ into the ground state $G$ can be accounted by the dissipators $T^{-}$ and $S^{-}$ with respective rates $\gamma_B$ and $\gamma_D$. We assume that the bright state B decays at a very high rate, while the dark state D is almost lossless. These assumptions are motivated by the experimental situation of \cite{Minev2019}, where $\gamma_B\gg\Omega_B\gg\Omega_D\gg\gamma_D$. We keep the same hierarchy throughout our manuscript. Due to the introduced dissipation, the dynamics of the system is non-Hermitian. Without quantum jumps, this evolution can be described by the effective non-Hermitian Hamiltonian: 
\begin{align}
 \label{Heff}   H_{eff}=H_{I}-\frac{\textit{i}}{2}\gamma_BT^{+}T^{-}-\frac{\textit{i}}{2}\gamma_DS^{+}S^{-}.
\end{align}
The full Lindbladian operator of the system includes this effective Hamiltonian part and the quantum jump operators. Since the bright state decay is constantly monitored, each decay from the bright state into the ground state, which is accompanied by emission of a photon, is detected. Such a transition is modeled by a quantum jump term, and therefore one has direct access to the statistics of the $B\rightarrow G$ quantum jumps throughout measurements. We add the (real) counting field $k$ to the corresponding jump term of the Lindladian to reproduce this experimentally accessible FCS. Strictly speaking, this new operator is not a Lindbladian, and the Liouvillian evolution with this operator does not preserve the trace of the density matrix for nonzero $k$. Nevertheless, it can be seen as a straightforward generalization of Liouvillian operators, analogous to the hybrid-Liouvillian evolution approach from \cite{Minganti2020}. In the superoperator notations (\cite{Schaller2014, Minganti2019}), this generalization of the Lindbladian gives the following operator:
\begin{align}
 \nonumber &\mathcal{L}_k=-\textit{i}\left\{\left(H_{I}-\frac{\textit{i}}{2}\Gamma^{\dagger}\Gamma\right)\otimes \mathbb{1}-\mathbb{1}\otimes \left(H_{I}^T+\frac{\textit{i}}{2}\Gamma^T\Gamma^*\right)\right\}&\\
 \label{fullL} &+\gamma_Be^{\textit{i}k}\, T^{-}\otimes T^{-}+\gamma_D S^{-}\otimes S^{-},&
\end{align}
where we denoted $\Gamma^{\dagger}\Gamma\equiv\gamma_BT^+T^-+\gamma_DS^+S^-$. $\otimes$ denotes the Kronecker product, superscripts $T$ and $*$ stand for transposition and complex conjugation. Note that this operator is periodic with respect to the counting field $k$, which is a direct consequence of an integer number of detected events.  We will further refer to $\mathcal{L}_k$ as a Lindbladian for simplicity, since it restores the Lindbladian dynamics at $k=0$, and EPs of this operator are directly related to the EPs of the original Lindbladian. With this, the density matrix evolves as
\begin{align}
    \rho_k(t)=e^{\mathcal{L}_kt}\rho_0,
\end{align}
where $\rho_k(t)$ is the dynamically evolving density matrix that depends of the counting field $k$. More precisely, $\rho_k(t)\equiv \sum_n e^{-i n k} \rho_n(t)$ is the Fourier image of $\rho_n(t)$, the density matrix of the system conditioned by $n$ quantum jumps exhibited over the measurement time $t$. Its trace gives the probability to observe exactly $n$ quantum jumps, $P_n(t)=Tr[\rho_n(t)]$. In principle, one can analyze this density matrix $\rho_k$ directly, performing the full state tomography. Instead, we focus on a situation that is common in many experiments with superconducting qubits (e.g. \cite{Vool2014, Minev2019, Spiecker2023}) - detecting only occurrences of the quantum jumps, without knowledge of the full structure of the density matrix.
\begin{figure*}[ht!]
    \begin{minipage}[t]{.3\linewidth}
        \centering
        \includegraphics[width=1.\linewidth]{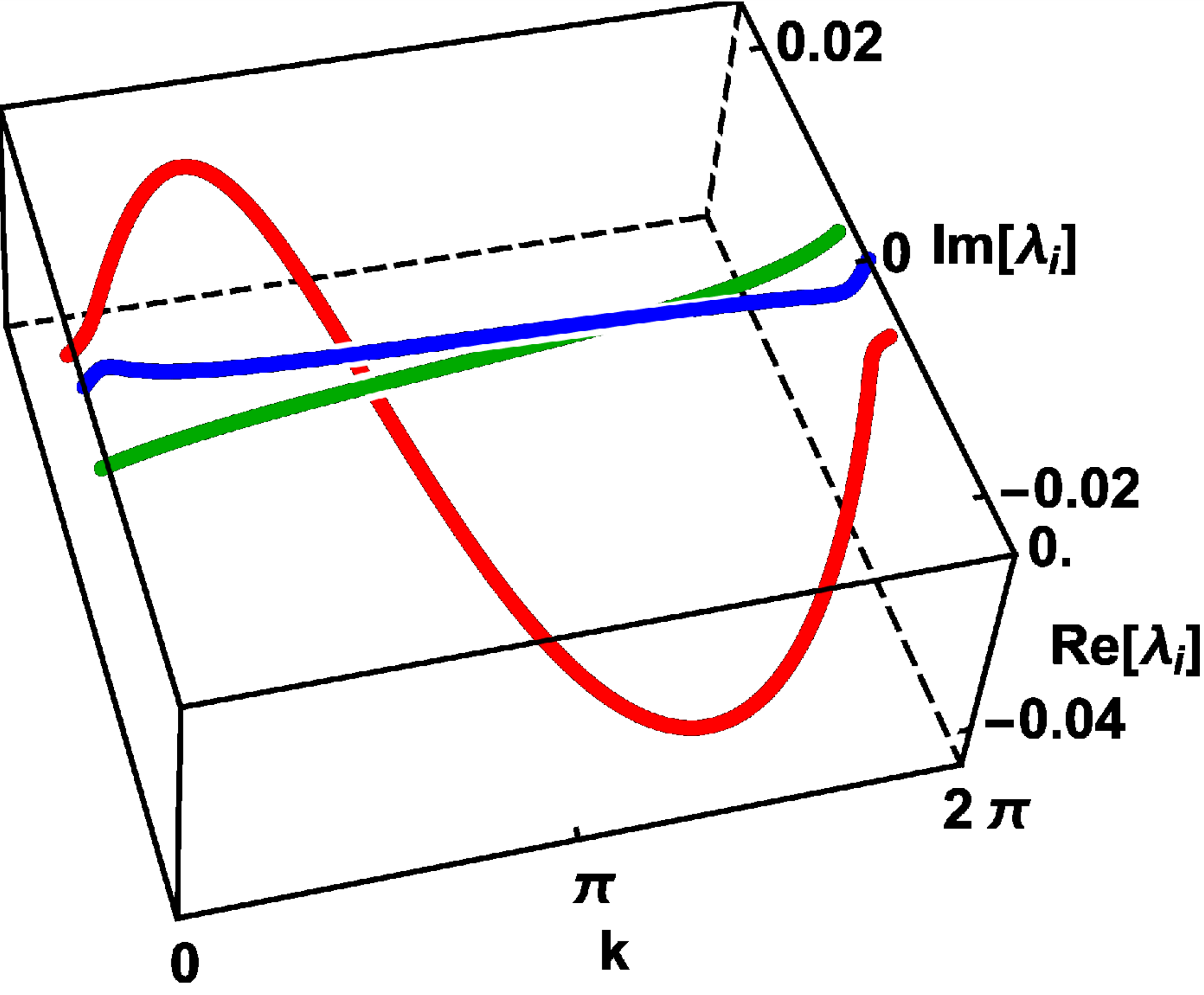}
        $\Omega_D<\Omega_3$
    \end{minipage}
    \hfill
    \begin{minipage}[t]{.3\linewidth}
        \centering 
        \includegraphics[width=1.\linewidth]{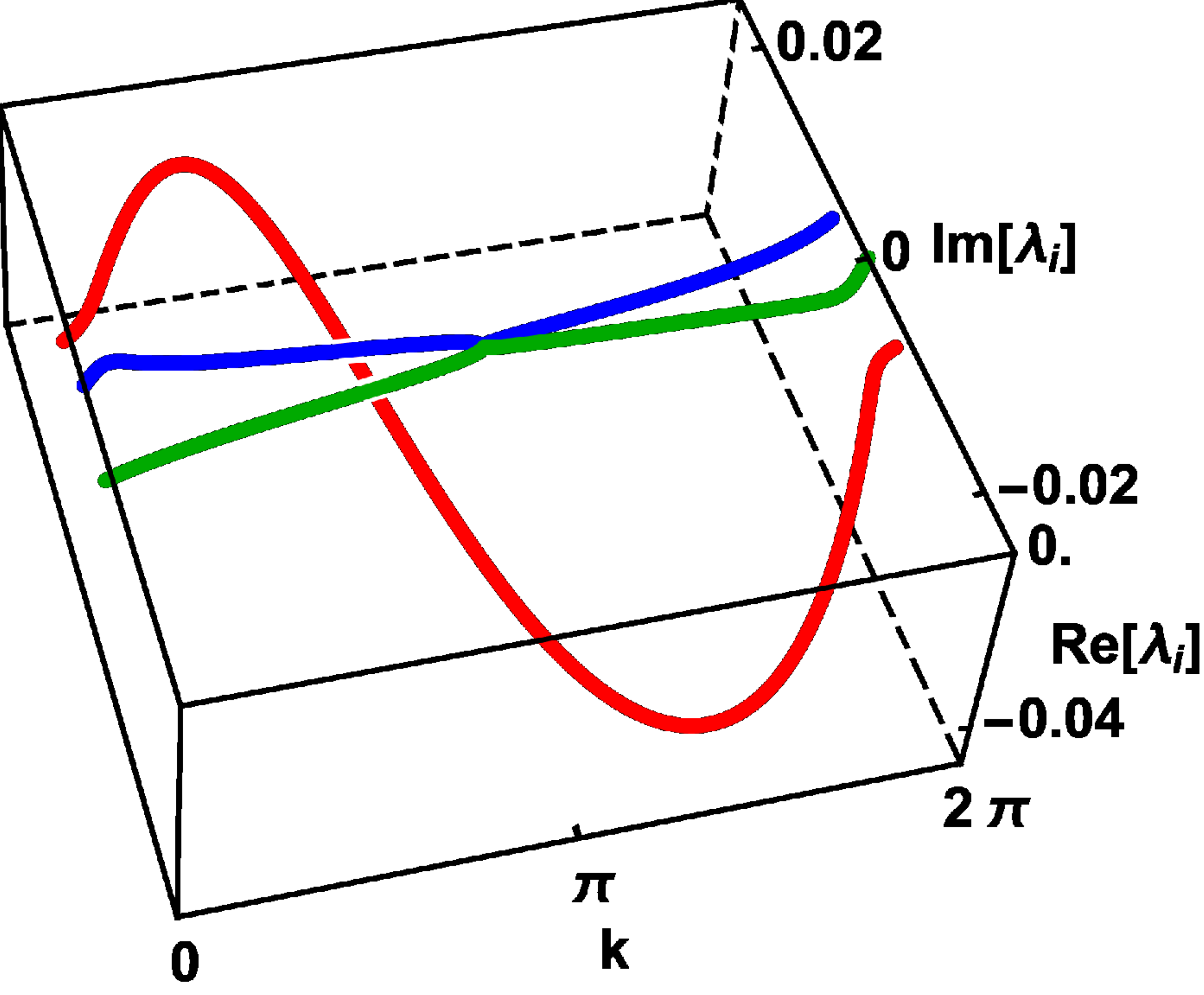}   
     $\Omega_D=\Omega_3$            
    \end{minipage}  
    \hfill
    \begin{minipage}[t]{.3\linewidth} 
    \centering
        \includegraphics[width=1.\linewidth]{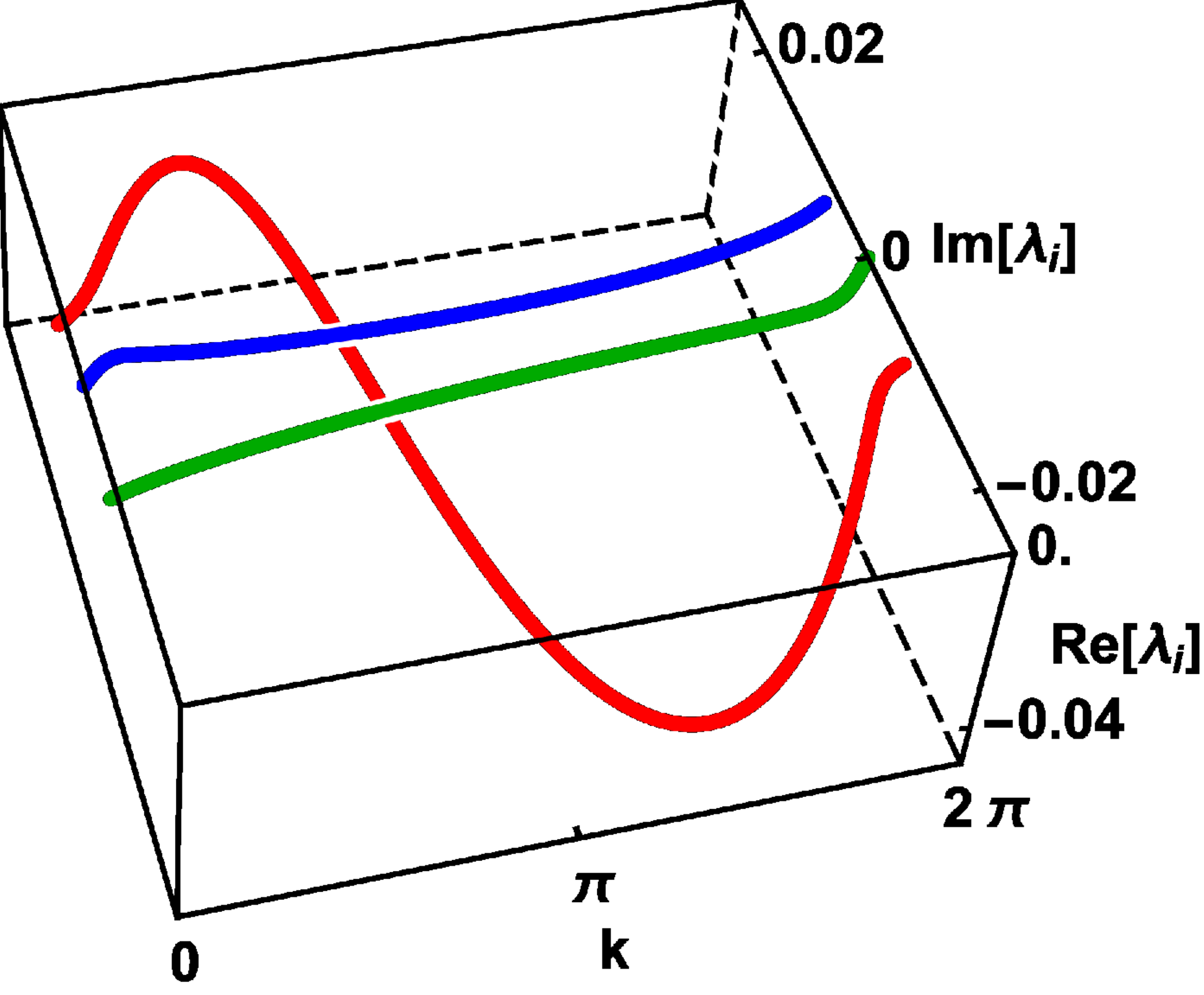}  
        $\Omega_D>\Omega_3$           
    \end{minipage}  
    \hfill
    \newline    
    \begin{minipage}[t]{.3\linewidth}
        \centering
        \includegraphics[width=1.2\linewidth]{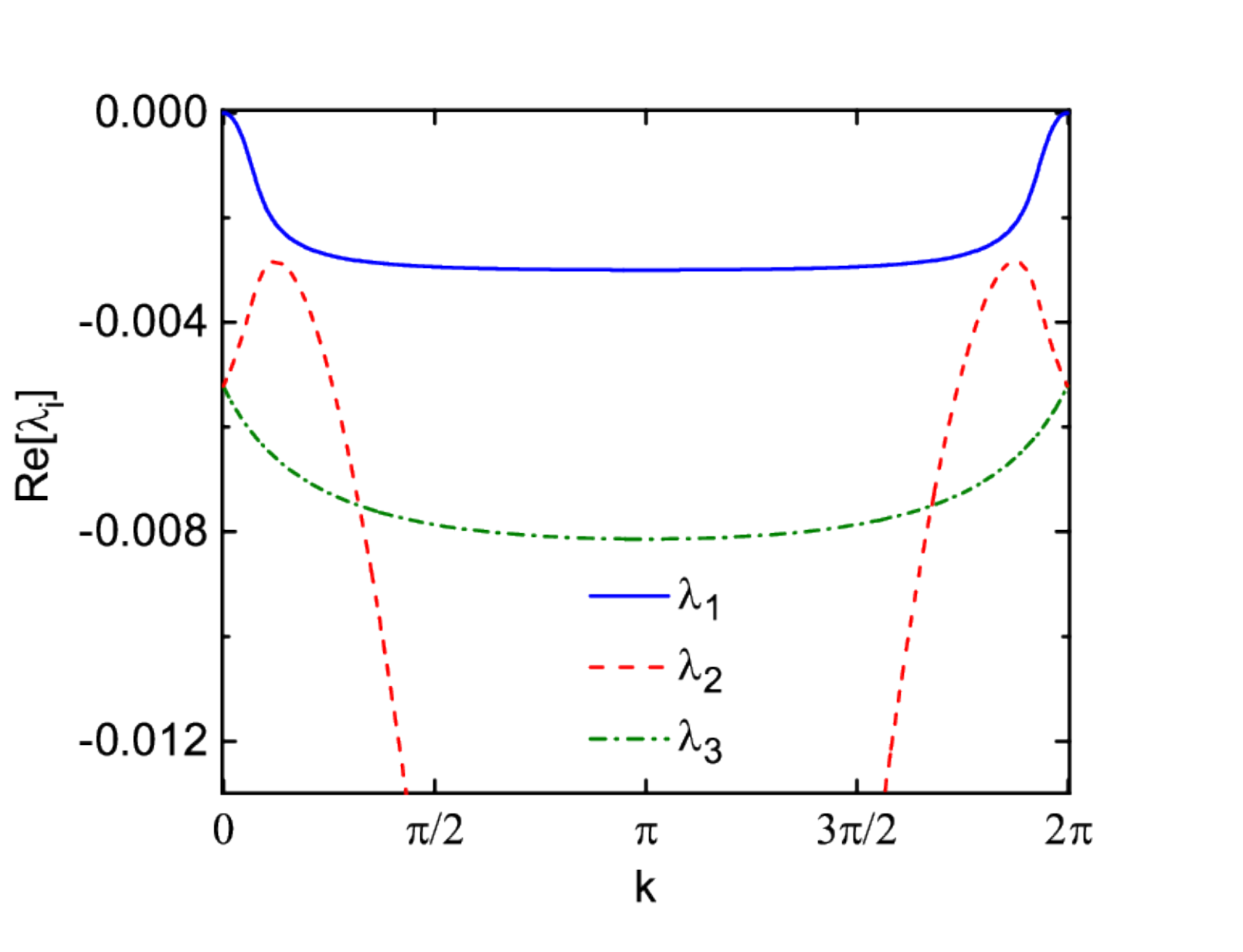}
    \end{minipage}
    \hspace{0.3cm}
    \begin{minipage}[t]{.3\linewidth}
        \centering 
        \includegraphics[width=1.2\linewidth]{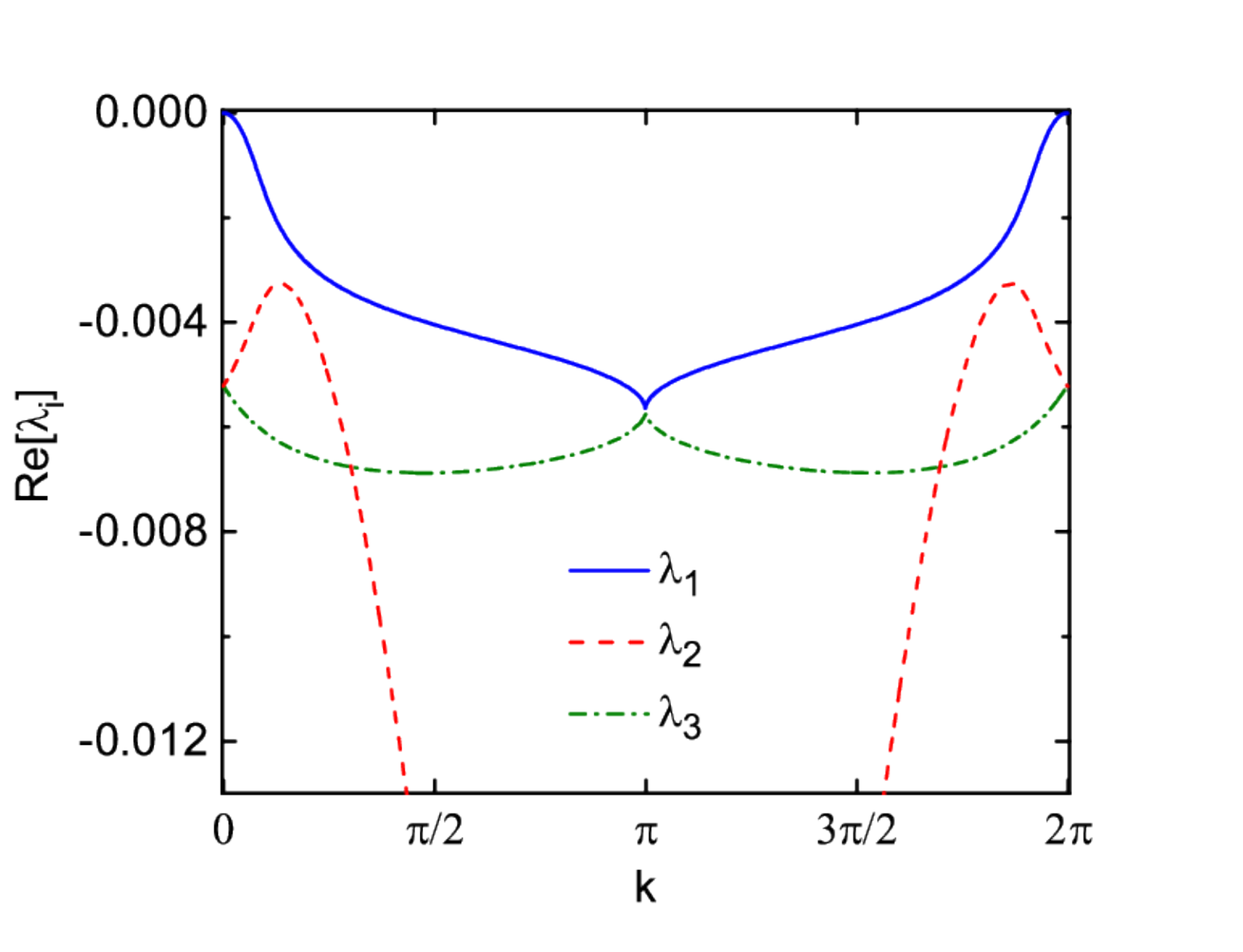}      
    \end{minipage}  
    \hspace{0.3cm}
    \begin{minipage}[t]{.3\linewidth} 
    \centering
        \includegraphics[width=1.2\linewidth]{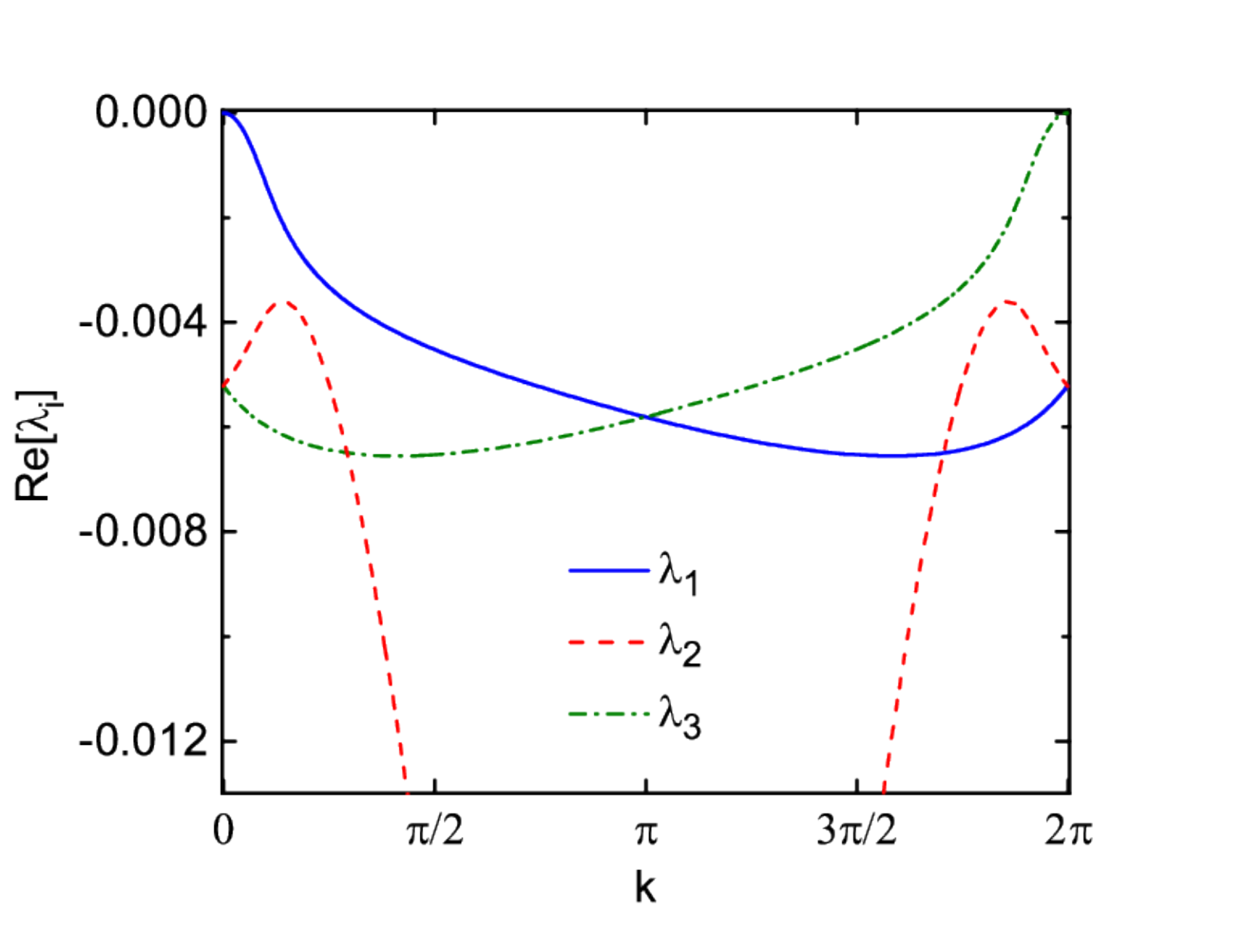}           
    \end{minipage}  
    \hfill
    \hspace{1cm}        
   \caption{Examples of two different eigenvalue braids and a transition between them. The transition between two classes results in an EP at $k=\pi$. Upper panel: Real and imaginary structure of the three least decaying eigenvalues of the Lindbladian Eq. (\ref{fullL}) as functions of the counting field $k$ at different values of the drive $\Omega_D$. Lower panel: Real parts of the corresponding eigenvalues. The transition occurs with variation of the drive $\Omega_D$. $\Omega_3$ is the critical frequency specified below (Table \ref{tab:EP}).}
   \label{fig:3Dstructure}
\end{figure*}

For further use, we introduce the Fourier transform $P_k(t)$ of the probability $P_n(t)$ to observe $n$ quantum jumps over time $t$:
\begin{align}
\label{Pkt} P_k(t) = \sum_n e^{ikn} P_n(t) = \left\langle e^{\textit{i}kn}\right\rangle(t) = Tr\left[e^{\mathcal{L}_kt}\rho_0\right],
\end{align}
\begin{align}
    P_n(t)=\int_0^{2\pi}\frac{dk}{2\pi}e^{-\textit{i}n k}P_k(t);
\end{align} 
these probabilities can be gathered during multiple repetitions of the experiment.

Note that $P_k(t)$ is exactly the generating function, known from the FCS formalism \cite{Gogolin2006}, and the $m$-th moment of the distribution of probability to observe $n$ events over observation time $t$ is given by $\langle n^m(t)\rangle=\left.\frac{\partial^m}{\partial (\textit{i}k)^m}P_k(t)\right\vert_{k=0}$. Each detection event is associated with a transition $\left\vert B\right\rangle\left\langle B\right\vert\rightarrow \left\vert G\right\rangle\left\langle G\right\vert$, so it is given by the jump operator $\mathcal{L}_J=\gamma_B T^-\otimes T^-$. Plugging Eq. (\ref{fullL}) into Eq. (\ref{Pkt}), we see that this property of FCS is indeed satisfied.\\
For simplicity, we set $\gamma_D=0$ throughout the paper, assuming that there is no decay from the dark state and $G\rightleftharpoons D$ transitions happen only due to the $\Omega_D$ drive. We discuss the $\gamma_D\neq 0$ case in Section \ref{sec:SD}. For numerical calculations, unless specified otherwise, we choose $\gamma_B=0.5, \, \Omega_B =0.1$ in dimensionless units (i.e., with respect to some characteristic frequency of the system). Some of the present results depend quantitatively on the choice of these parameters, while others are universal, as we specify below.
\label{sec:invariants}
\section{Topological invariants}
Since the real counting field $k$ acts as a phase for the jump term in Eq. (\ref{fullL}), the Lindblad operator is $2\pi$ periodic over $k$, $\mathcal{L}_k=\mathcal{L}_{k+2\pi}$. The same applies to its spectrum. 
Nevertheless, a particular eigenvalue of the Lindbladian is not necessarily $2\pi$ periodic, as eigenvalues can swap their positions during the $2\pi$ circling. Moreover, even if the eigenvalues are $2\pi$ periodic, they still may wind around each other. A qualitative change in this winding structure means a change in the system's topology. Such structures and transitions between them can be described by means of the braid theory. One can switch between different braids by a proper change of the system's parameters \cite{Wang2021, Zhang2023, Zhang2023b, Konig2023}. 
For Hamiltonian systems stochastically interacting with a detector, the topological braids can appear the same way in full counting statistics \cite{Ren2013, Li2014, Riwar2019, Kleinherbers2023}, and EPs can emerge at nonzero values of the counting field \cite{Ivanov2010}. Below, we demonstrate that the same applies to the Lindbladian EPs. Moreover, the change in eigenvalues' braid structure with respect to the counting field necessarily results in a change in the system's topology even if periodicity of the eigenvalues is unaffected. Similar to the results of the topological band theory, the qualitative changes in braid structures, and the subsequent topological transitions, occur when the system passes through an EP.\\
We consider braids of eigenvalues of the Lindbladian Eq. (\ref{fullL}) as functions of the counting field $k$. In particular, out of nine eigenvalues of this Lindbladian, we focus on three eigenvalues with smallest decay rates
(i.e., three eigenstates with smallest real parts at $k=0$). The fourth one does not contain dependence on $k$ and does not change its braiding with respect to other eigenvalues at any parameters; the other five eigenvalues have much higher decay rates, so they are separated from the former ones by a large gap, which prevents them from participating in braids with the long-living eigenstates. Braids of these three eigenvalues correspond to $B_3$ braid group. The periodicity of the eigenvalues over $k$ means that the braids are closed, and the knot theory provides an equivalent description \cite{Hu2021, Patil2022, Guria2024}. We will not delve into details of the braid and knot theories, referring to reviews on the subjects \cite{Kassel2008, Birman2008, Aldrovandi2021}, but rather use some basic results from them, which are summarized in Appendix \ref{sec:BrKn}.\\
\begin{figure}[!ht]
\begin{minipage}{.49\linewidth} 
\begin{tikzpicture}[scale=0.5] \hspace{-1.8cm}
\braid[line width=3pt, rotate=90, number of strands=3,
style strands={1}{OliveGreen},
style strands={2}{red}, style strands={3}{blue}] (braid) 1 a_2^{-1} a_1^{-1} a_1^{-1} a_2^{-1} 1;
\hspace{1.5cm}
\draw[line width=0.5mm] node[below]{\Large{\textit{k}}}(-3,0)   node[above]{$\,\,\,k_0$}(0.5,0) (-3.,0) -- (3.2,0) node[above]{$k_0+2\pi$}(0.5,0)  node[
	sloped,
	pos=1,
	allow upside down]{\arrowIn};
\end{tikzpicture}
\end{minipage}%
\begin{minipage}{.49\linewidth} \hspace{-0.7cm}
\includegraphics[width=4.8cm]{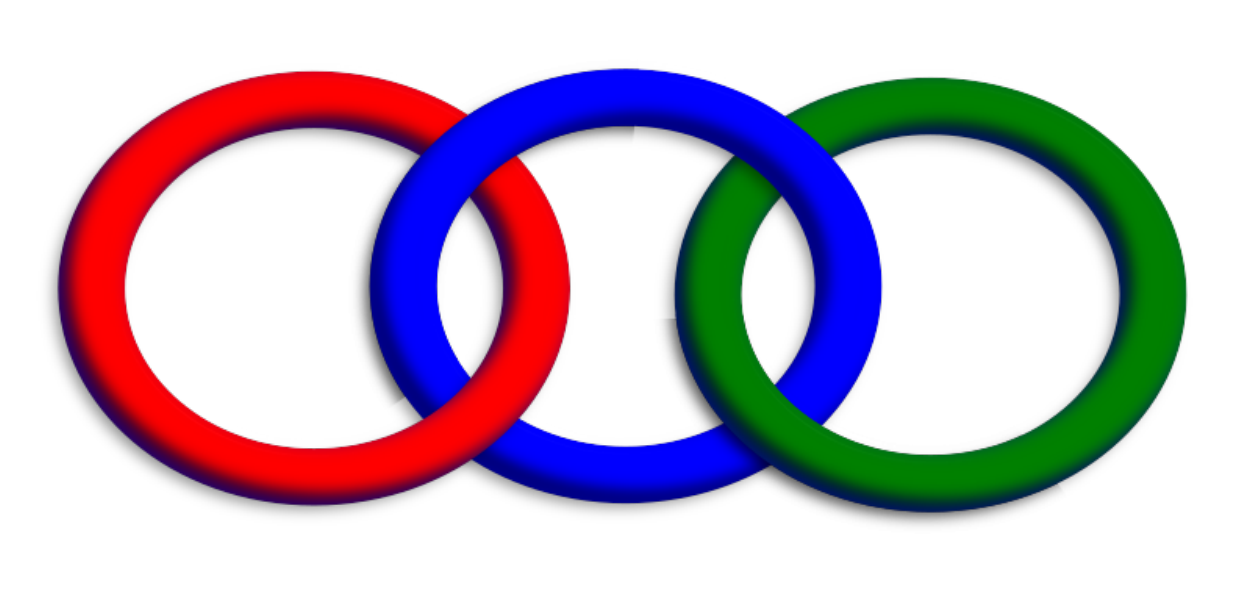}
\vspace{0.0cm}%
\end{minipage}
\begin{FlushLeft}
Class I, $\Omega_1>\Omega_D:$ $\nu=4:$ $\nu_{1;2}=2$, $\nu_{1;3}=2$, $\nu_{2;3}=0.$ Connected sum of two Hopf links. Braid word: $\sigma_1\sigma_2 \sigma_2\sigma_1$.
\end{FlushLeft}
\vspace{.0cm}%
\begin{minipage}{.49\linewidth} 
\vspace{.3cm}%
\begin{tikzpicture}[scale=0.5] \hspace{-1.8cm}
\braid[line width=3pt, rotate=90, number of strands=3, 
style strands={1}{OliveGreen}, style strands ={2}{red}, style strands={3}{blue}, style strands={4}{blue}] (braid) 1 a_1^{-1} 1 1  a_1^{-1} 1;
\hspace{1.5cm}
\draw[line width=0.5mm] node[below]{\Large{\textit{k}}}(-3,0)   node[above]{$\,\,\,k_0$}(0.5,0) (-3.,0) -- (3.2,0) node[above]{$k_0+2\pi$}(0.5,0)  node[
	sloped,
	pos=1,
	allow upside down]{\arrowIn};
\end{tikzpicture}
\end{minipage}
\begin{minipage}{.49\linewidth} \vspace{.5cm} \hspace{-0.7cm}
\includegraphics[width=4.8cm]{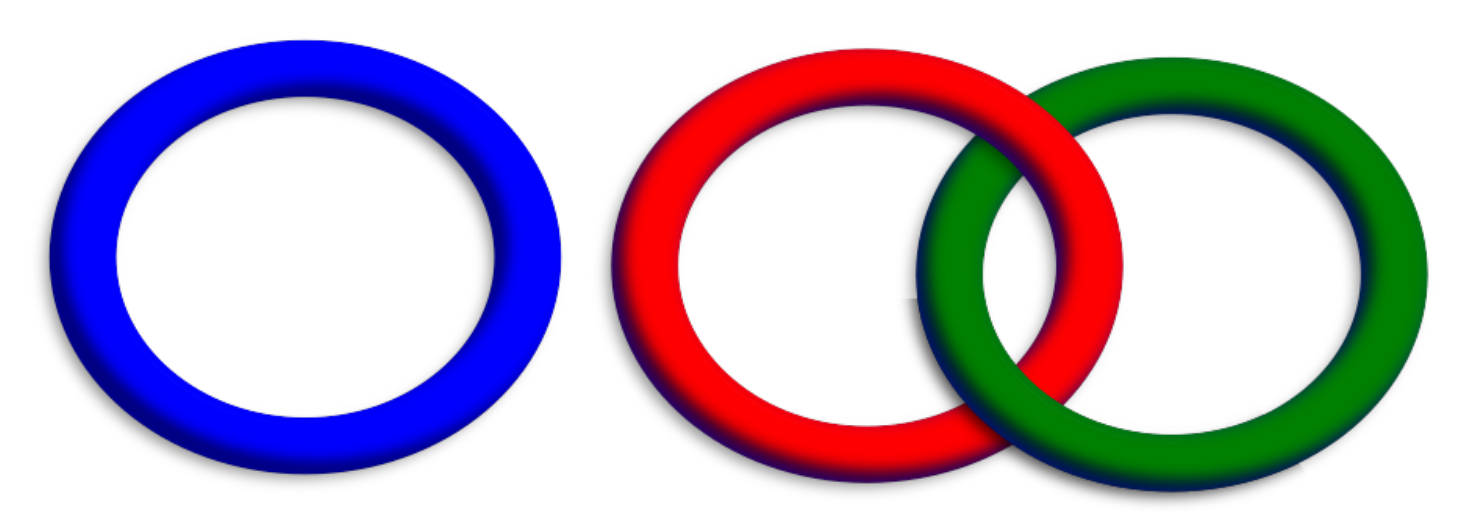}
\vspace{0.5cm}
\end{minipage}
Class II, $\Omega_2>\Omega_D>\Omega_1:$ $\nu=2:$ $\nu_{1;2}=0$, $\nu_{1;3}=0$, $\nu_{2;3}=2.$ Hopf link and unlink. Braid word: $\sigma_2\sigma_2$.
\newline
\begin{minipage}{.49\linewidth} 
\vspace{0.5cm}
\begin{tikzpicture}[scale=0.5] \hspace{-1.8cm}
\braid[line width=3pt, rotate=90, number of strands=3, 
style strands ={1}{OliveGreen}, style strands={2}{red}, style strands={3}{blue}] (braid) 1 1 a_1^{-1} 1 1 1;
\hspace{1.5cm}
\draw[line width=0.5mm] node[below]{\Large{\textit{k}}}(-3,0)   node[above]{$\,\,\,k_0$}(0.5,0) (-3.,0) -- (3.2,0) node[above]{$k_0+2\pi$}(0.5,0)  node[
	sloped,
	pos=1,
	allow upside down]{\arrowIn};
\end{tikzpicture}
\end{minipage}
\begin{minipage}{.49\linewidth} \vspace{0.3cm}\hspace{-0.7cm}
\includegraphics[width=4.cm]{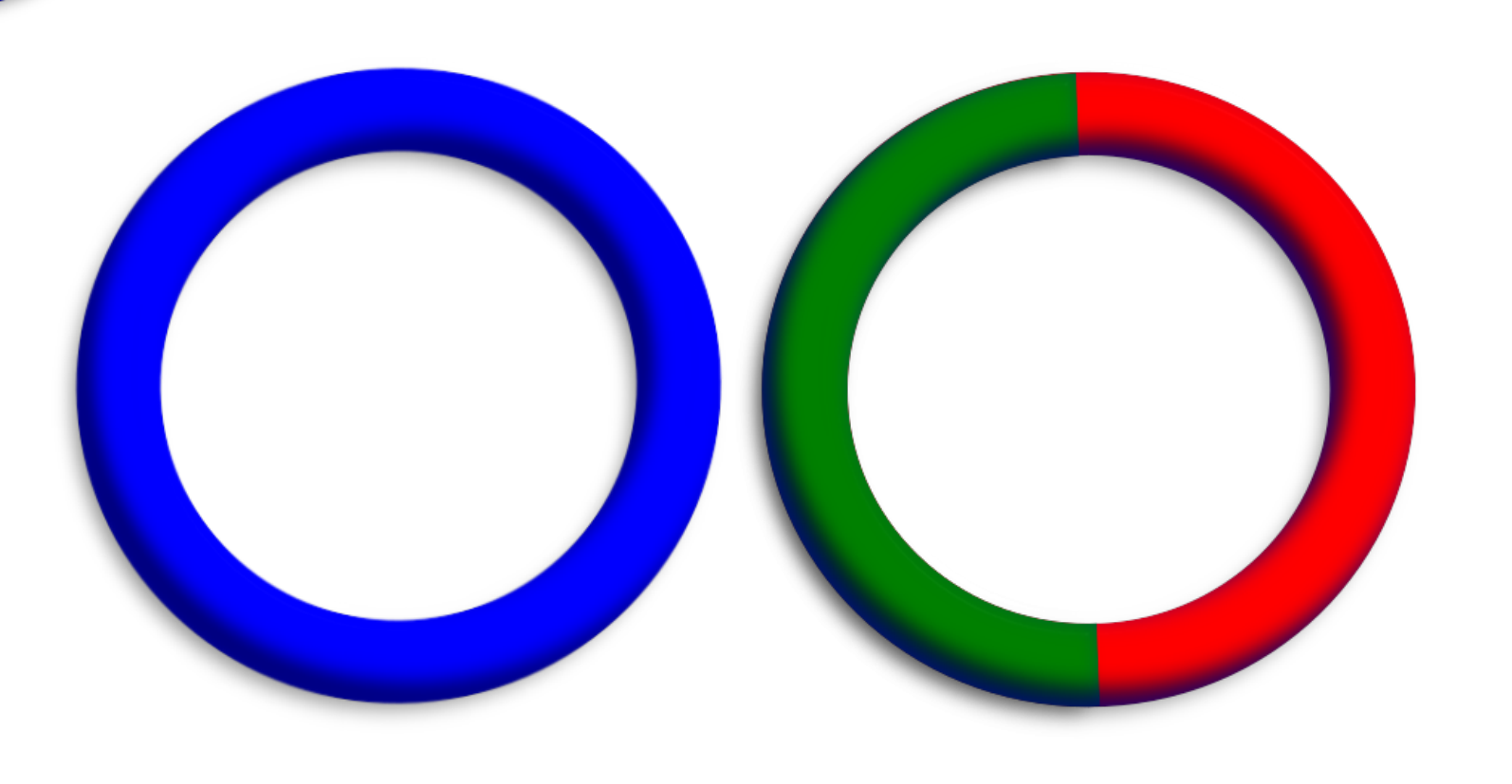}
\vspace{0.5cm}
\end{minipage}
\begin{FlushLeft}
Class III, $\Omega_3>\Omega_D>\Omega_2:$ $\nu=1:$ $\nu_{1;2,3}=0$, $\nu_{2;3}=1$. Unknot and unlink. Braid word: $\sigma_2$.
\end{FlushLeft}
\begin{minipage}{.49\linewidth} \vspace{0.3cm}
\begin{tikzpicture}[scale=0.5] \hspace{-1.8cm}
\braid[line width=3pt, rotate=90, number of strands=3, 
style strands ={1}{OliveGreen}, style strands={2}{red}, style strands={3}{blue}] (braid) 1 1 a_1^{-1}  a_2^{-1} 1 1;
\hspace{1.5cm}
\draw[line width=0.5mm] node[below]{\Large{\textit{k}}}(-3,0)   node[above]{$\,\,\,k_0$}(0.5,0) (-3.,0) -- (3.2,0) node[above]{$k_0+2\pi$}(0.5,0)  node[
	sloped,
	pos=1,
	allow upside down]{\arrowIn};
\end{tikzpicture} 
\end{minipage}
\begin{minipage}{.49\linewidth} \vspace{0.3cm}\hspace{-0.7cm}
\includegraphics[width=2.0cm]{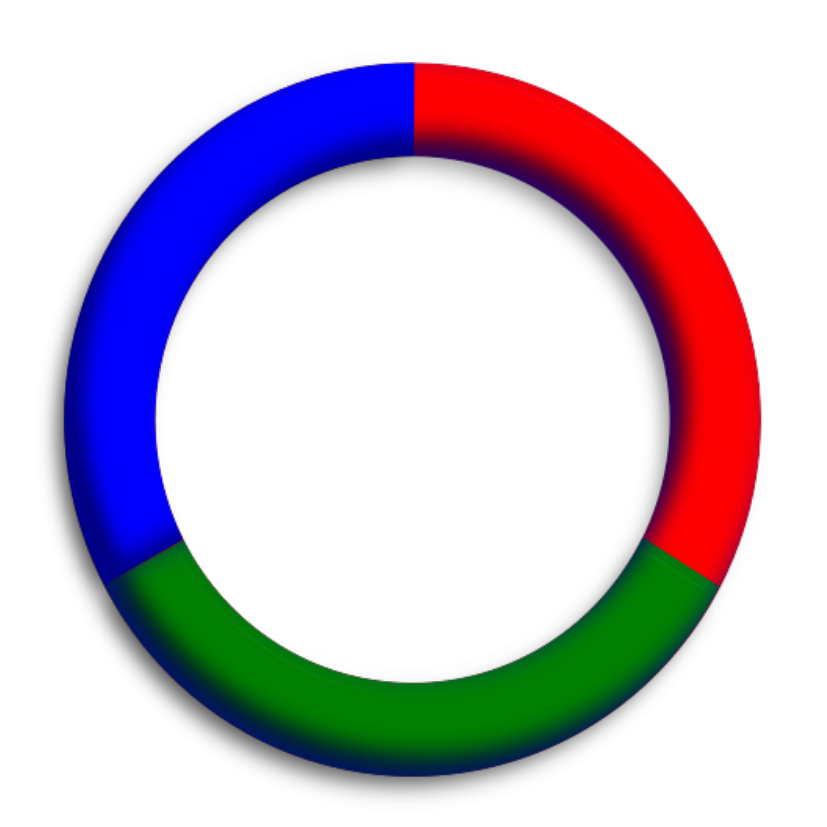}
\vspace{0.5cm}
\end{minipage}
\begin{FlushLeft}
Class IV, $\Omega_4>\Omega_D>\Omega_3:$ $\nu=2$. Three-component unknot. Braid word: $\sigma_2\sigma_1$.
\end{FlushLeft}
\begin{minipage}{.49\linewidth} 
\begin{tikzpicture}[scale=0.5] \vspace{0.5cm}\hspace{-1.8cm}
\braid[line width=3pt, rotate=90, number of strands=3, 
style strands ={1}{OliveGreen}, style strands={2}{red}, style strands={3}{blue}] (braid) 1 1 1 1 1 1;
\hspace{1.5cm}
\draw[line width=0.5mm] node[below]{\Large{\textit{k}}}(-3,0)   node[above]{$\,\,\,k_0$}(0.5,0) (-3.,0) -- (3.2,0) node[above]{$k_0+2\pi$}(0.5,0)  node[
	sloped,
	pos=1,
	allow upside down]{\arrowIn};
\end{tikzpicture}
\end{minipage}
\begin{minipage}{.49\linewidth} \vspace{-0.3cm}\hspace{-0.9cm}
\includegraphics[width=5.0cm]{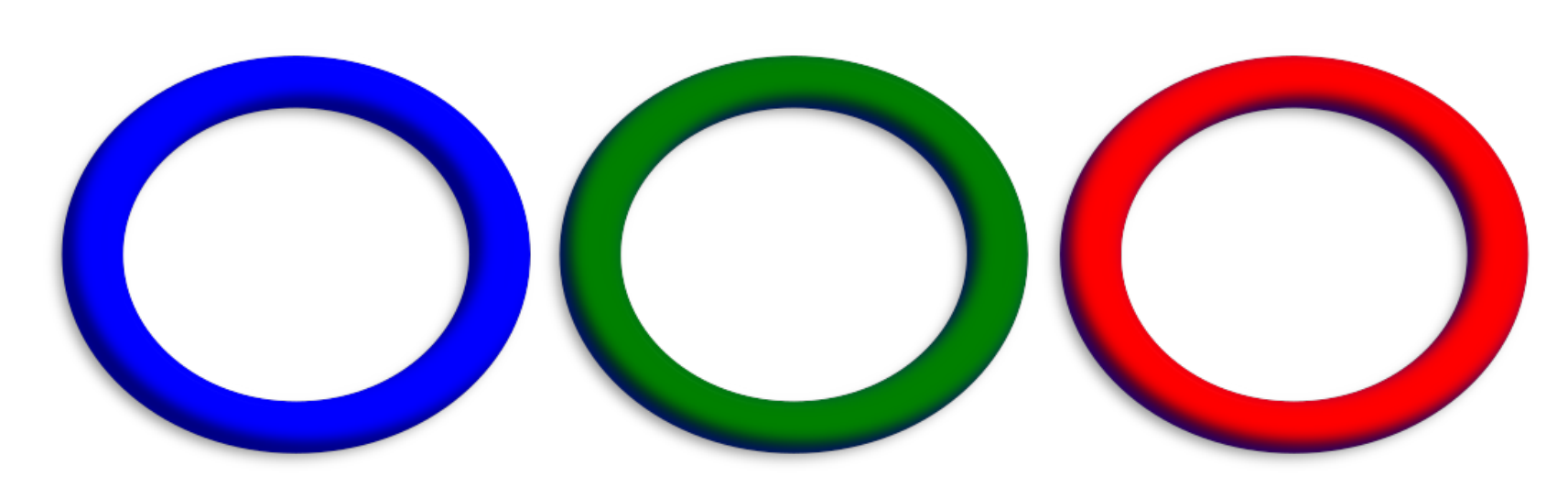}
\vspace{0.2cm}
\end{minipage}
\begin{FlushLeft}
Class V, $\Omega_D>\Omega_4:\,\,\, \nu=0.$ Three unlinks. Braid word: $\mathbb{1}$.
\end{FlushLeft}
\vspace{-0.3cm}
\caption{Different topological classes and their characterization. The braid diagrams are shown on the left. The braiding is performed along the $2\pi$ change of the counting field $k$. Transitions between different braids occur with variation of the $\Omega_D$ parameter. Coincidentally, these diagrams schematically represent the structure of the real parts of the Lindbladian eigenvalues $Re[\lambda_i(k)]$. The equivalent knot representations for each class of the braids are shown on the right. All colors correspond to particular eigenvalues and their $2\pi$ variation over the counting field starting from some initial $k_0$.}  
\label{fig:phases}
\vspace{-0.1cm}
\end{figure}
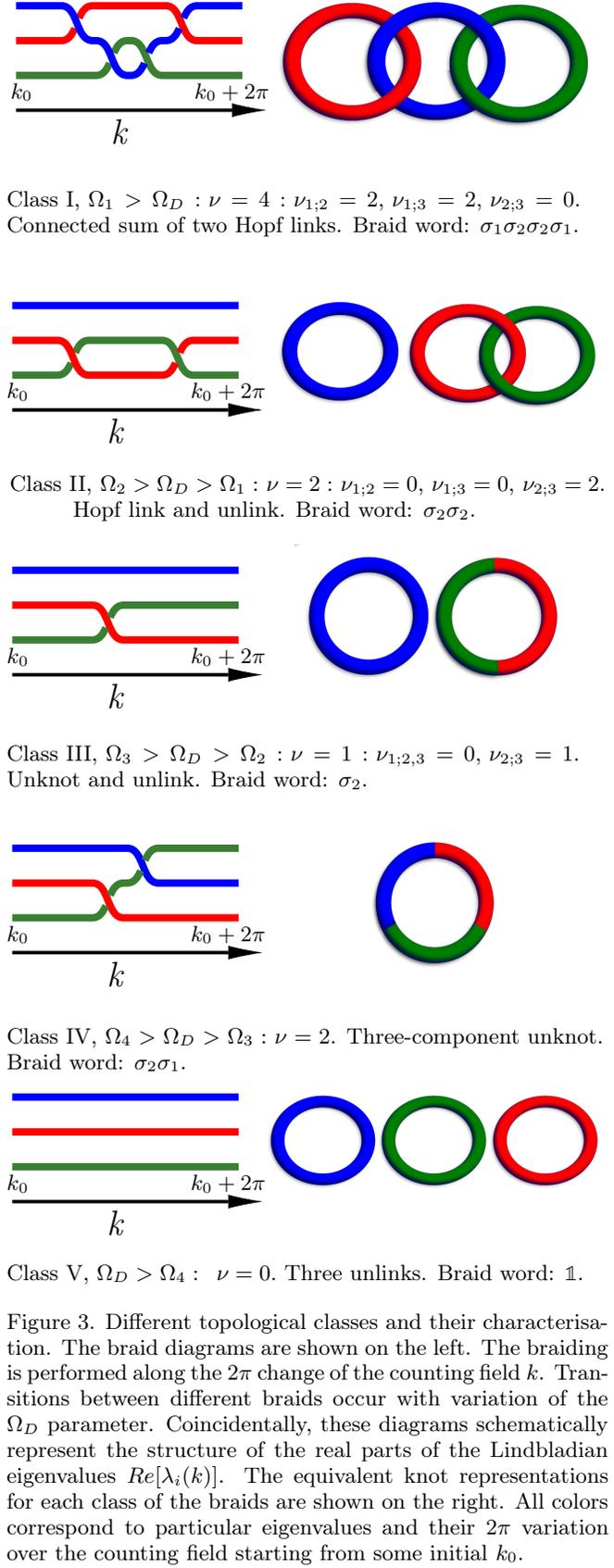
We illustrate the braid structure of the considered eigenvalues in Fig. \ref{fig:3Dstructure}, where the upper panels show how a change in the external parameter $\Omega_D$ results in a change in periodicity of the complex eigenvalues. This change happens through an EP at the particular drive value $\Omega_D=\Omega_3$ (specified below). The lower panels show the real parts of the eigenvalues, corresponding to the eigenstates' decay rates.

Let us introduce the braid index for two Lindbladian complex eigenvalues $\lambda_a$ and $\lambda_b$, similar to the braid index used for characterization of topological bands \cite{Wang2021, Zhang2023}:
\begin{align}
    \nu^{(a,b)}=\int_0^{2\pi}\frac{dk}{2\pi \textit{i}}\frac{d}{dk}\log \left[\frac{\lambda_a(k)-\lambda_{b}(k)}{2}\right].
\end{align}

We denote further
\begin{align}
    &\nu_{a;b}=\nu^{(a,b)}+\nu^{(b,a)};&\\
    &\nu_{a,b;c}=\nu^{(a,c)}+\nu^{(c,a)}+\nu^{(b,c)}+\nu^{(c,b)}; &\\
    &\nu_{a;b,c}=\nu^{(a,b)}+\nu^{(b,a)}+\nu^{(a,c)}+\nu^{(c,a)}; &\\
 \label{nu_tot}   &\nu=\sum_{i<j}\left(\nu^{(i,j)}+\nu^{(j,i)} \right). &
\end{align}

The integer topological invariant $\nu$ counts the total braid degree of the class. This index can coincide for different topologically distinct classes, so other indices allow for more nuanced description. This is aligned with the fact that $B_3$ braid group is non-Abelian and hence cannot be characterized by a single parameter. $\nu_{a;b}$ shows the relative braid degree between eigenvalues $a$ and $b$.
$\nu_{a, b;c}$ shows the braid degree of two eigenvalues $a$ and $b$ with respect to the third eigenvalue $c$ ($\nu_{a;b,c}$ acts in the similar way, showing the braid degree of one eigenvalue with regard to the other two). 
The indices accounting for braiding of more than two eigenvalues become important when periodicity of some eigenvalues differs from $2\pi$ and these eigenvalues braid with $2\pi$-periodic eigenvalues, since $\nu_{a;b}$ is unable to provide a meaningful result in this case accounting for integration over only a part of the full period of the eigenvalues (though this particular situation does not happen in our system for the three chosen eigenvalues).\\
\begin{table}[!t]
    \centering
    \begin{tabular}{c| c | c | c}
   $\,$ class $\,$  & $\,\,\,\nu\,\,\,$ & $\nu_{a;b}$ & braid word \\ \hline
     I & 4 &\, $\nu_{1;2}=2$, $\nu_{1;3}=2$ \,&\, $\sigma_1\sigma_2 \sigma_2\sigma_1$\, \\ \hline
     II & 2 & $\nu_{2;3}=2$ & $\sigma_2 \sigma_2$  \\ \hline
     III & 1 & $\nu_{2;3}=1$ & $\sigma_2$  \\ \hline
     IV & 2 & - & $\sigma_2 \sigma_1$  \\ \hline
     V & 0 & - & $\mathbb{1}$ 
    \end{tabular}
    \caption{Different topological classes corresponding to different braids of eigenvalues, Eq. (\ref{fullL}), with respect to the counting field $k$. Each class is characterized by its topological index $\nu$ and a set of its invariants, which can be related to a unique braid word.}
    \label{tab:phases}
\end{table}
\begin{table}[!t]
    \centering
    \begin{tabular}{c|c|c}
    transition  & $\Omega_{i}$ ($i=1..4$) &\, $k$ \, \\ \hline
     I - II &\, 0.00230 \,&\, 0.02825*\, \\ \hline
     II - III &\, 0.00521 \,&\, 0 \,\\ \hline
     III - IV &\, 0.00779 \,&\, $\pi$ \,\\ \hline
     IV - V &\, 0.03147 \,&\, 2.70492*\,
    \end{tabular}
    \caption{Parameters of the exceptional points. Each EP corresponds to a specific change in topology of the system. There are four EPs separating five different classes. $^*$ denotes that for a given frequency $\Omega_{D}=\Omega_i$, the same exceptional point is present also at $2\pi-k$, since the eigenvalues are $2\pi$-periodic in $k$ and invariant under translations $k\rightarrow -k$. }
    \label{tab:EP}
\end{table}
Naturally, the $B_3$ braid group has infinitely many elements, as there infinitely many non-equivalent braid words. Since we consider a particular model with a fixed Lindbladian, only few of the elements appear in our consideration.
For chosen parameters (see Section \ref{sec:Model}), we identify five different topological classes listed in Table \ref{tab:phases} and shown in Fig. \ref{fig:phases}, where we draw the braid diagrams of each class along with a knot equivalent to this periodic braid diagram. Each color represents variation of a particular eigenvalue over $2\pi$. These classes are separated by four exceptional points listed in Table \ref{tab:EP}. Note that, similarly as it happens with non-Hermitian eigenvalues braids of Bloch bands \cite{Zhang2023}, our braids of eigenvalues can be defined on a period $[k_0, k_0+2\pi]$ of the counting field with an arbitrary initial point $k_0$, so the braid words given in Table \ref{tab:phases} are defined up to cyclic permutations. As one should expect from the topological nature of these classes, they are protected against infinitesimal perturbations of the external parameters of the system. Namely, while exact values of $\Omega_D$ and $k$ are listed in Table \ref{tab:EP}, where the EPs emerge, strongly depend on all parameters' values, the overall structure of these transitions remains robust. For instance, it takes a sufficient change in $\gamma_D$ to remove the class I from the system, though other phases are robust even to a large change of this parameter, as is discussed in more details in Sec. \ref{sec:SD}.
\begin{figure*}
    \begin{minipage}[!ht]{.32\linewidth}
        \centering        \includegraphics[width=1.1\linewidth]{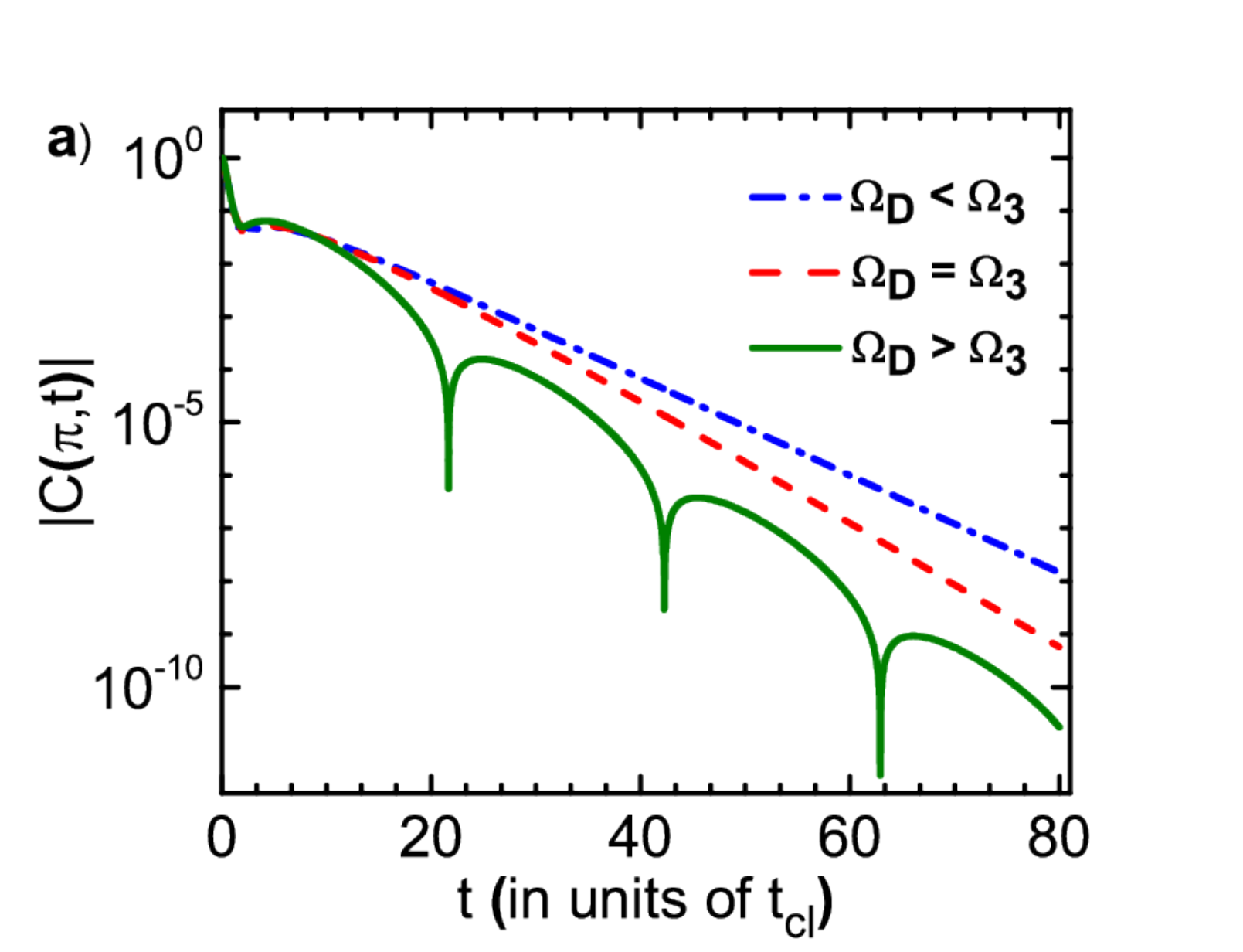}
    \end{minipage}
    \hspace{-0.3cm}
    \begin{minipage}[!ht]{.32\linewidth}
        \centering         \includegraphics[width=1.1\linewidth]{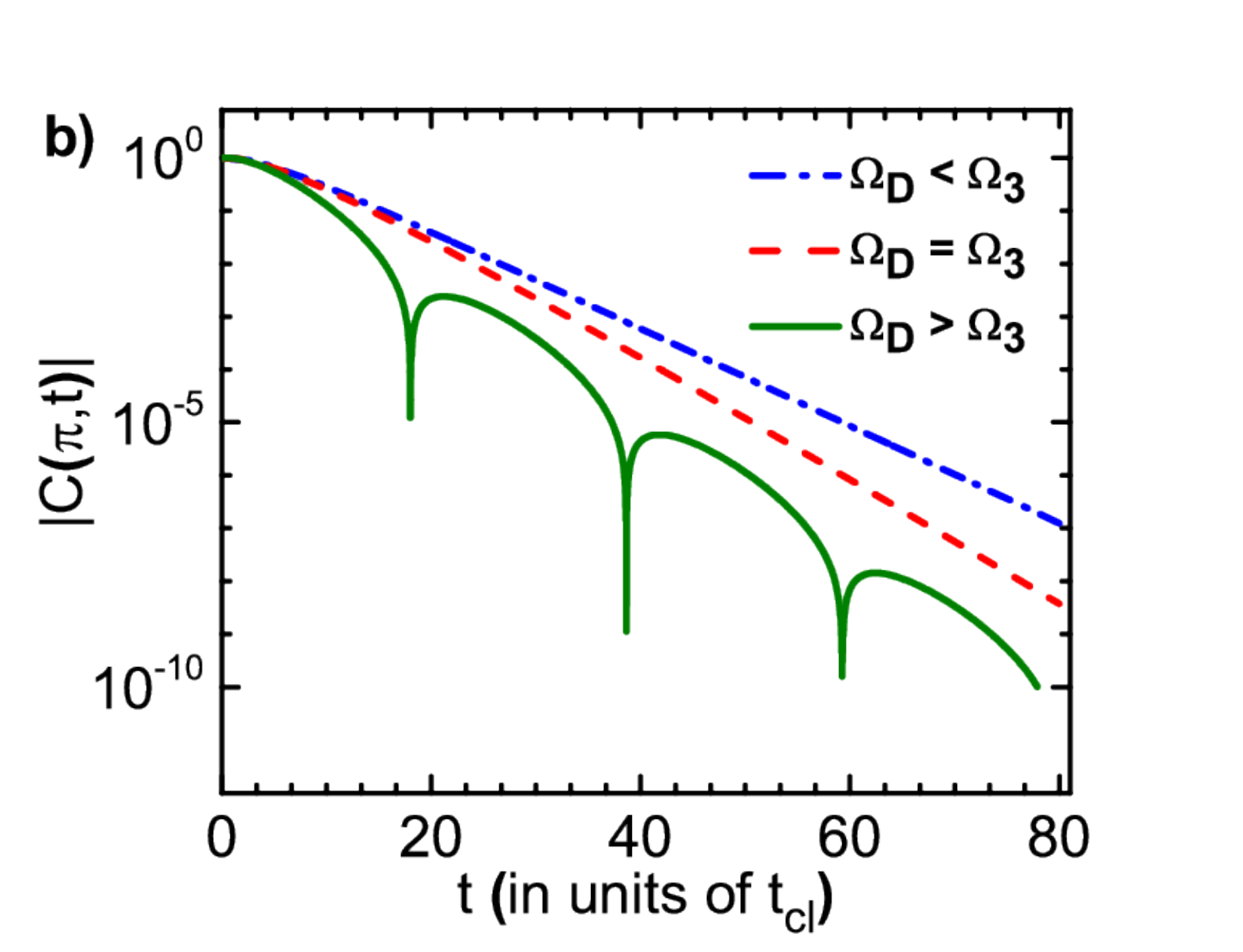}   
    \end{minipage}  
    \begin{minipage}[!ht]{.32\linewidth} 
    \centering       \includegraphics[width=1.1\linewidth]{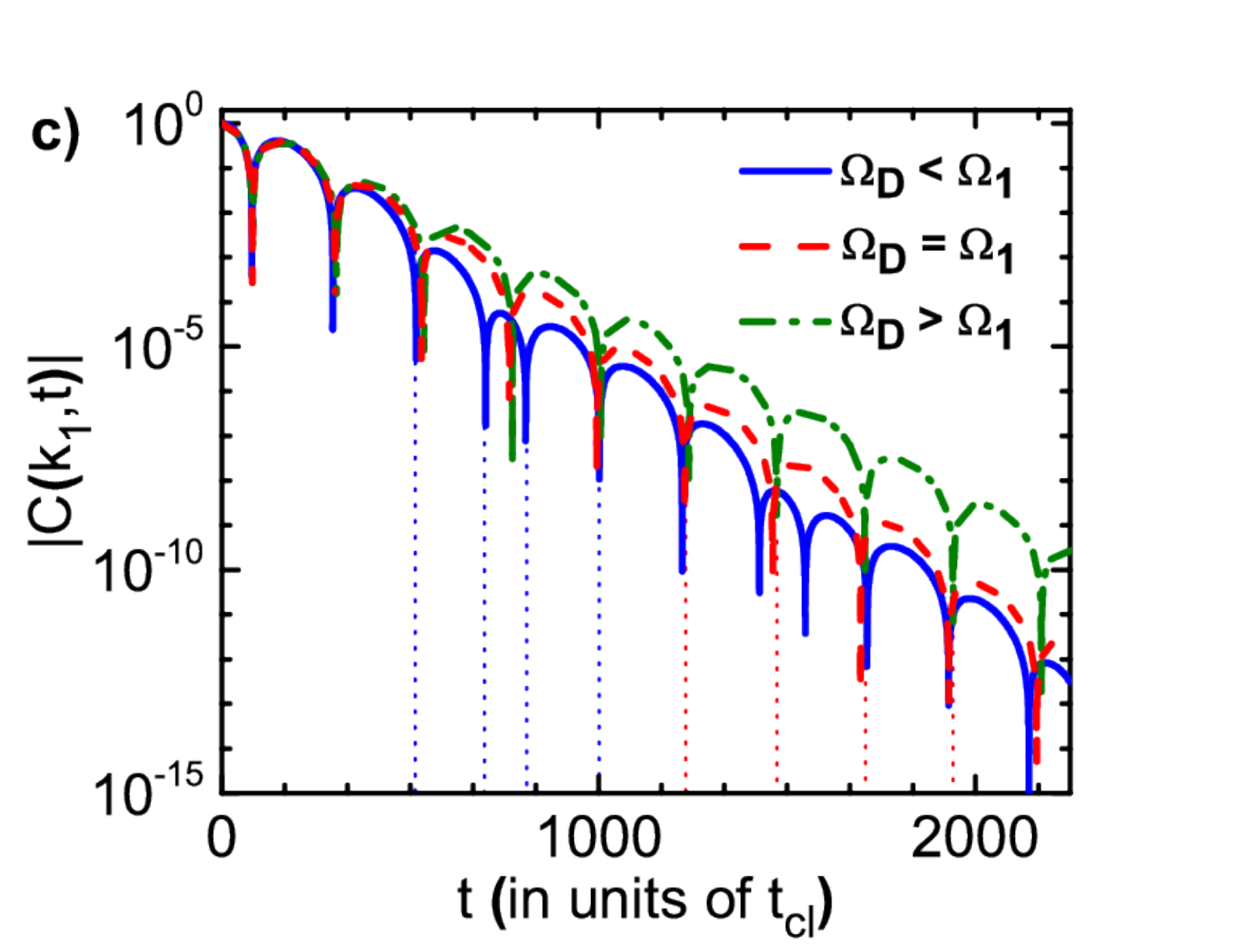}
    \end{minipage}       
   \caption{$C(k, t)$ and transitions of its dynamics corresponding to topological transitions between different topological classes. (a) and (b): $k=\pi$, the transition between $\textit{III}$ and $\textit{IV}$ classes of Table \ref{tab:phases}. Passing through the EP, $C(\pi,t)$ changes its behavior between purely exponential decay and exponential decay with oscillations (note the $\log$ scale). Dash-dotted blue line - $\Omega_D<\Omega_3$ ($\Omega_D=0.0075$), i.e. the system is in the class $\textit{III}$; solid green line - $\Omega_D>\Omega_3$ ($\Omega_D=0.009$), i.e. the system is in the class $\textit{IV}$; dashed red line - $\Omega_D=\Omega_3$, exceptional point. (a): The system is initially prepared in the ground (G) state; (b): The evolution starts from the dark (D) state. (c): Transition between classes $\textit{I}$ and $\textit{II}$. $k_1=0.02825$ is taken from Table \ref{tab:EP}, the system is initially prepared in the ground state. Blue solid line - $\Omega_D<\Omega_1$ ($\Omega_D=0.00225$) ; red dashed line - $\Omega_D=\Omega_1$, green dash-dotted line - $\Omega_D>\Omega_1$ ($\Omega_D=0.00235$). $C(k_1,t)$ changes its behavior at the EP between decay with harmonic oscillations and decay with a beating pattern; the vertical blue and red dotted lines show the change in the oscillations periodicity. Time t is in units of the inverse transition rate to the bright state $t_{cl}=\Gamma_B^{-1}$.}
   \label{fig:dynamicalObs}
\end{figure*}
\section{Observable dynamical properties}
\label{sec:observable}
After we established the topological nature of the considered system, we are interested in identifying a physical observable that is affected by these topological features, so one can identify the occurring topological transitions in experiments.
Usually, braiding occurs when some Hamiltonian parameters are varied (e.g., driving frequencies) so that the direct observation of the associated dynamical phase transitions or encircling the associated EP is possible. In contrast, in our case, there is no way to directly change the counting field $k$ which has a statistical nature. Despite this, we show that one still can choose a time-dependent observable that distinguishes different classes (at least some of them) in experiments.\\
Let us consider the Fourier transform $P_k(t)$ of the probability to observe $n$ quantum jumps over time $t$, which is given by Eq. (\ref{Pkt}).
Expanding in the basis of eigenvectors of $\mathcal{L}_k$, and assuming that we are not at an EP (though we may be infinitesimally close to it), we obtain
\begin{align}
\label{PFourier} P_k(t) = Tr\left[\sum_je^{\lambda_k^{(j)}t}\ket{\chi_k^{(j),R}} \braket{\chi_k^{(j),L}}{\rho_0}
\right],
\end{align}
where $\ket{\chi_k^{(j), R}}$ and $\bra{\chi_k^{(j), L}}$ are the right and left eigenvectors corresponding to the complex eigenvalue $\lambda^{(j)}_k$ of the Lindbladian $\mathcal{L}_k$ given by Eq. (\ref{fullL}) at fixed $k$ (they are well-defined as long as we are not exactly at an EP). Above, we used
\begin{align}
\label{LRbasis}  \braket{\chi_k^{(i),L}}{\chi_k^{(j),R}}=\delta_{ij},\,\,\,\sum_i \ket{\chi_k^{(i),R}} \bra{\chi_k^{(i),L}}=1.
\end{align}
While taking the trace in Eq. (\ref{PFourier}), one should convert $\ket{\chi_k^{(j),R}}$ from superoperator notations back to the matrix representation. In what follows we will investigate 
\begin{align}
\label{Ccorr} &C(k, t)=
Re[P_k(t)]=\left\langle \cos\left(kn\right)\right\rangle(t)\ .
\end{align} 

In analogy to the qualitative analysis of the Lindblad operator in Ref.~\cite{Minev2019}, we introduce the characteristic rate of the 
$G \rightleftharpoons B$ transitions, $\Gamma_B$, so that the characteristic time of one quantum jump (click) observation is given by $t_{cl}=\Gamma_B^{-1}$. The characteristic decay rate of the longest living state is denoted by $\Gamma_D$. One easily obtains
\begin{align}
    \Gamma_B=\frac{\Omega^2_B}{\gamma_B}, \,\,\, \Gamma_D=\frac{\Omega^2_D}{\Gamma_B}=\gamma_B\frac{\Omega^2_D}{\Omega^2_B}.
\end{align}
As we will argue below, our formalism allows observing the EP transitions if the EP involves the longest lived states. In Fig. \ref{fig:dynamicalObs} we show two such transitions.   
The exponential decays present in Fig. \ref{fig:dynamicalObs} behave as $e^{-\alpha \Gamma_D t}$, with $\alpha$ being a non-universal number ${\cal O}(1)$. As we show in Fig. \ref{fig:dynamicalObs}, crossing an EP that involves the longest living eigenstate of the Lindbladian results in a drastic change of the observable $C(k,t)$. The transition $\textit{III-IV}$ shown in Figs. \ref{fig:dynamicalObs}(a) and \ref{fig:dynamicalObs}(b) corresponds to the transition illustrated in Fig. \ref{fig:3Dstructure}, namely, the braid of two eigenvalues with smallest decay rates changes there, changing the periodicity of these eigenvalues. This immediately results in a transition of $C(\pi,t)$ from pure exponential decay (Class $\textit{III}$) into decay with oscillations (Class $\textit{IV}$). The physical meaning of $C(\pi,t)$ can be simply understood.  Since $P_k(t)$ is the Fourier transform of $P_n(t)$, which is the probability to observe $n$ jumps over time $t$, $C(0,t)=1$ has a simple interpretation: $P_{k=0}(t)=\sum_{n=0}^{\infty}P_n(t)=1$, namely, the probabilities to observe all numbers of quantum jumps are summed into $1$. At $k=\pi$, $C(\pi, t)$ also has a straightforward interpretation
\begin{align}
    P_{k=\pi}(t)=\sum_{n=0}^{\infty}e^{\textit{i}\pi n}P_{n}(t)=\sum_{n=even}P_n(t)-\sum_{n=odd}P_n(t),
\end{align}
i.e., it is the difference between probabilities to observe even and odd number of jumps over time $t$.\\
In class $\textit{IV}$, $C(\pi,t)$ slowly changes its sign, so at different times it is more probable to observe either even or odd number of quantum jumps. This nontrivial structure of the quantum jump probabilities originates from the non-Poissonian distribution of quantum jumps.  We analyze this distribution in more details in Appendix \ref{sec:QJ}. In short, the interplay between $G \rightleftharpoons B$ and $G \rightleftharpoons D$ transitions, which independently would be Poissonian and characterized by $\Gamma_B$ and $\Gamma_D$ rates, creates this effective non-Poissonian distribution with parity of jumps that oscillates in time. Further, Fig. \ref{fig:dynamicalObs}(c) shows the $\textit{I}- \textit{II}$ transition from Table \ref{tab:EP}. Note that in this transition none of the involved eigenvalues changes its periodicity. Nevertheless, the braid of the eigenvalues changes, and the changed topology (namely, a transition between a connected sum of two Hopf links to a Hopf link with an unlink, specified in Fig. \ref{fig:phases}) manifests in the change of the behavior of $C(k,t)$. At the EP, in addition to a usual exponential decay with the $\Gamma_D$ rate, the time evolution of $C(k,t)$ switches between harmonic oscillations and a beating pattern.\\ 
Note that $C(k,t)$ does not require any postselection, as all occurred jumps contribute to this observable. Nevertheless, $C(k, t)$ at $k=\pi$ and $k_1$ (given in the first row of Table \ref{tab:EP}) identify the EP transitions at arbitrary long times. Practically, the visibility of these transitions decrease, as $C(k,t)$ decays exponentially with time, nevertheless, the EP-related dynamics always dominates this observable.
Other transitions listed in Table \ref{tab:EP} cannot be clearly observed through the dynamical behavior of $C(k,t)$, as they involve states with higher decay rates and the EP dynamical transition becomes masked by the dynamics of the longer-lived states, the same way as it normally happens for the Lindbladian dynamics at $k=0$. So, in this case the EP-related behavior can give only exponentially small subleading contribution to $C(k,t)$. In Sec. \ref{sec:SD}, we show that, nevertheless, there is an innate connection between the ``unobservable'' EP
at $k=0$ and the one at $k=\pi$ which we analyzed above.  

\section{Retrieving the topology from experimental jump distributions}
\label{sec:Exp}
\begin{figure}[!t]
    \centering    \includegraphics[width=1.\linewidth]{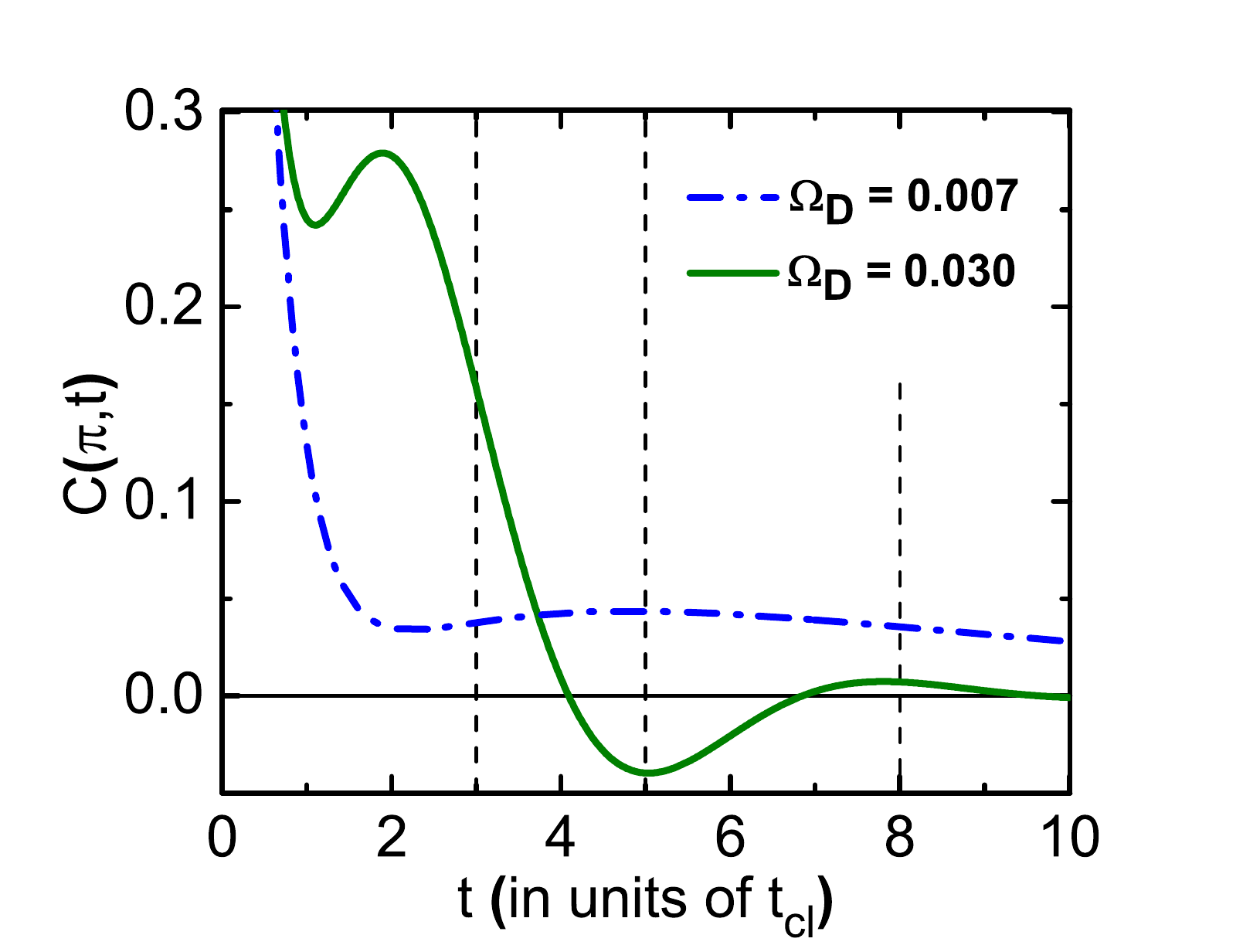}
    \caption{$C(\pi, t)$ on two sides of the $\textit{III}-\textit{IV}$ transition. Dotted-dashed blue line: pure decay at $\Omega_D<\Omega_3$ ($\Omega_D=0.007$); green solid line: oscillatory behavior at $\Omega_D>\Omega_3$ ($\Omega_D=0.030$). The initial state is G.}
    \label{fig:JhistC}
\end{figure}
\begin{figure*}[!ht]
    \begin{minipage}[t]{1.\linewidth}
    \includegraphics[width=0.32\linewidth]{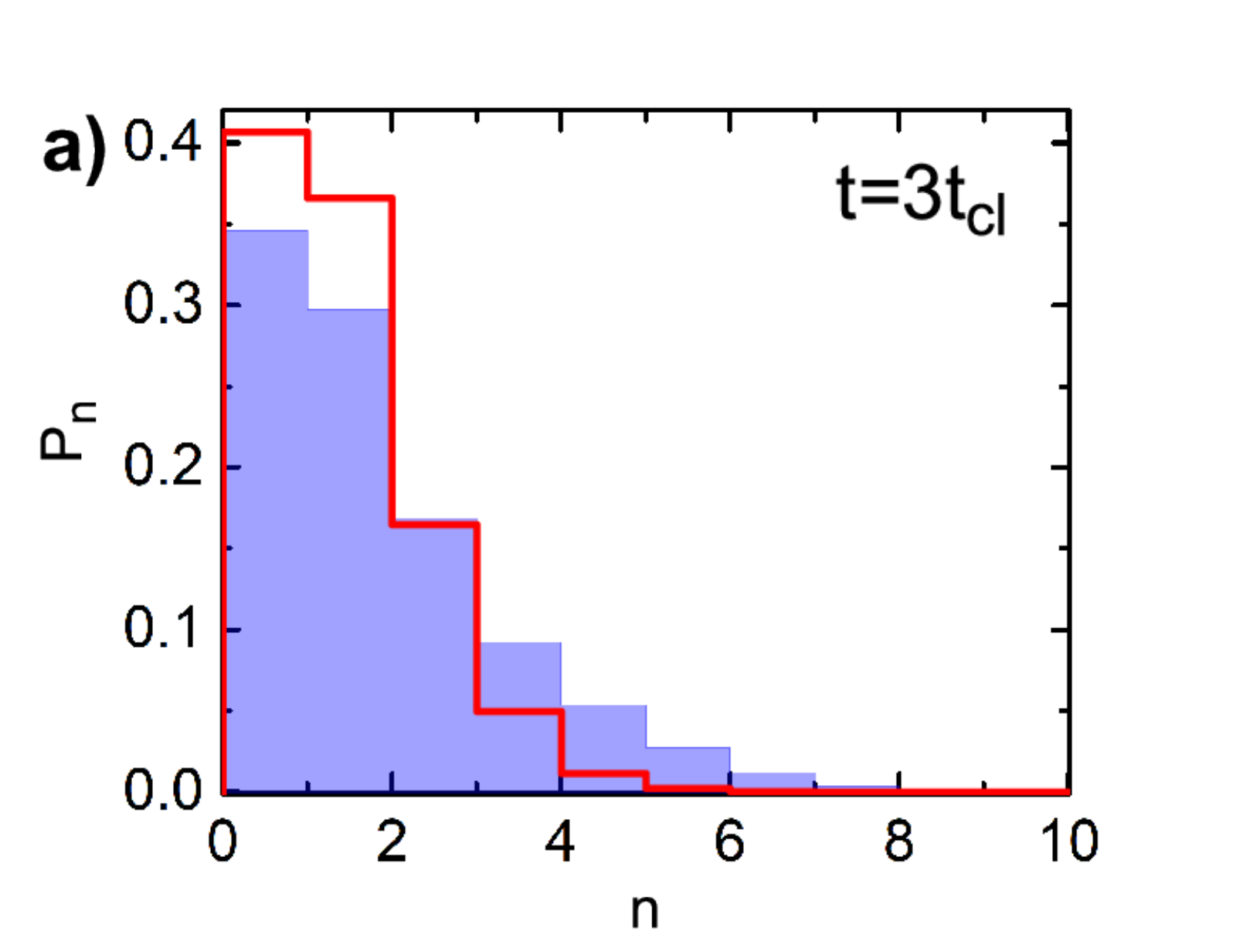}
    \includegraphics[width=0.32\linewidth]{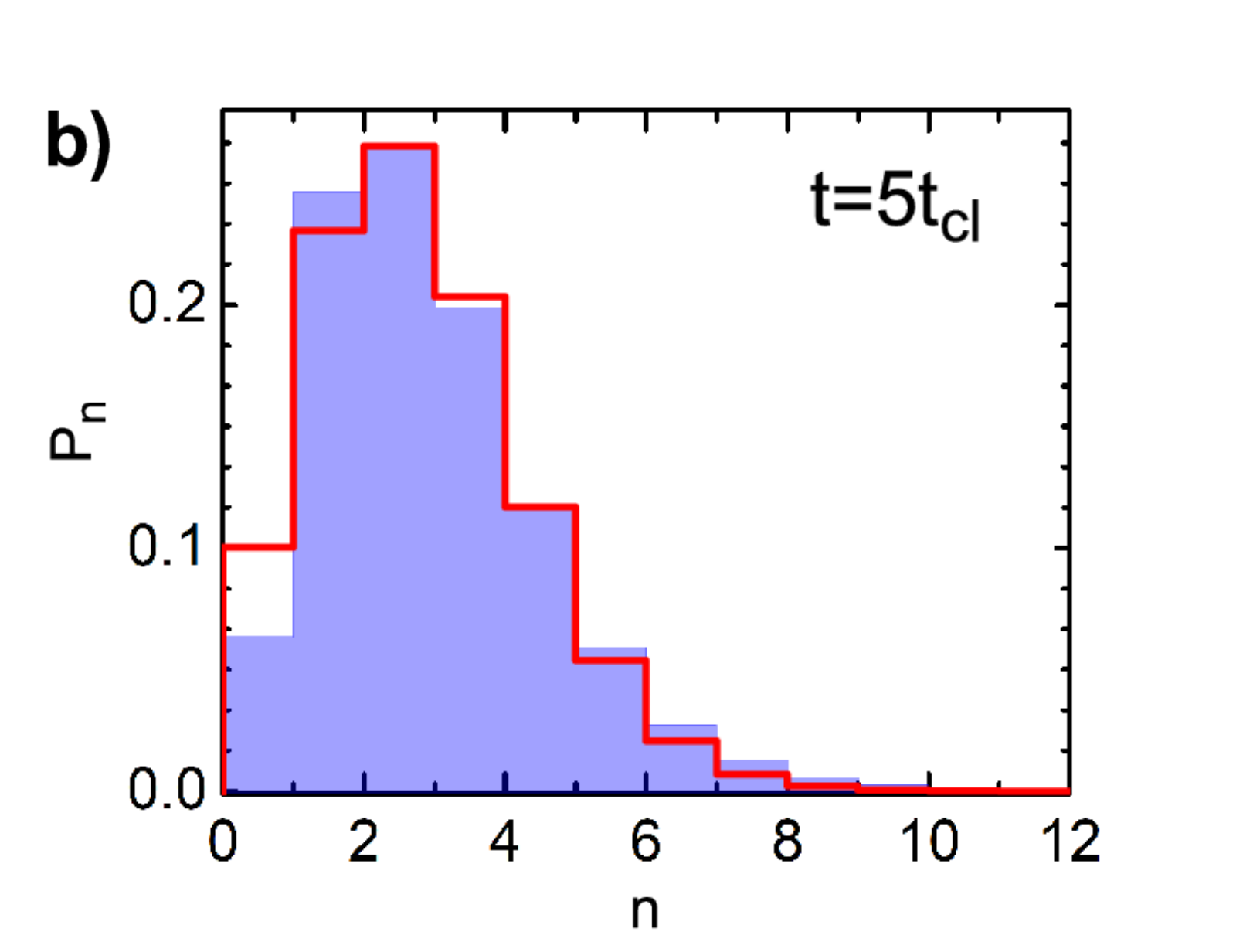}
    \includegraphics[width=0.32\linewidth]{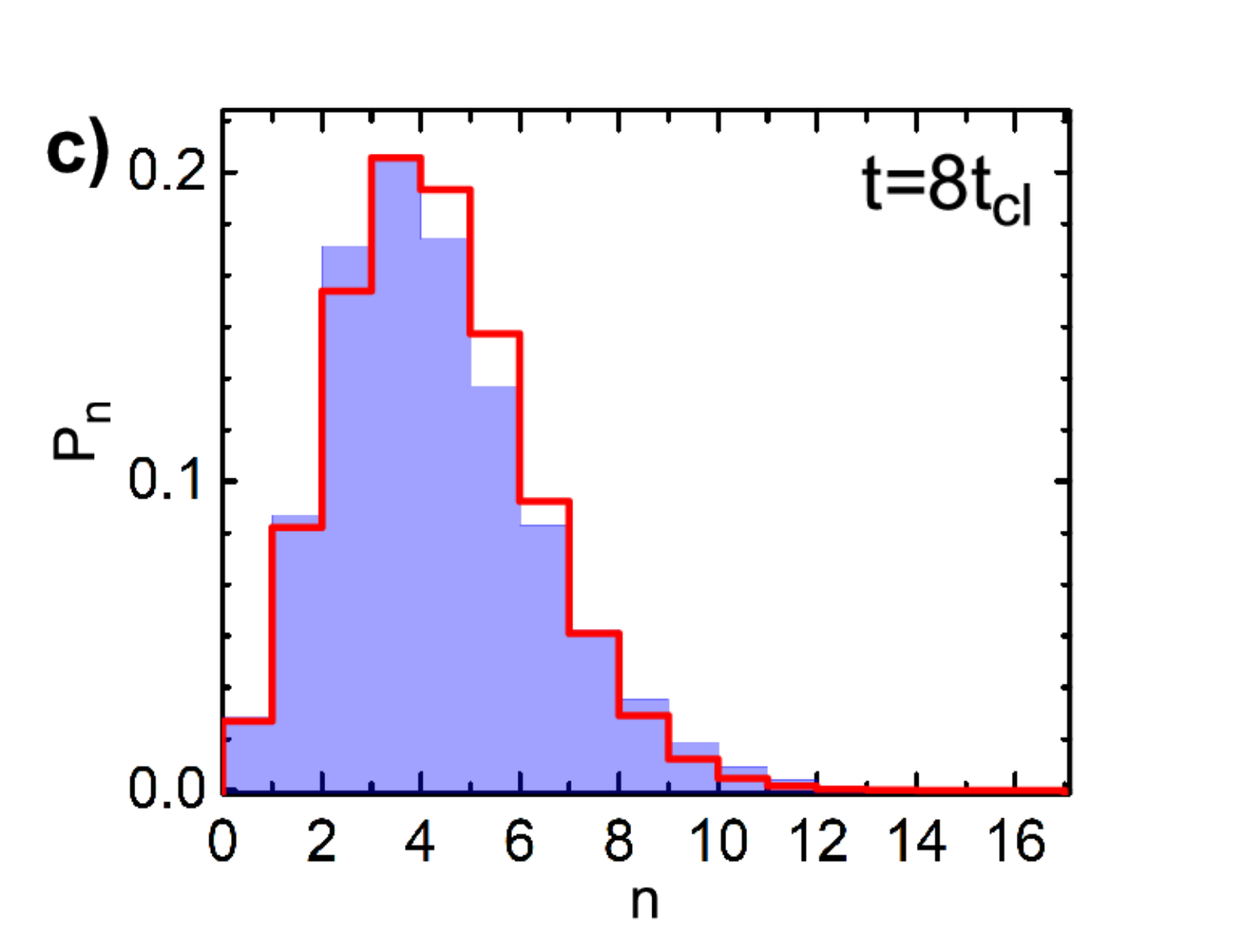} 
    \includegraphics[width=0.32\linewidth]{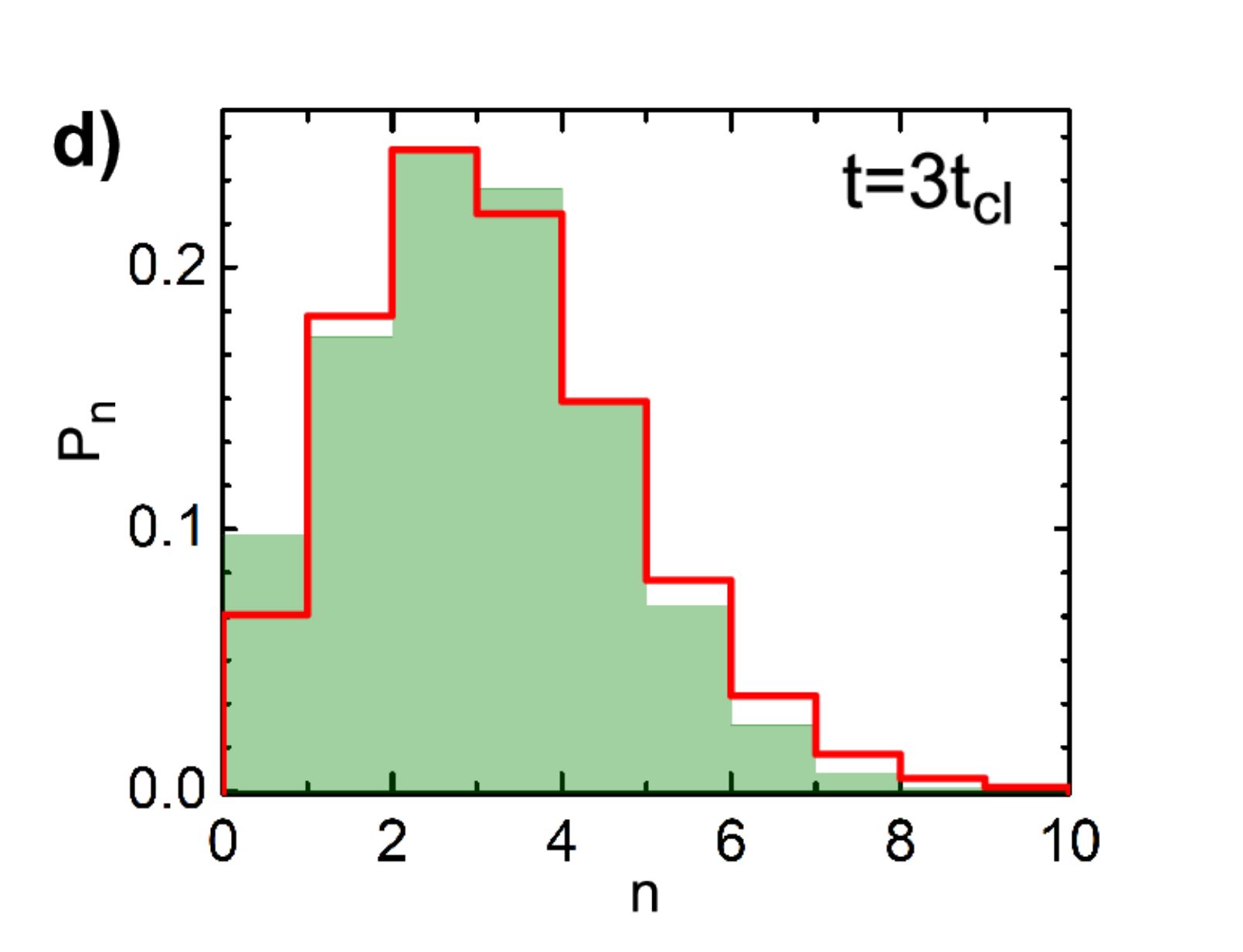}
    \includegraphics[width=0.32\linewidth]{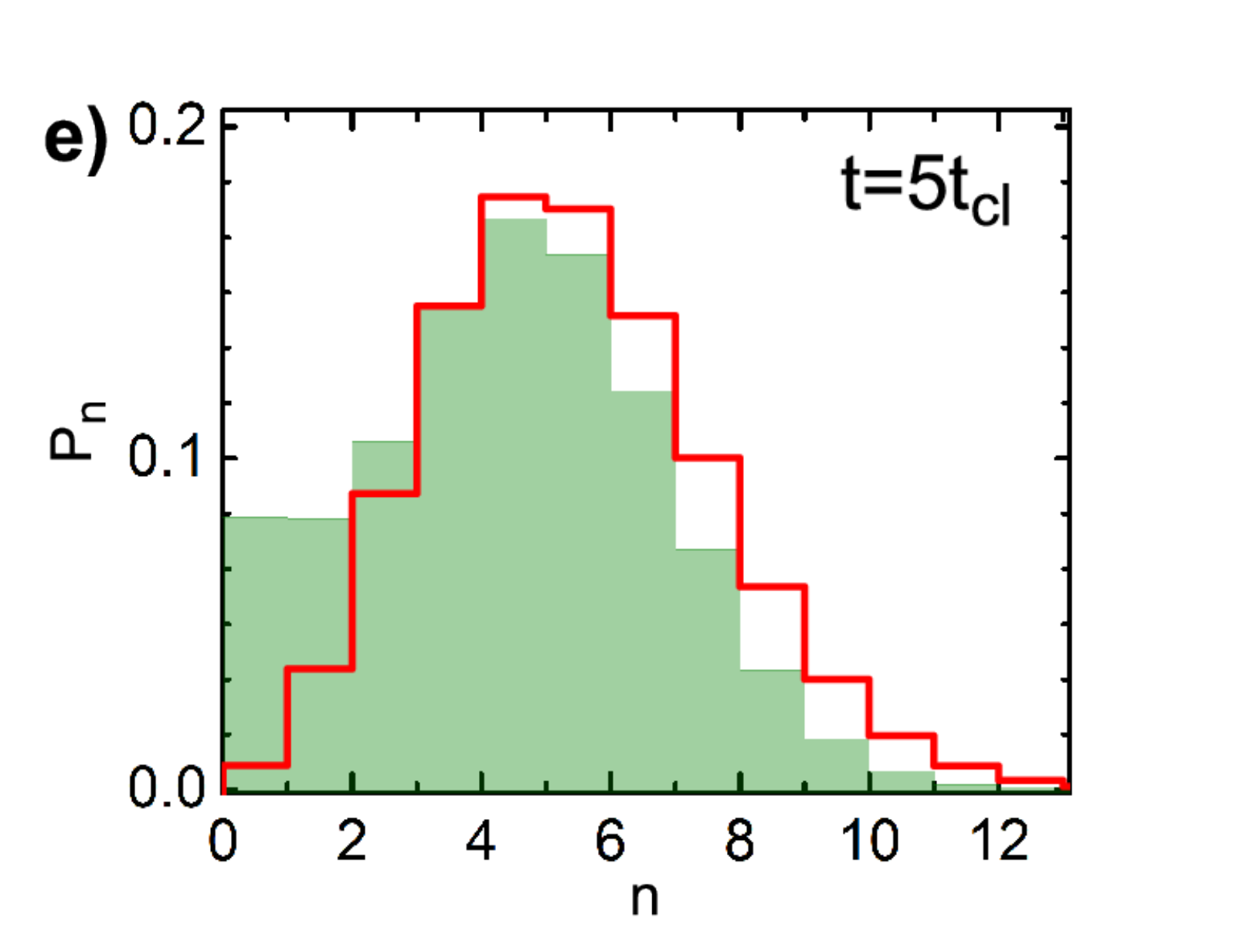}  
    \includegraphics[width=0.32\linewidth]{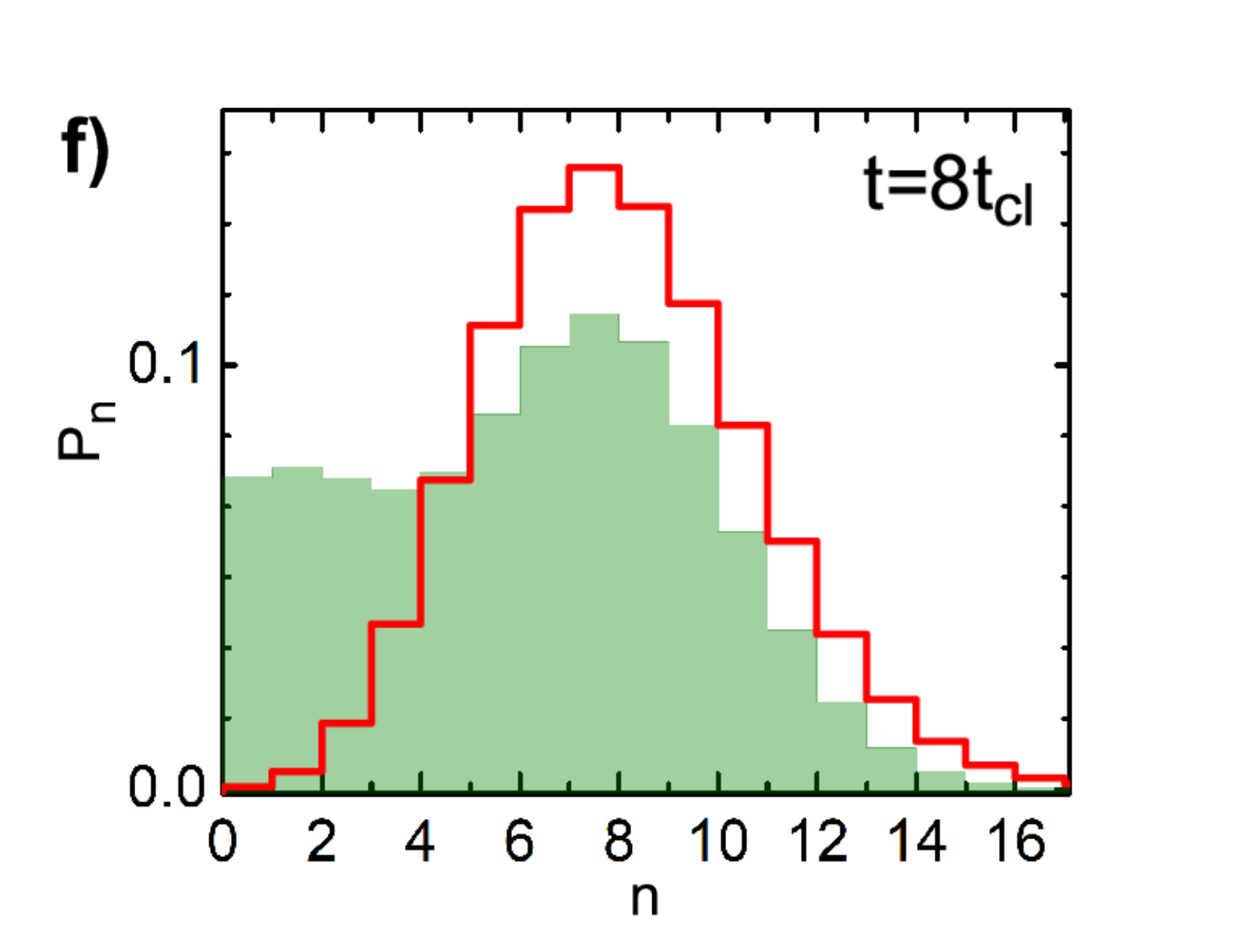}  
    \end{minipage}
    \caption{Distributions of quantum jumps at fixed times $3t_{cl}$ (a), (d), $5t_{cl}$ (b), (e), $8t_{cl}$ (c), (f) (vertical dashed lines in Fig. \ref{fig:JhistC}). (a)-(c): $\Omega_D=0.03\gg \Omega_3$, oscillatory regime. (d)-(f): $\Omega_D=0.007<\Omega_3$, pure decay. Red histograms show closest Poissonian distributions.}
    \label{fig:Jhist}
\end{figure*}
\begin{figure*}[!ht]
    \hspace{-1cm}
    \begin{minipage}[t]{.49\linewidth}
    \centering    \includegraphics[width=1.2\linewidth]{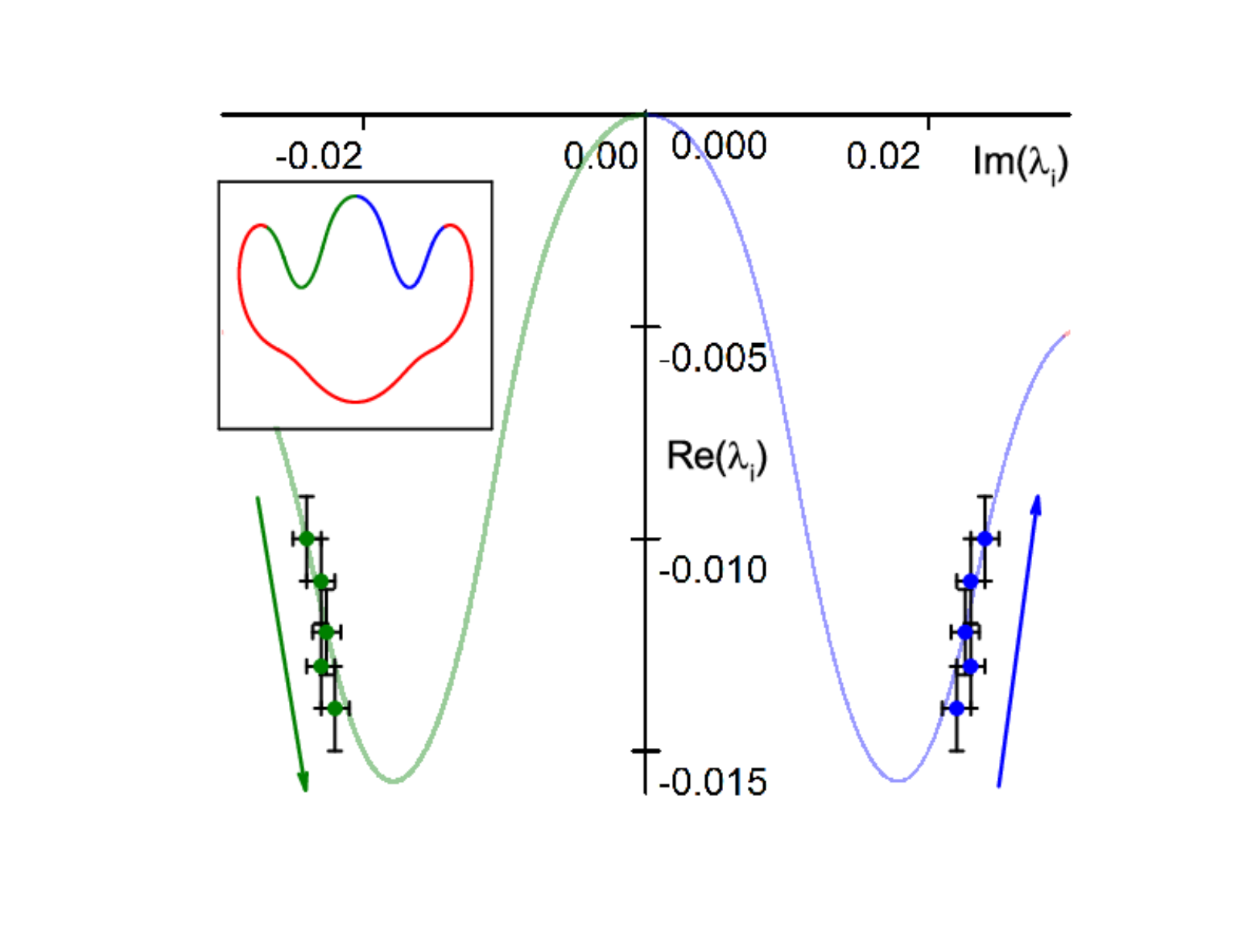}
    \end{minipage}
    \begin{minipage}[t]{.49\linewidth}
    \centering
    \includegraphics[width=1.2\linewidth]{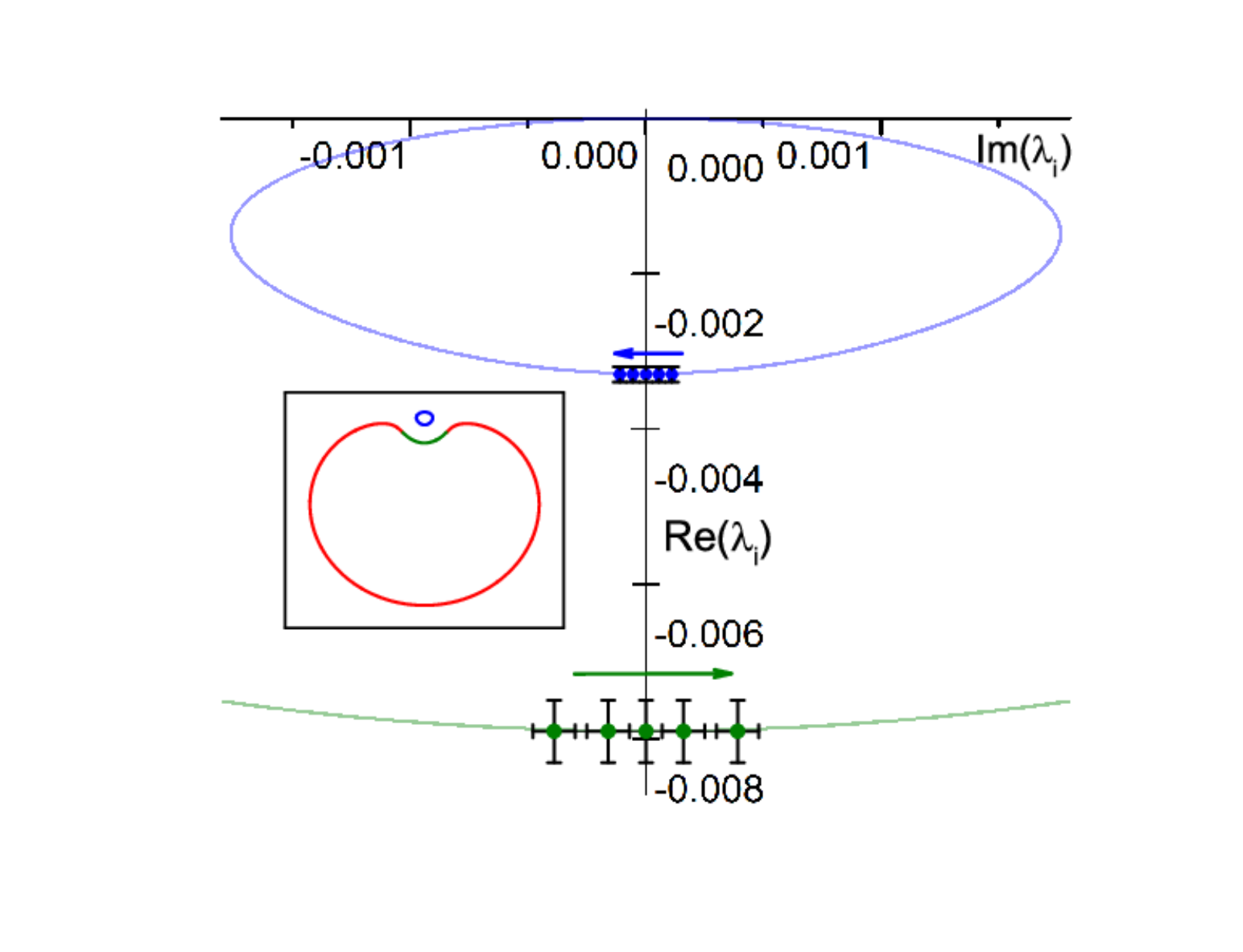}   
    \end{minipage}
    \caption{Parametric plots of imaginary ($x$ axis) and real ($y$ axis) parts of eigenvalues $\lambda_k^{(1)}$, $\lambda_k^{(2)}$ depending on $k$. Left panel: $\Omega_D=0.03$; right panel: $\Omega_D=0.007$. Blue and green dots are fits for $P_k(t)$ through Eqs. (\ref{LogPk}) and  (\ref{loglogP}). Background solid blue and green lines: exact values for these eigenvalues. Values of $k$ are $\pi-0.4, \, \pi-0.2,\, \pi,\, \pi+0.2,\, \pi+0.4$, arrows show movement directions of the eigenvalues with increase of $k$. Insets: The same, zoomed-out, parametric plots showing topological structures for the three leading eigenvalues as $k$ changes from $0$ to $2\pi$.}
    \label{fig:parpl}
\end{figure*}
In this section we discuss in more details how 
a change of the underlying topology of the system can be spotted from $P_k(t)$, that is obtained from experimentally gathered distributions of quantum jump probabilities.
The Lindbladian eigenvalues separated from the longest-living state by a nonzero Liouvillian gap give exponentially small contributions to Eq. (\ref{Pkt}) comparing to the leading terms. As a result, restoring all the eigenvalues from
the jump distributions gathered during the experiment would require fitting $P_k(t)$ by nine complex decaying exponents and nine complex prefactors, which does not seem plausible. Nevertheless, one can still distinguish different topological structures
of Fig. \ref{fig:phases} whenever at least the two longest decaying eigenstates are topologically different. 
We focus on the transition between classes $\textit{III}$ and $\textit{IV}$, as the most robust case. This particular transition is also most suited for experimental studies since it always happens at the fixed $k=\pi$ (as we discuss in the next section), rather than at some accidental value of $k$. We choose the ground state G ($\rho_0 = \ket{G}\otimes\ket{G}$) as the initial state. This is a reasonable choice, since immediately after each jump the system is reliably in the G state. Every such point can be chosen as the starting time for the observations. \\
Since many quantum jumps contribute to the observable $C(\pi,t)$ and the resulting sum decays in time exponentially, from the practical perspective, the observation time should be kept relatively small ($t/t_{cl}\sim 10^0-10^2$). 
We illustrate it in Fig. \ref{fig:JhistC}, which reflects the same transition as in Fig. \ref{fig:dynamicalObs}(a), namely, the regime $\Omega_D=0.03$ which belongs to class $\textit{IV}$ far away from the exceptional point [so the oscillations of $P_k(t)$ are easier to spot] and the regime $\Omega_D=0.007$ of Class $\textit{III}$, where the sign of $P_k(t)$ is always positive. 
Away from the EP, the difference of imaginary parts of the eigenvalues forming that EP grows, increasing the oscillations' frequency, and making them easier to spot. In Fig. \ref{fig:Jhist}, we plot the distributions of quantum jumps taken at three different times, corresponding to vertical dashed lines in Fig. \ref{fig:JhistC}. It is worth noting here that in experiments with superconducting qubits such statistics can be easily gathered for an arbitrary number of points in time and with at least $\sim 10^5$ repetitions \cite{Spiecker2024} at each point in time, so the finite size effects are almost negligible for the histograms in Fig. \ref{fig:Jhist}. 
We compare these distributions with best fit Poissonian distributions (red lines) to highlight the non-Poissonian character of jumps in both regimes. In the oscillatory regime (upper panels), despite the fact that the distributions follow the overall shape of the Poissonian distribution, they are qualitatively different. Indeed, small deviations from the Poissonian statistics accumulate to the oscillatory behavior of the staggered probability $C(\pi,t)$. In contrast, the Poissonian statistics always produces monotonous positive staggered probabilities $C^{P}(\pi,t)$. In particular, we have $C(\pi, 3t_{cl})\simeq 0.158$, $C(\pi, 5t_{cl})\simeq -0.040$, $C(\pi, 8t_{cl})\simeq 0.007$, and $C^P(\pi,3t_{cl})\simeq 0.165$, $C^P(\pi, 5_{t_{cl}})\simeq 0.010$, $C^P(\pi, 8t_{cl})\simeq 0.005$. In the pure decay regime ($\Omega_D<\Omega_3$), the distributions are clearly non-Poissonian, as they are biased towards small $n$ contributions. Nevertheless, this regime always produces positive staggered probabilities $C(\pi,t)$: $C(\pi, 3t_{cl})\simeq 0.038$, $C(\pi, 5t_{cl})\simeq 0.043$, $C(\pi, 8t_{cl})\simeq 0.035$. 

We now discuss how the complex eigenvalues of the Lindbladian can be retrieved from the quantum jump probability distributions collected at different times. 
Taking the Fourier transform of the distributions $P_n(t)$ with respect to $n$, we obtain the probabilities $P_k(t)$ depending on the counting field $k$. For $k=\pi$, they reduce to functions plotted in Fig. \ref{fig:JhistC}. Using Eq.  (\ref{PFourier}), we can write

\begin{align}
    \label{PkFit}
    P_k(t)=\sum_iC^{(j)}_ke^{\lambda_k^{(j)}t},
\end{align}
where $C_k^{(j)}$ are complex constants that depend on the eigenvectors and the initial conditions. We neglect all contributions from the fast decaying modes and approximate this expression as
\begin{align}
  \label{PkAppr}
  P_k(t)\simeq C^{(1)}_ke^{\lambda_k^{(1)}t}+C^{(2)}_ke^{\lambda_k^{(2)}t}.
\end{align}
Assuming that the second exponent has a larger negative real part, we see that the first term dominates at longer times, so we can approximate
\begin{align}
    \label{LogPk}
    \log P_k(t)\simeq \log C^{(1)}_k+\lambda_k^{(1)}t+\frac{C_k^{(2)}}{C_k^{(1)}}e^{(\lambda_k^{(2)}-\lambda_k^{(1)})t}.
\end{align}
\begin{figure*}[!ht]
    \hspace{-1cm}
    \begin{minipage}[t]{.49\linewidth}
    \centering    \includegraphics[width=1.2\linewidth]{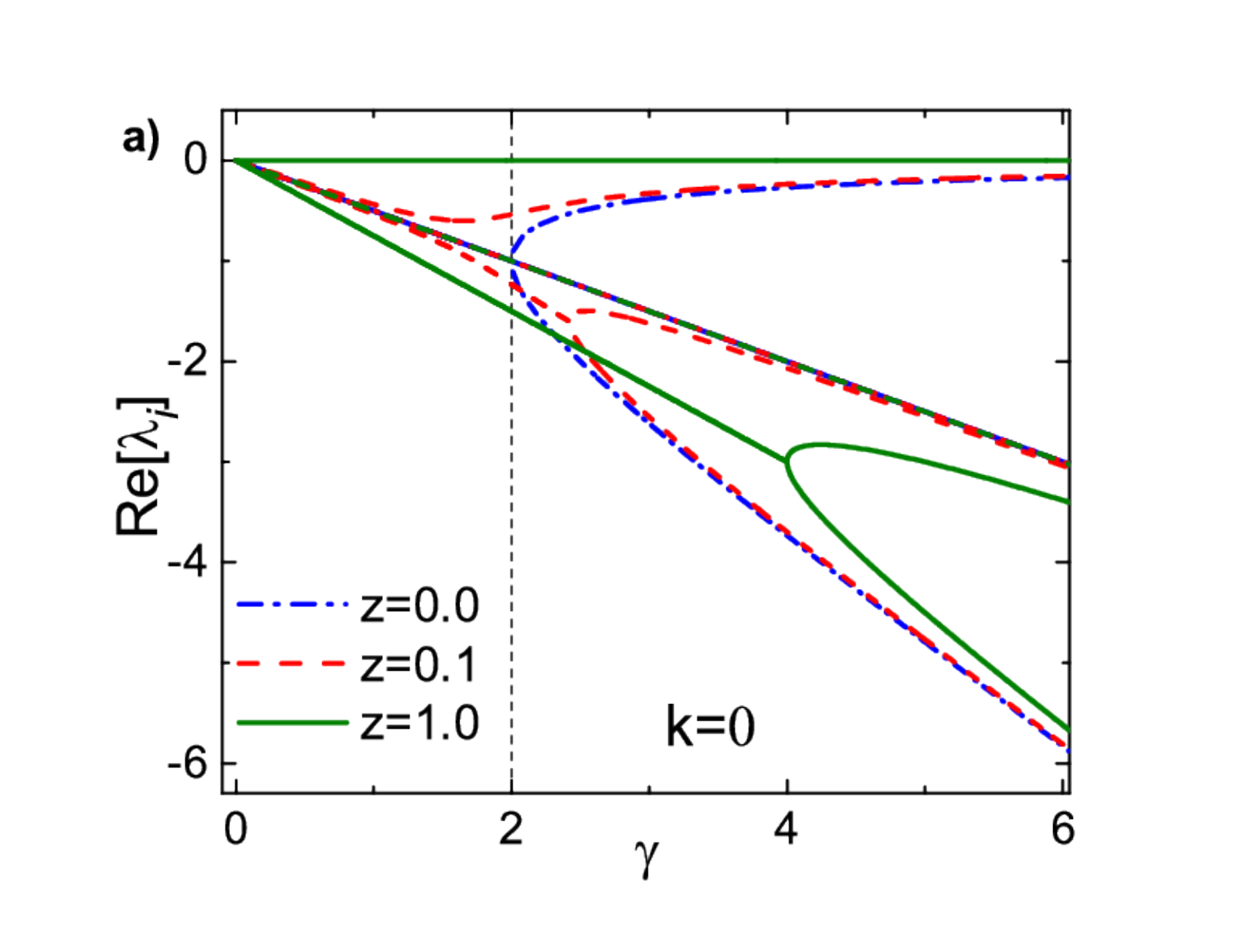}
    \end{minipage}
    \begin{minipage}[t]{.49\linewidth}
    \centering
    \includegraphics[width=1.2\linewidth]{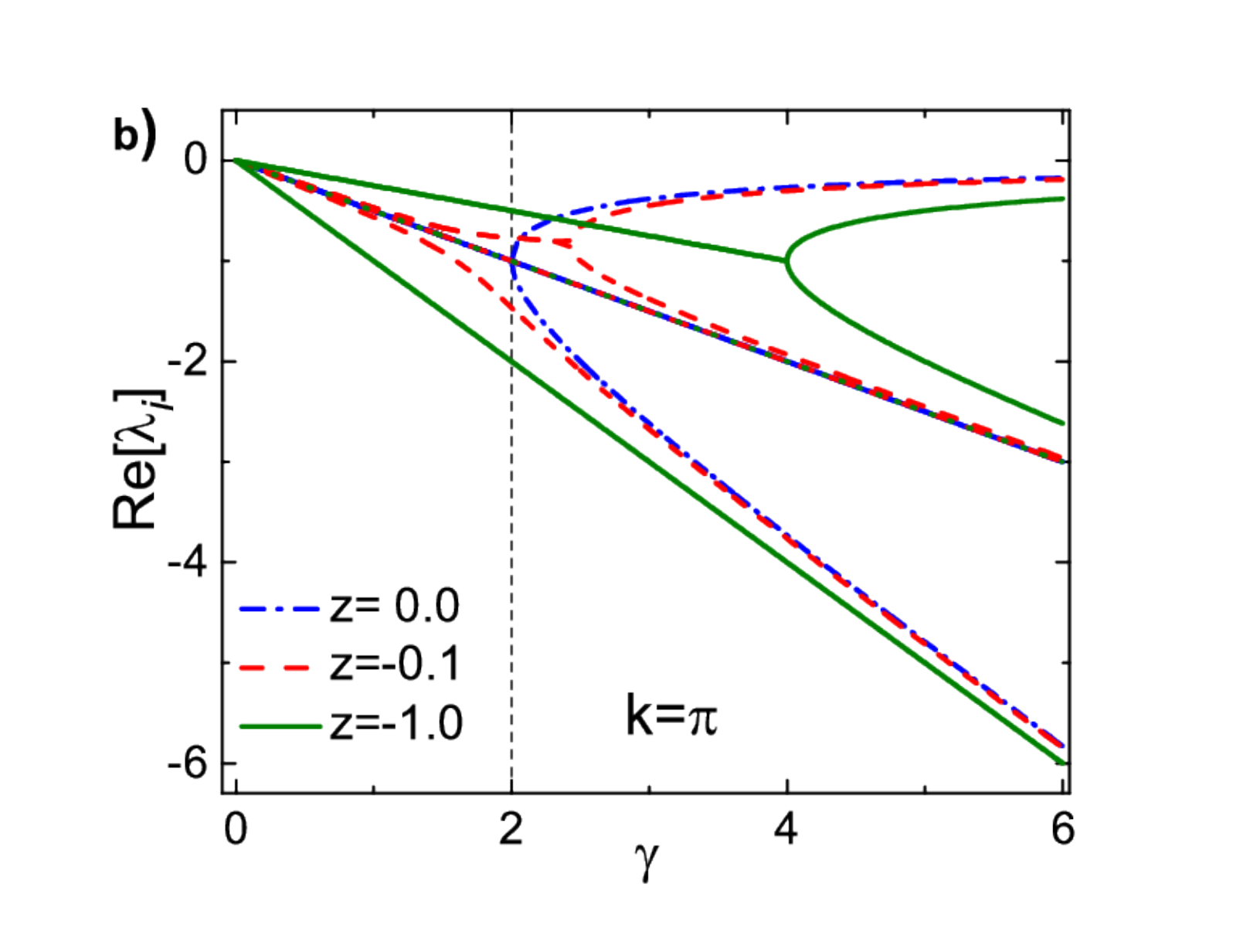}   
    \end{minipage}
    \caption{Exceptional point of the Lindbladian $L_0$ perturbed by the jump term $L_J$, Eq. (\ref{L0J}) (this particular example is given by Eq. (\ref{ModelEff}) with phases (a) $k=0$ and (b) $k=\pi$. Real parts of the Lindbladian eigenvalues $\lambda_i$ are plotted as functions of the parameter $\gamma$. $z=re^{\textit{i}k}$. Blue dotted-dashed line is the unperturbed exceptional point structure $r=0$, red dashed line corresponds to the suppressed jumps case $r=0.1$, green solid line denotes the full jump term $r=1$. Vertical dashed black line highlights the critical value of $\gamma$ for the unperturbed exceptional point.}
    \label{fig:pert}
\end{figure*}
The last term is exponentially small comparing to the linear one, so we can neglect it for now and approximate $\log P_k(t)$ by a linear fit, finding real and imaginary parts of $\lambda_k^{(1)}\pm\delta \lambda^{(1)}_k$ and $C_k^{(1)}\pm \delta C_k^{(1)}$ for a chosen $k$. $\delta \lambda_k^{(1)}$ and $\delta C_k^{(1)}$ are small errors accumulated during this procedure. Next, we find the difference between two eigenvalues, present in the exponent Eq. (\ref{LogPk}), by subtracting the established leading contributions
\begin{align}
 \label{loglogP}    \log \left[ \log \frac{P_k(t)}{C_k^{(1)}}-\lambda_k^{(1)}t\right]\simeq\log \frac{C_k^{(2)}}{C_k^{(1)}}+(\lambda_k^{(2)}-\lambda_k^{(1)})t\\
     \label{loglogPErr}      +\frac{\delta C_k^{(1)}+\delta\lambda_k^{(1)}t }{\log P_k(t)-\log C_k^{(1)}-\lambda_k^{(1)}t}-\delta\lambda_k^{(1)}t-\frac{\delta C_k^{(1)}}{C_k^{(k)}}.
\end{align}
Here Eq. (\ref{loglogP}) allows finding $\lambda_k^{(2)}$ from a linear fit of the left-hand side, while Eq. (\ref{loglogPErr}) accounts for errors accumulated through this procedure due to $\delta \lambda_k^{(1)}$ and $\delta C_k^{(1)}$, while it leads to increased errors in $\delta \lambda_k^{(2)}$ and $\delta C_k^{(2)}$. We implement this procedure and plot the retrieved values $\lambda_k^{(1)}$ and $\lambda_k^{(2)}$ for various $k$ in Fig. \ref{fig:parpl} for both regimes to illustrate that it indeed allows distinguishing the underlying topology of the eigenvalues. The background lines in Fig. \ref{fig:parpl} are parametric plots for real and imaginary parts of the eigenvalues (we plotted the zoomed-in area which involves the transition between two leading eigenvalues), same as the eigenvalues shown in Fig. \ref{fig:3Dstructure}. Discrete points with error bars show the eigenvalues obtained from the fitting procedure explained above. The insets show the parametric plots with full structures of all three eigenvalues, corresponding to Fig. \ref{fig:3Dstructure}, for which the zoom-ins are plotted (axes are not shown).

\section{Duality between Exceptional Points in Full Counting Statistics}
\label{sec:SD}
As proven in \cite{Minganti2019}, the stationary state of the full Lindbladian operator cannot participate in the formation of an EP. This theorem relies on the fact that the trace of this state must be conserved. However, this reasoning can no longer be applied to the Lindbladian taken at nonzero counting field $k\neq 0$, since none of its eigenstates preserves the full trace anymore. We show in this section that in some cases one can use the so-called staggered cumulants \cite{Ivanov2010} with $k=\pi$, or more generally $C(k,t)$, to unveil EPs which are normally hidden in the fast decaying modes at $k=0$. This duality can be established analytically for two-level systems in the following way. \\
Let us consider an arbitrary 
two-level system with a non-Hermitian Hamiltonian $H_{nH}=H-\frac{i}{2}\Gamma^{\dagger}\Gamma$ (here $H$ is Hermitian and $\Gamma$ is an arbitrary dissipator) which depends on a parameter $\gamma$ (e.g., it may be included in the dissipator $\Gamma$) and exhibits an exceptional point at some critical value of this parameter, $\gamma_C$. We introduce a modified Lindblad operator that takes form
\begin{align}
\label{L0J} &\mathcal{L}=L_0+z L_J,& \\
\label{L0def} &L_0= -\textit{i}\left\{\left(H-\frac{\textit{i}}{2}\Gamma^{\dagger}\Gamma\right)\otimes \mathbb{1}-\mathbb{1}\otimes \left(H^T+\frac{\textit{i}}{2}\Gamma^T\Gamma^*\right) \right\}. &
\end{align}
Here $L_0$ is the Lindbladian without quantum jumps, $L_J=\Gamma \otimes \Gamma^*$ is the jump term of the Lindbladian (here we use the superoperator notations), $z$ is a complex parameter with $|z|\in [0,1]$ that suppresses the weight of the jump contribution. $z=0$ corresponds to the purely Hamiltonian evolution without quantum jumps, $z=1$ restores the full Lindbladian evolution. $0<|z|<1$ corresponds to a hybrid-Liouvillian with suppressed contribution of quantum jumps that accounts for an imperfect detector \cite{Minganti2020}. Additionally, this parameter can have a nonzero phase, which corresponds to the nonzero counting field, similar to Eq. (\ref{fullL}). Let us put $|z|\ll 1$, so the jump term $L_J$ in the Lindbladian Eq. (\ref{L0J}) can be considered as a small perturbation for $L_0$. The second-order EP of the non-Hermitian Hamiltonian $H_{nH}$ occurring at some parameter's value $\gamma_C$ translates into the third-order EP of the Lindbladian $L_0(\gamma_C)$, as shown in Appendix \ref{sec:EP2}.

The effects of a small perturbation on the Lindbladian eigenvalues in the vicinity of the third-order EP can be tracked analytically. Following the analysis of \cite{Seyranian2003, Heiss2008, Demange2011}, we show in Appendix \ref{sec:EP2} how a general perturbation at $k=0$ splits three degenerate eigenvalues into a pair forming a perturbed EP of the second order and an isolated eigenvalue that does not participate in any EP. This isolated eigenvalue is the longest living, i.e., it has the smallest negative real part (smallest decay rate) of all eigenvalues, and becomes stationary in the limit $z=1$. By changing the sign of the jump term (i.e., adding the counting field with $k=\pi$), one inverses this order. We stress out that this behavior of EP perturbations is general for dissipative two-level systems and does not rely on any additional assumptions.
\begin{figure}[!t]
    \centering    \includegraphics[width=1.1\linewidth]{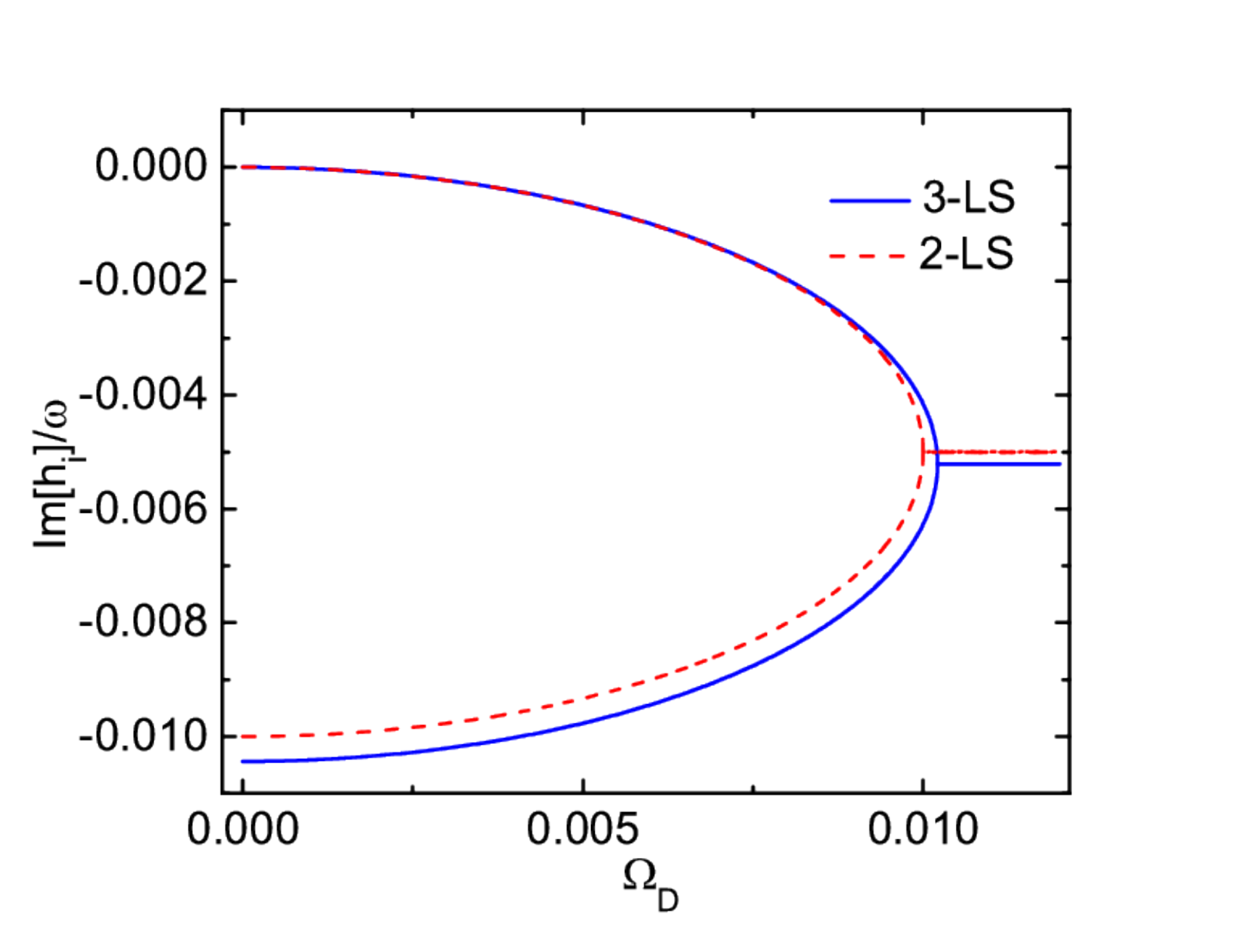}
    \caption{Eigenvalues $h_i$ (imaginary parts) of the three-level system Hamiltonian Eq. (\ref{Heff}) (blue solid lines, 3-LS) and eigenvalues of the effective G-D two-level system's Hamiltonian obtained via adiabatic elimination (red dashed line, 2-LS) as functions of the driving frequency $\Omega_D$. For the three-level system, only two relevant eigenvalues (long-living states forming an exceptional point) are plotted. The reduced system retains the EP of the same structure in close proximity to the original EP.}
    \label{fig:Hreduced}
\end{figure}

We illustrate this principle for a dissipative two-level system
\begin{align}
\label{ModelEff}    H_{nH}=\frac{\omega}{2}\sigma^x-\textit{i}\frac{\gamma}{2}\sigma^+\sigma^-, \,\,\, L_J=\gamma \sigma^-\otimes \sigma^-,
\end{align}
which was used in \cite{Minganti2020} to study effects of perturbations on the EP by partially suppressed quantum jumps for the case of zero phase $k=0$. The behavior of the eigenvalues is shown in Fig. \ref{fig:pert}(a). On the contrary, the order of eigenvalues is inverted at $k=\pi$, so the same Liouvillian exceptional point forms now between two least decaying states, as shown in Fig. \ref{fig:pert}(b). This allows observing the EP in the decaying modes, which is normally hidden by the stationary mode. We stress out that although at $k=\pi$ the eigenstates hosting the EP are exponentially decaying, their decay is slowest comparing to all other modes, so they give the leading contribution to $C(k,t)$ (with other modes decaying exponentially faster), and transitions of the $C(k,t)$ dynamics [Eq. (\ref{Ccorr})], similar to those in Fig. \ref{fig:dynamicalObs}, can be observed.  
\begin{figure}[!t]
    \centering     \includegraphics[width=1.1\linewidth]{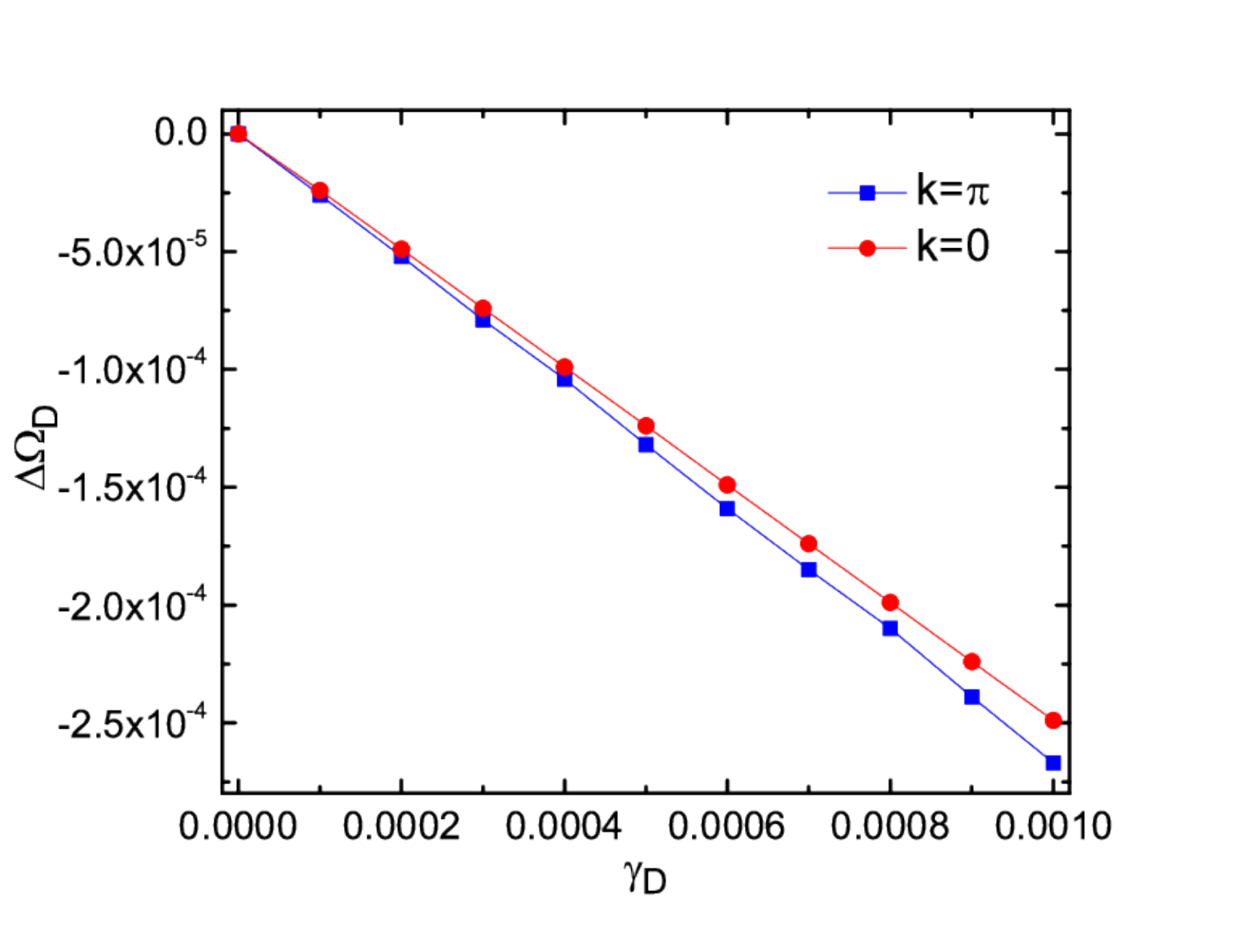}
    \caption{Change in the critical driving frequency $\Omega_D$ for the EP of the full Lindbladian Eq. (\ref{fullL}) as a function of the dark stay decay $\gamma_D$ for $k=0$ (blue squares) and $k=\pi$ (red dots). The $y-$axis shows change in this frequency $\Delta\Omega_D=\Omega_D(\gamma_D)-\Omega_D(0)$, where the critical driving frequencies $\Omega_D(0)$ for the corresponding $k$ are given in Table \ref{tab:EP}.} 
    \label{fig:gammaD}
\end{figure}

For a system with more than two states, the situation becomes more complicated and the $0 - \pi$ duality of the EPs in different modes, strictly speaking, does not necessarily hold. Nevertheless, when the non-Hermitian Hamiltonian of the system hosts a second-order EP in its two least decaying states and all other states are substantially separated from them, one can use the adiabatic elimination method \cite{Shnirman1998, Makhlin2000, Brion2007, Mirrahimi2009, Reiter2012, Azouit2017, Finkelstein2020} to reduce the problem to an effective two-level system preserving the EP. 
We discuss this procedure in details in Appendix \ref{sec:AE}. For the three-level system under consideration, with one state exhibiting very fast decay, the reduced two-level system faithfully reproduces the EP of the non-Hermitian Hamiltonian Eq. (\ref{Heff}), as illustrated at Fig. \ref{fig:Hreduced}. The same procedure can be performed on the full Lindbladian, effectively reducing the description from nine to four eigenstates. The quantum jump term of the full Lindbladian perturbs the reduced Lindbladian eigenvalues in a nontrivial way. Nevertheless, a general expansion in terms of the jump amplitude discussed in Appendix \ref{sec:EP2} still holds, so at least for $|z|\ll 1$ case the stricture of the perturbed eigenvalues must be preserved. As we show numerically in Appendix \ref{sec:AE}, the perturbed eigenvalues still exhibit the $0-\pi$ duality in $k$ for the EPs, though now they do not necessarily emerge at the same values of the driving frequency $\Omega_D$. This duality is the reason why in Table \ref{tab:EP} there is a pair of EPs corresponding to $\textit{II}-\textit{III}$ and $\textit{III}-\textit{IV}$ transitions exactly at $k=0$ and $k=\pi$. This is illustrated in Fig. \ref{fig:gammaD}, where we consider various dark state decay rates $\gamma_D$ - though the change of this parameter shifts the frequencies $\Omega_D$ for the EP, the values $k$ of the two EPs remain fixed - there are always EPs at $k=0$ and $k=\pi$. This is not true for two other EPs listed in Table \ref{tab:EP}, where the values of $k$ are incidental and very sensitive to any changes in the parameters of the problem. For instance, the class $\textit{I}$ (and the subsequent transition $\textit{I} - \textit{II}$) completely disappears for substantially large $\gamma_D$ values, so the transitions that do not involve changes of the eigenvalues periodicity show lesser resilience to perturbations.

\section{Discussion and perspectives}
\label{sec:conslusion}
We have analyzed the dissipative Lindbladian dynamics of a V-shaped three-level system comprised of the ground, dark, and bright states with monitored $B \rightarrow G$ quantum  jumps. By drawing an analogy with topological band theory, we identify distinct topological classes of FCS,  corresponding to different braids of complex eigenvalues of the Lindbladian.
The qualitatively different quantum jump distributions of these classes are topologically protected, as infinitely small perturbations of the system's parameters do not affect the braid structure. It takes a finite change of the system's parameters (comparable to the scale of the distance between the eigenvalues) to change the integer braid index of the system and induce a topological transition.
Transitions between the topological classes necessarily happen through exceptional points emerging at specific values of the counting field. 

We have introduced the dynamical observable $P_k(t)$ which allows to identify transitions between different topological classes at arbitrary times as long as the corresponding EP involves the eigenvalues with the smallest decay rates. Such analysis of the EPs at nonzero values of the counting field $k$ provides an exception to the theorem prohibiting formations of EPs in the Lindbladian eigenvalues that dominate at long times \cite{Minganti2019}.
Some of these EPs are dual, i.e., they are located at special values of $k$ and necessarily emerge in pairs ($0$ - $\pi$). This duality allows for observation of the EP at $k=\pi$ at long times. Moreover, it allows for establishing an immediate relation to the $k=0$ EP of the conventional trace-preserving Lindbladian, which is concealed at long times by the presence of a trace preserving eigenstate. This approach has an advantage of neither requiring any postselection techniques nor monitoring particular eigenstates nor employing other highly selective measurements of the system, relying solely on counting the number of quantum jumps as a function of time. 

Following the measurements, the histograms describing the numbers of quantum jumps that occur over a given time can be analyzed by means of the Fourier transform. This analysis provides meaningful information about the underlying EP structure and the topological class of the system. In other words, by appropriately fitting $P_k(t)$ as a function of time, one can restore the leading eigenvalues and their topological structure. We have demonstrated that for realistic, experiment-limited data sets, such retrieval of the topological structures can be done with reasonable accuracy provided sufficient resolution in time and sufficiently long-time statistics of the measurements. Both of these requirements are not a problem for state-of-the-art experimental setups of superconducting qubits, where quantum jump traces can be collected at high measurement rates and up to arbitrarily long times over multiple experimental repetitions. Nevertheless, as large numbers of quantum jumps contribute to the $P_k(t)$ observable at long times, the noise and measurement imperfections have higher impact on the precision of the fit in this case. Hence, it is preferable to retrieve the topological structures of the Lindbladian eigenvalues not too close to the exceptional points, so that the distinct topology can be clearly seen already at relatively early observation times.

We argue that the developed approach to the dynamics of quantum jumps is quite general and can be applied to an analysis of \textit{any} system exhibiting quantum jumps. It would be interesting to see how this approach can be related to other monitored quantum system, e.g., \cite{Snizhko2020, Poperl2024}, and how EP structures of the counting fields alter with change of monitoring protocols \cite{Minganti2022}. In this work we concentrated on the second-order EPs, as they are most common among all EP types. The overall reasoning about transitions in topological structures and the dynamics of the system should be applicable to higher-order EPs and even to exceptional lines and exceptional surfaces. If there are several types of quantum jumps in a system, one can introduce several different counting fields and employ the same analysis for higher-dimensional topological bands. 

Our approach connects dynamical phase transitions to underlying topological transitions in a wide class of dissipative systems. It may turn out to be particularly useful for quantum state-engineering \cite{Verstraete2009} and for understanding emergent non-Poissonian errors in quantum processors, which is crucial for error correction protocols \cite{Fowler2012, Caruso2014, Martinis2015}. In many instances, the noise acting on qubits in quantum computers may be approximated by the Pauli-Lindblad model \cite{vandenBerg2023, Koyluoglu2024}, so errors occurring due to such noise are mimicked by Lindbladian jump terms \cite{Guimaraes2023, Guimaraes2024}, with dissipators modeled from the experimentally identified noise channels. The effective Lindbladians obtained within such an approach can be further investigated for the underlying topology when counting fields are inserted into their jump terms. Understanding the Lindbladian eigenvalues' topology in such systems may allow controlling transitions between Poissonian and non-Poissonian distributions of quantum jumps. Namely, one can focus on suppressing particular error channels that favor topological phases corresponding to non-Poissonian jumps. Hence, potential opportunities emerge for topologically protecting the Poissonian structure of quantum jumps (and therefore ensuring the uncorrelated nature of errors) against emergent correlated errors.
\section*{Acknowledgments}
The initial stages of this work were motivated by the Master's thesis of Jonathan Daniel. 
We are thankful to Mathieu Féchant, Philipp Lenhard, Ioan Pop and Martin Spiecker for discussions on experimental observations of quantum jumps. We thank Mikhail Kiselev for a discussion on full counting statistics in mesoscopic transport.\\
This work was supported by the DFG Grant SH 81/8-1.
A.I.P. and A.S. were supported by the German Ministry of Education and Research (BMBF) within the project QSolid (FKZ: 13N16151). Y.G. was supported by the DFG grant EG 96/13-1 and NSF-BSF 2023666. Y.G. acknowledges support as an incumbent of InfoSys chair.

\appendix
\section{Effective three-level system}
\label{sec:RWA}
The system under consideration can be constructed with two physical qubits. We use Pauli matrices $\vec{\sigma}$ to denote the first qubit, and $\vec{\tau}$ to denote the second qubit. Their frequencies are $\varepsilon$ and $\omega$ correspondingly. The qubits are coupled longitudinally with the coupling constant $\chi$,
\begin{align}
 \label{H0}  H_0=\frac{1}{2}\left(\varepsilon\sigma^z+\omega\tau^z+\chi \sigma^z\tau^z\right).
\end{align}
We encode the four states written in $\{\ket{\sigma, \tau}\}$ basis as
\begin{align}
 \label{fstates}   \ket{G}=\ket{\downarrow\downarrow}, \, \ket{D}=\ket{\uparrow\downarrow}, \,
    \ket{B}=\ket{\downarrow\uparrow}, \, \ket{F}=\ket{\uparrow\uparrow}.    
\end{align}
The state $\ket{F}$ is an auxiliary one, corresponding to both qubits being excited. In principle, it can be used to account for all higher energy excited states of the two qubit system in experimental realizations \cite{Minev2019}.
The eigen-energies of the Hamiltonian (\ref{H0}) (counting from the ground state, so $E_G\equiv 0$) are given by
\begin{align}
   E_D=\varepsilon-\chi, \, E_B=\omega-\chi, \, E_F=\varepsilon+\omega. 
\end{align}
One applies harmonic Rabi drives with frequencies $\varepsilon-\chi$ and $\omega-\chi$ to induce Rabi transitions $\ket{G}\rightleftharpoons\ket{D}$ and $\ket{G}\rightleftharpoons\ket{B}$. With an appropriate choice of $\chi$, the state $\ket{F}$ can be sufficiently detuned, so it does not participate in these transitions. The full Hamiltonian reads as
\begin{align}
 \label{Htot}   &H=H_0+H_{drive},&\\
 \label{Hdrive}   &H_{drive}=\Omega_{D}\cos\left((\varepsilon-\chi )t\right)\sigma^x+\Omega_{B}\cos\left((\omega-\chi )t\right)\tau^x.&
\end{align}
The possible phase shifts of 
both drives are ignored as they  do not affect the results.
Now, we obtain the effective three-level system Hamiltonian. For this, we transform to the interaction picture
\begin{align}
    H_I=e^{\textit{i}H_0t}H_{drive}e^{-\textit{i}H_0t}.
\end{align}
Using
\begin{align}
 \label{int_rep}  \sigma^x_I=e^{\textit{i}(\varepsilon+\chi \tau^z)t}\sigma^x, \,\,\, \tau^x_I=e^{\textit{i}(\omega+\chi \sigma^z)t}\tau^x,
\end{align}
and \ref{Hdrive}, we obtain
\begin{align}
\nonumber    H_I&=\Omega_D\cos\left((\varepsilon-\chi )t\right)e^{\textit{i}(\varepsilon+\chi \tau^z)} \sigma^x +& \\
\label{HIfin}    &+\Omega_B\cos\left((\omega-\chi )t\right)e^{\textit{i}(\omega+\chi \sigma^z)}\tau^x.&
\end{align}
Operators \ref{int_rep} act on our four states \ref{fstates} in the following way:
\begin{align}
  & e^{\textit{i}(\varepsilon+\chi \tau^z)t}\sigma^x\ket{G}=e^{\textit{i}(\varepsilon-\chi)t} \ket{D},& \\
  & e^{\textit{i}(\varepsilon+\chi \tau^z)t}\sigma^x\ket{D}=e^{-\textit{i}(\varepsilon-\chi)t} \ket{G},&\\ 
  & e^{\textit{i}(\varepsilon+\chi \tau^z)t}\sigma^x\ket{B}=e^{\textit{i}(\varepsilon+\chi)t} \ket{F},& \\
  & e^{\textit{i}(\varepsilon+\chi \tau^z)t}\sigma^x\ket{F}=e^{-\textit{i}(\varepsilon+\chi)t} \ket{B},& \\
  & e^{\textit{i}(\omega+\chi \sigma^z)t}\tau^x\ket{G}=e^{\textit{i}(\omega-\chi)t} \ket{B},& \\
  & e^{\textit{i}(\omega+\chi \sigma^z)t}\tau^x\ket{B}=e^{-\textit{i}(\omega-\chi)t} \ket{G},& \\
  & e^{\textit{i}(\omega+\chi \sigma^z)t}\tau^x\ket{D}=e^{\textit{i}(\omega+\chi)t} \ket{F},& \\
  & e^{\textit{i}(\omega+\chi \sigma^z)t}\tau^x\ket{F}=e^{-\textit{i}(\omega+\chi)t} \ket{D}.&  
\end{align}
Substituting these expressions into Eq. (\ref{HIfin}), we see that there are time dependent oscillating terms and time-independent ones. We apply the rotating-wave approximation to keep only the latter, which brings us to the effective Hamiltonian written in Eq. (\ref{Hint}) with operators
\begin{align}
    S^{+}=\ket{D}\bra{{G}},\,\,\,T^{+}=\ket{B}\bra{{G}},\\
    S^{-}=\left(S^{+}\right)^{\dagger},\,\,\, T^{-}=\left(T^{+}\right)^{\dagger},
\end{align}
which can be represented in the form of Eqs. (\ref{STp}) and (\ref{STm}).
Note that these terms do not allow transitions involving the auxiliary state $F$, effectively reducing the four level system to the three-level $V$-shaped one. The non-Hermitian terms in Eq. (\ref{Heff}) commute with $H_0$, so they do not change in the interaction picture.

\section{Basics of braid and knot theories}
\label{sec:BrKn}
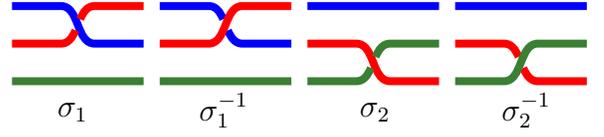
\begin{figure}
\begin{tikzpicture}[scale=0.5] 
\braid[line width=3pt, rotate=90, number of strands=3, 
style strands={1}{OliveGreen}, style strands={2}{red}, style strands={3}{blue}] (braid) 1 a_2^{-1} 1;
\end{tikzpicture}
\begin{tikzpicture}[scale=0.5]
\braid[line width=3pt, rotate=90, number of strands=3,
style strands={1}{OliveGreen}, style strands={2}{red}, style strands={3}{blue}] (braid) 1 a_2 1;
\end{tikzpicture}
\begin{tikzpicture}[scale=0.5] 
\braid[line width=3pt, rotate=90, number of strands=3,
style strands={1}{OliveGreen}, style strands={2}{red}, style strands={3}{blue}] (braid) 1 a_1^{-1} 1;
\end{tikzpicture}
\begin{tikzpicture}[scale=0.5] 
\braid[line width=3pt, rotate=90, number of strands=3,
style strands={1}{OliveGreen}, style strands={2}{red}, style strands={3}{blue}] (braid) 1 a_1 1;
\end{tikzpicture}
\begin{minipage}{.22\linewidth}
\vspace{0.13cm} \large{$\sigma_1$}  
\end{minipage}
\begin{minipage}{.22\linewidth}
\vspace{0.1cm} \large{$\sigma_1^{-1}$}   
\end{minipage}
\begin{minipage}{.22\linewidth}
\vspace{0.13cm} \large{$\sigma_2$}  
\end{minipage}
\begin{minipage}{.22\linewidth}
\vspace{0.1cm} \large{$\sigma_2^{-1}$}   
\end{minipage}
\caption{Elementary braid elements of $B_3$ braid group for three strands.}
\label{fig:braids}
\end{figure}

Here we collect some basic results and methods from the theory of closed braids that we use in the main part. 
Throughout the paper, we consider braids composed of three strands, their braid group is $B_3$, though all our results can be straightforwardly applied to $B_n$ braid group describing $n$ strands. The nontrivial elements of $B_3$ braid group are given in Fig. \ref{fig:braids}: each $\sigma_i$ element accounts for braiding between $i$ and $i+1$ strands (counted from above). It is $\sigma_i$ if the strand $i$ goes above $i+1$ and $\sigma_i^{-1}$ otherwise. In our case, there are only four braid elements: $\sigma_1$, $\sigma_1^{-1}$, $\sigma_2$, $\sigma_2^{-1}$ (and additionally the identity element $\mathbb{1}$ when no braiding is performed). A sequential application of these elements allows creating any possible braid pattern for three strands. A resulting sequence of braidings is called the braid word.\\
\begin{figure}
\begin{tikzpicture}[scale=0.5]
\braid[line width=3pt, rotate=90, number of strands=2, 
style strands={1}{red}, style strands={2}{blue}] (braid) 1  a_1^{-1} 1 a_1^{-1} 1;
\draw[dotted, line width=1.5pt, blue] (braid-rev-2-s) -- ++(0,1) -- ++(5.5,0) -- (braid-rev-2-e);
\draw[dotted, line width=1.5pt, red] (braid-rev-1-s) -- ++(0,-1) -- ++(5.5,0) -- (braid-rev-1-e);
\end{tikzpicture}\hspace{1cm}
\begin{tikzpicture}[scale=0.5]
\braid[line width=3pt, rotate=90, number of strands=2,
style strands={1}{red}, style strands={2}{blue}] (braid) 1 1 a_1 1 1;
\draw[dotted, line width=1.5pt, red] (braid-rev-2-s) -- ++(0,-1) -- ++(2.73,0);
\draw[dotted, line width=1.5pt, blue] (2.77,0) -- ++(2.73,0) -- (braid-rev-1-e);
\draw[dotted, line width=1.5pt, blue] (braid-rev-1-s) -- ++(0, 1) --  ++(2.73,0);
\draw[dotted, line width=1.5pt, red] (2.77,3) -- ++(2.73,0) -- (braid-rev-2-e);
\end{tikzpicture}
\begin{minipage}{.45\linewidth}
\vspace{0.13cm} {\large $\sigma_1^2$}:  Hopf link
\end{minipage}
\begin{minipage}{.45\linewidth}
\vspace{0.1cm} {\large $\sigma_1$}: Unknot   
\end{minipage}
\caption{Knot representations of two closed braids. Dotted lines connect associated points.}
\label{fig:knots}
\end{figure}
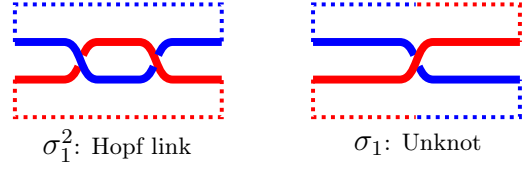
The theory allows for homological representations, so the braid elements can be written in matrix form. In the reduced Burau representation \cite{Kassel2008} of $B_3$ braid group, the nontrivial elements are given by matrices
\begin{align}
 \label{reducedBurau}   \sigma_1=\begin{pmatrix} t & 1 \\ 0 & 1
    \end{pmatrix}, \,\,\, \sigma_2=\begin{pmatrix}
        1 & 0 \\ -t & t
    \end{pmatrix}.
\end{align}
We note in passing that this homological representation is known to be faithful for $B_3$ (i.e. it is a one-to-one representation of the braid group), so it suffices for us. For extension of this analysis to $B_n \, (n>3)$, other representations, although more complicated, may be preferable.\\
In Eq. (\ref{reducedBurau}), we can choose $t\in (0,1)$ to reflect the fact that our $B_3$ braid group is not cyclic:
\begin{align}
    \forall m \in \mathbb{Z}: \,\,\,\sigma_i^m\neq \mathbb{1},\,\,\, (\sigma_i\sigma_j\sigma_i)^{4m}\neq \mathbb{1}.
\end{align}
Other rules for braid operations are as follows: 
\begin{align}
\label{BIdent}    ...\sigma_i\sigma_i^{-1}...=...\mathbb{1}...\,,
\end{align}
\begin{align}
   \label{BnA} ...\sigma_1\sigma_2...\neq ...\sigma_2\sigma_1...\,,
\end{align}
\begin{align} \label{Brel}  ...\sigma_1\sigma_2\sigma_1...=...\sigma_2\sigma_1\sigma_2...\,.
\end{align}
Eq. (\ref{BIdent}) is a trivial property of inverse group elements. Eq. (\ref{BnA}) reflects the fact the $B_3$ group is non-Abelian, so its elements are non-commutative. Nevertheless, Eq. (\ref{Brel}) is a universal property of neighboring strands of braid groups, applicable even to non-Abelian braids. These properties can be straightforwardly verified either by using the diagrammatic representation of Fig. \ref{fig:braids} or matrices Eq. (\ref{reducedBurau}). Combining these properties together, one can derive all possible nontrivial relations between $B_3$ braids and find equivalent braid words.\\
\begin{figure*}[!th]
\setcounter{section}{3}
    \hspace{-1cm}
    \begin{minipage}[t]{.49\textwidth}
    \centering    \includegraphics[width=1.1\textwidth]{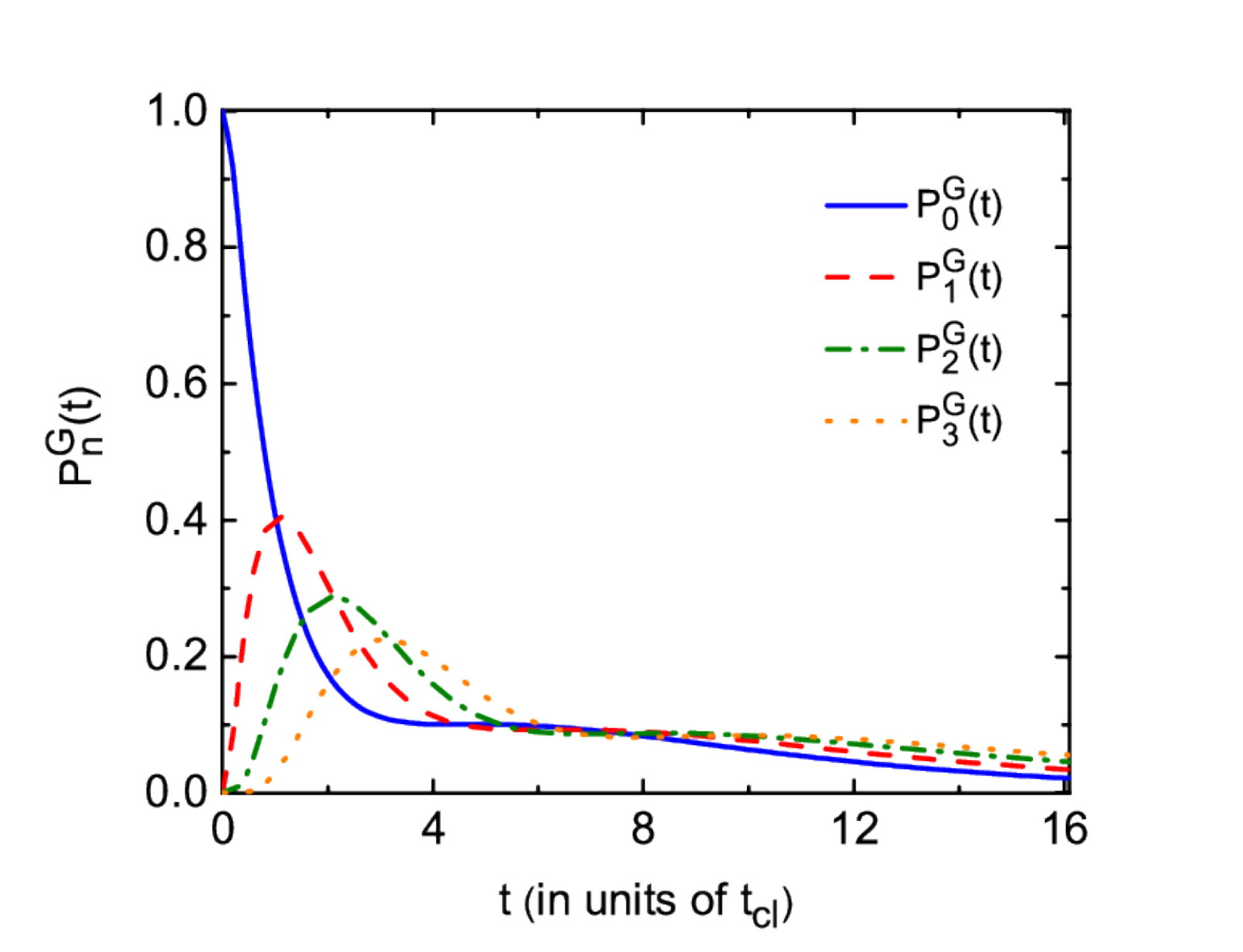}
    \end{minipage}
    \begin{minipage}[t]{.49\textwidth}
    \centering    \includegraphics[width=1.1\textwidth]{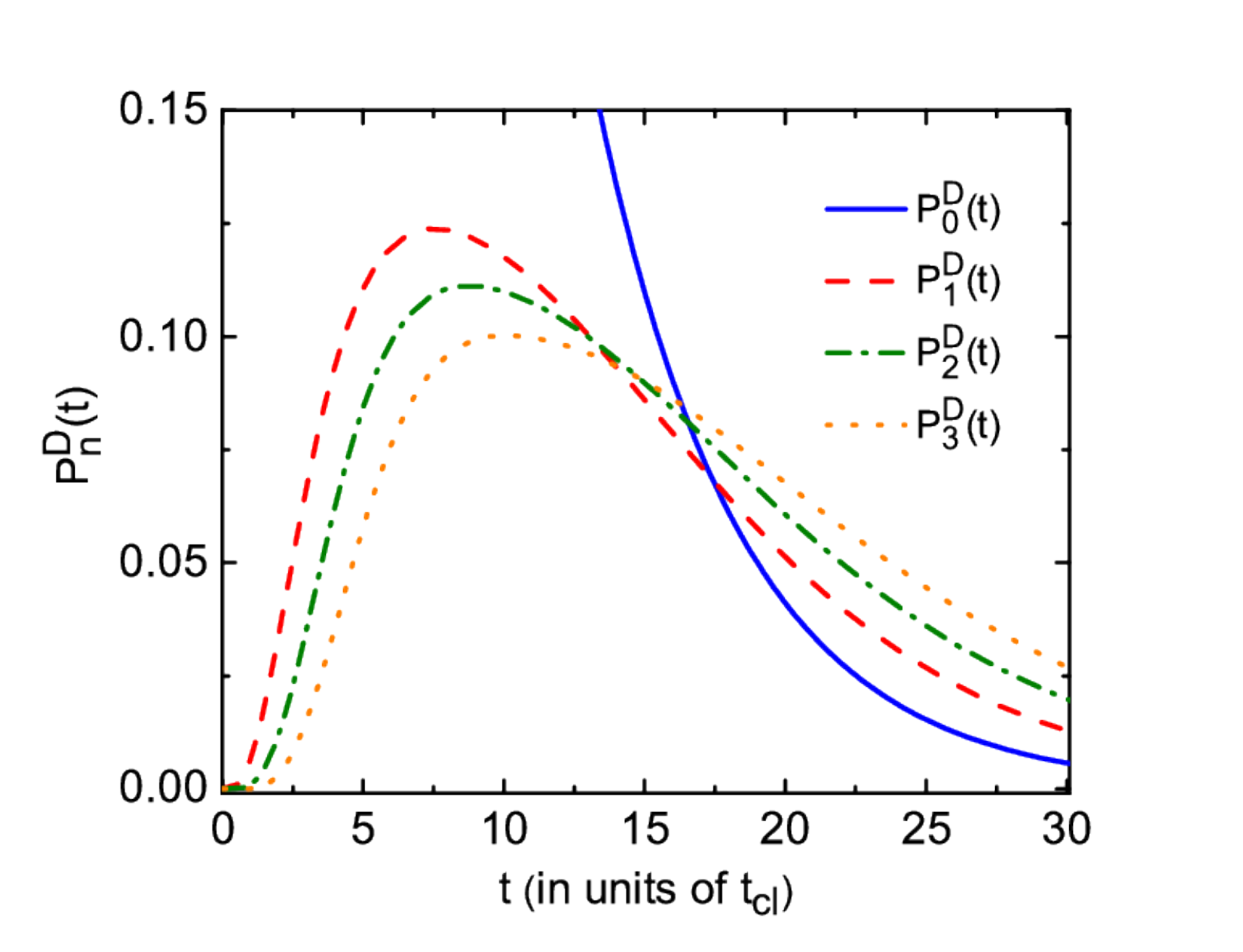}
    \end{minipage}
    \caption{Probabilities to have $n$ clicks during the observation starting from the ground state ($P_n^G(t)$, left panel) or the dark state ($P_n^D(t)$, right panel) as functions of time. $\Omega_D=0.008$.} 
    \label{fig:PGDt}
\setcounter{section}{2}    
\end{figure*}
Now let us account for the periodicity of the braids. All strands in our case are defined on the interval $[k_0,k_0+2\pi]$ with an arbitrary real $k_0$. It means that the starting and ending points in $k$ should be associated, so the braids are closed. These strands exist in three-dimensional space $\mathbb{R}^3$ formed by axes $k, Re[\lambda], Im[\lambda]$, and therefore the resulting closed braids are isotopic to links, which in turn form knots \cite{Kassel2008}. These links are oriented in our case, meaning that $\sigma_i\neq\sigma_i^{-1}$. This idea is illustrated in Fig. \ref{fig:knots} for two strands, where we connect the starting and ending points of strands by dotted lines. The left diagram (braid word $\sigma_1^2$) forms two intersecting rings, this structure is known as the Hopf link. On the right diagram (braid word $\sigma_1$), moving along solid and dotted lines one arrives at the same point, so this structure effectively forms a single circle, known as unknot. By shrinking the dotted lines and continuously deforming the solid lines, we obtain diagrams similar to the ones shown on the right side of Fig. \ref{fig:phases}. Due to the arbitrary choice of $k_0$ and closed nature of the braids, the condition Eq. (\ref{BnA}) is relaxed; although elements $\sigma_1$ and $\sigma_2$ are still non-commutative, braid words are now defined up to cyclic permutations. It means that if two braid words can be obtained one from the other by applying Eqs. (\ref{BIdent}) and (\ref{Brel}) and cyclic permutations, they belong to the same topological class. The braid index Eq. (\ref{nu_tot}) counts the number of $\sigma_i$ elements in the braid word minus the number of $\sigma_i^{-1}$ elements, while other braid indices introduced in Section \ref{sec:invariants} count the number of braid elements (contributing with appropriate signs) encountered while moving along a particular closed path, similar to Fig. \ref{fig:knots}.

\section{Quantum jumps distribution}
\label{sec:QJ}
Here we analyze the time-dynamics of the quantum jumps distribution, in particular, the oscillating in time parity of the number of jump reported in Section \ref{sec:observable}. As is evident from Figs. \ref{fig:dynamicalObs}(a) and  \ref{fig:dynamicalObs}(b), the initial condition plays an important role for the dynamics of the system at short times ($\approx 3t_{cl}$), while afterwards this dynamics becomes universal. Hence, we start by considering probability to observe $n$ quantum jumps during the observation time $t$ either starting from the ground state or the dark state: $P^G_n(t)$, $P^D_n(t)$. For both cases we chose $\Omega_D=0.008$, corresponding to class IV, where $C(\pi,t)$ shows harmonic, sign-changing oscillations in time. These probability distributions can be analyzed the same way for other classes. We plot both distributions for $n=0,...,3$ in Fig. \ref{fig:PGDt}. Evidently, the jump probability distributions are non-Poissonian. By starting from the ground state (Fig. \ref{fig:PGDt}, left panel), one can observe distributions for $n$ jumps that initially look as Poissonian for the decay rate $\Gamma_B$ (note that each $P_n^{G}(t)$ peaks at $t_n=n \Gamma_B^{-1}$), but have an additional (very broad) peak in their long-time tails with very slow decay. We interpret this as follows - starting from the ground state, the system is most probably trapped within $G\rightleftharpoons B$ transitions, but has a small probability of escaping into the dark state. Such an event would drastically reduce the number of observed jumps. Moreover, if the system does not exhibit $\approx n$ jumps over time $n \Gamma_B^{-1}$, this indicates that the system has most probably escaped into the dark state and will likely stay there even longer (since $\Gamma_D\ll \Gamma_B$), further increasing the relative weights of small-$n$ observations over long times.
This provides the timescale for the second maximum in the probability distributions, $t_n\simeq n \Gamma_B^{-1}+\Gamma_D^{-1}$. Note that it holds for $n>0$, as $P_0(t)$ is monotonous and has a broad plateau rather than a peak. The situation is reversed when the system is initially prepared in the dark state (Fig. \ref{fig:PGDt}, right panel). In addition to the jump probabilities governed by the dark state decay rate, there is a chance that the system escapes into the bright sector, leading to a substantial shift of the jump probability distributions towards shorter times as compared to what  
one would expect from the $\Gamma_D$ rate [the plotted distributions $P_n^{D}(t)$ have peaks at $t_n\approx n \Gamma_B^{-1}+\Gamma_D^{-1}$].\\
Next we analyze the oscillations of the ``staggered'' probability distribution $C(\pi,t)$. For that, we introduce the partial sum of the jump probabilities $P^{(N)}(t)$, which sums probabilities to observe from $0$ to $N$ quantum jumps during observation time $t$, weighted with signs according to their parities
\begin{align}
  \label{PartN}  P^{(N)}(t)=\sum_{l=0}^N (-1)^lP_l^{G}(t).
\end{align}
\begin{figure}[!t]
    \centering
    \hspace*{-0.35cm}
    \includegraphics[width=1.15\linewidth]{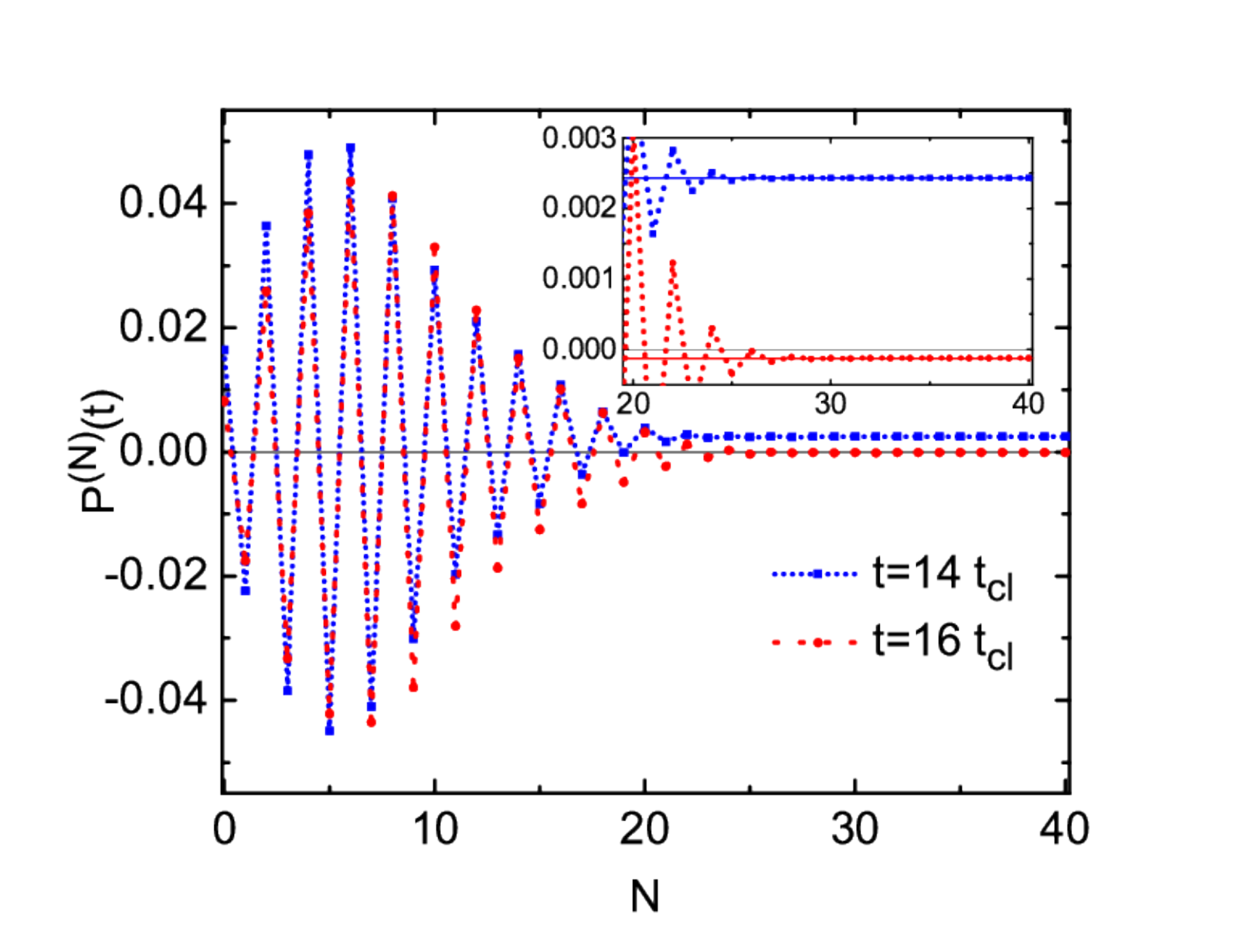}
    \caption{Partial sum $P^{(N)}(t)$ as a function of $N$ for $t=14 t_{cl}$ (blue points) and $t=16t_{cl}$ (red points). $\Omega_D=0.01$. Inset:  Late time saturation of the corresponding values to positive $C(\pi,14t_{cl})$ (blue line) and negative $C(\pi, 16t_{cl})$ (red line).}    \label{fig:PNt}
\end{figure}

For $N\rightarrow \infty$, this expression simply turns into $C(\pi,t)$, which is the Fourier transform of the jump probabilities taken at $k=\pi$:
\begin{align}
    \lim_{N\rightarrow\infty}P^{(N)}(t)=C(\pi,t).
\end{align}
Note that in Eq. (\ref{PartN}) we chose probabilities for a system initially prepared in the ground state, though this choice between G and D states is arbitrary and simply accounts for the difference between Figs. \ref{fig:dynamicalObs}(a) and \ref{fig:dynamicalObs}(b). 

The analysis of the partial sum $P^{(N)}(t)$ allows us to discern the main processes responsible for oscillations of $C(\pi,t)$ in class IV. Namely, one immediately can see if a certain number of jumps strongly dominates (e.g., if $n=0$ or $1$ is the most probable and defines the whole structure of the distribution), or if it is rather a collective effect arising from interplay of multiple possible outcomes. As can be seen from Fig. \ref{fig:PNt}, the latter scenario is correct: probabilities for many different $n$ sum up, so $P^{(N)}(t)$ is strongly oscillating before it saturates to $C(\pi,t)$. In Fig. \ref{fig:PNt} (we chose there $\Omega_D=0.01$ to make the transitions easily visible), it saturates at time $t=14t_{cl}$ to $C(\pi,14t_{cl})=0.0024$ (blue line) and at time $t=16t_{cl}$ to $C(\pi,16t_{cl})=-0.00012$ (red line), note that the amplitudes of the transient oscillations are  much larger than the eventual saturation values. In both cases, approximately $N\approx 2t/t_{cl}$ jumps make important contributions. This is because probabilities with relatively small numbers of jumps still give substantial contributions even at long times (see large tails in Fig. \ref{fig:PGDt}). On the other hand it is very unlikely to observe numbers of jumps considerably exceeding those predicted by the Poissonian distribution with the largest (of two) rate $\Gamma_B$ (so probabilities for $n\gg t/t_{cl}$ are strongly suppressed).\\

\section{Perturbations of an exceptional point}
\label{sec:EP2}
Here we consider a general Hamiltonian of a two-level system exhibiting an exceptional point. The presence of the exceptional point means that this Hamiltonian is non-diagonalizable, and (after a proper re-scaling) can be represented there in a Jordan form
\begin{align}
    H_{EP}=\begin{pmatrix}
        \alpha-\textit{i}\beta & 1\\
        0 & \alpha-\textit{i}\beta
    \end{pmatrix},
\end{align}
where $\alpha$ and $\beta$ are real and imaginary parts of the degenerate eigenvalue. 
At the EP there is no basis of eigenvectors, instead, there exists the Jordan chain $\ket{\psi_1},\ket{\psi_2}$ such that 
\begin{align}
\nonumber    &H_{EP}\ket{\psi_1}=(\alpha-\textit{i}\beta)\ket{\psi_1},&\\
 \label{JchH}   &H_{EP}\ket{\psi_2}=(\alpha-\textit{i}\beta)\ket{\psi_2}+\ket{\psi_1}&.
\end{align}
Using Eqs. (\ref{JchH}) and (\ref{L0def}), we can construct the corresponding Jordan chain of the Lindbladian 
\begin{align}
L_{EP}=-\textit{i}\left\{H_{EP}\otimes \mathbb{1}-\mathbb{1}\otimes \left(H_{EP}^\dag \right)^T\right\}\ ,
\end{align}
in which the quantum jumps have been omitted for now. Defining $\rho_{ij}\equiv\ket{\psi_i}\bra{\psi_j}$,
we obtain
\begin{align}
    &L_{EP}\left[\frac{1}{2}\left(\rho_{12}+\rho_{21}\right)\right]=-2\beta\left[\frac{1}{2}\left(\rho_{12}+\rho_{21}\right)\right], &\\
    &L_{EP}\rho_{11}=-2\beta\rho_{11},&\\
    &L_{EP}\left[\frac{1}{2\textit{i}}\left(\rho_{12}-\rho_{21}\right)\right]=-2\beta \left[\frac{1}{2\textit{i}}\left(\rho_{12}-\rho_{21}\right)\right] +\rho_{11}, &\\
    &L_{EP}\left[\frac{1}{2}\rho_{22}\right]=-2\beta \left[\frac{1}{2}\rho_{22}\right]+\frac{1}{2\textit{i}}\left(\rho_{12}-\rho_{21}\right).&
\end{align}
\begin{figure*}[!ht]
    \hspace{-1.6cm}
    \begin{minipage}[t]{.49\linewidth}
    \centering
    \includegraphics[width=1.2\linewidth]{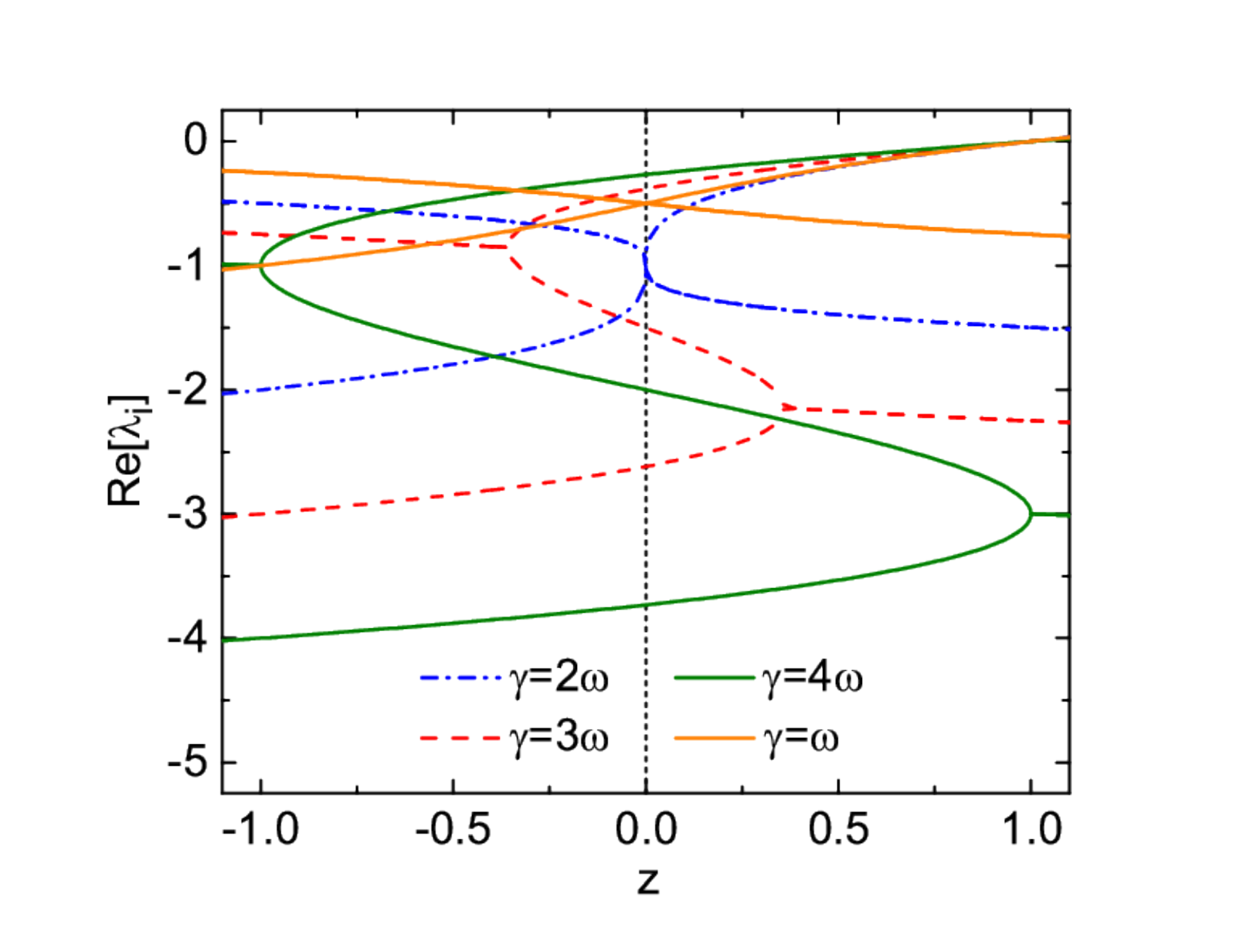}
    \end{minipage}
    \hspace{0.5cm}
    \begin{minipage}[t]{.49\linewidth}
    \centering
    \includegraphics[width=1.2\linewidth]{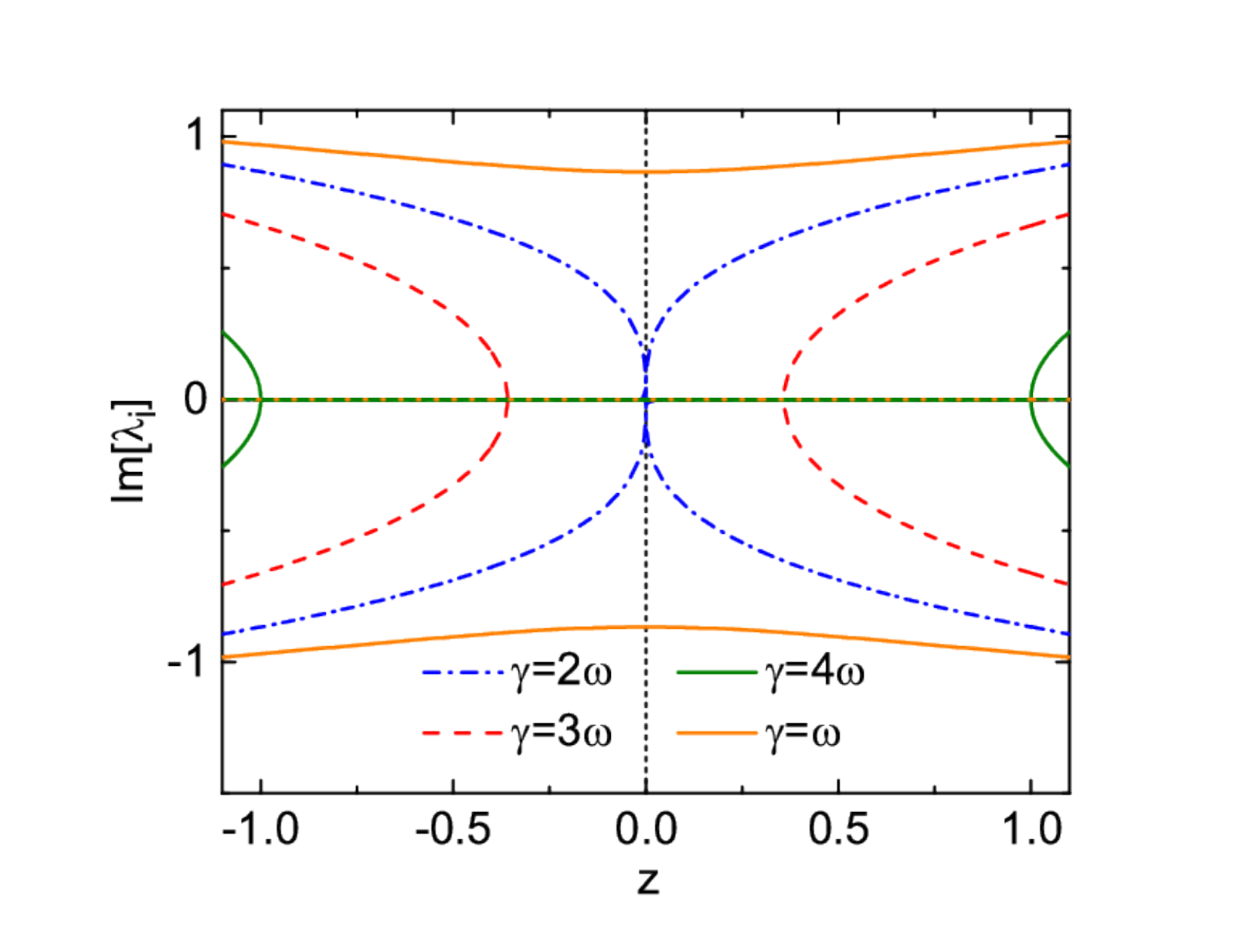}   
    \end{minipage}
    \caption{Exceptional point of the Lindbladian $L_0$ perturbed by the jump term $L_J$, Eq. (\ref{fig:3Dstructure}) [numerical parameters are given in Eq. (\ref{Example1})] with phase $k=0$ ($z>0$) and $k=\pi$ ($z<0$), $z=re^{\textit{i}k}$. Real (left panel) and imaginary (right panel) parts of the Lindbladian eigenvalues (in units of $\omega$) are plotted as functions of $z$ at fixed parameter $\gamma$. Blue dot-dashed line corresponds to the critical frequency where there is an exceptional point at $r=0$, while red dashed line, green solid line and orange solid line show the same system with changed parameter $\gamma$, they do not have an exceptional point at $r=0$, but the perturbed system acquires pairs of exceptional points at some values of the parameters $\gamma$ and $r$ (red and green lines) as the original third order exceptional point splits. Dashed black line is used as a guide for eyes and shows the case of absent quantum jumps.}
    \label{fig:LzGamma}
\end{figure*}
As we see, the Jordan chain is formed by the following three vectors: $u_1=\rho_{11}$, $u_2=\frac{1}{2\textit{i}}(\rho_{12}-\rho_{21})$, $u_3=\frac{1}{2}\rho_{22}$, while the other eigenvector, $e=\frac{1}{2}(\rho_{12}+\rho_{21})$, does not participate in the EP, although its eigenvalue is degenerate with that of $u_1$.  

The Jordan form of the Lindbladian $L_{EP}$ written in the basis $\{e, u_1, u_2, u_3\}$ takes the form
\begin{equation}
\label{LEP} L_{EP}=\begin{tikzpicture}[
    node distance=1mm and 0mm,
    baseline]
\matrix (M1) [matrix of nodes,{left delimiter=(},{right delimiter=)}]
{
        $-2\beta$ & 0 & 0 & 0 \\
        0 & $-2\beta$ & 1 & 0 \\
        0 & 0 & $-2\beta$ & 1\\
        0 & 0 & 0 & $-2\beta$\\
};
\draw[red,dashed, thick] 
        (M1-2-2.north west) -| (M1-4-4.south east) -| (M1-2-2.north west);
\end{tikzpicture}.
\end{equation}
The dashed red box shows the Jordan form related to the third order EP. For simplicity, in what follows we consider only this reduced $3\times 3$ matrix.    
Next we consider a perturbed Liouvillian $L_{EP}+z L_J$, where, e.g., $L_J$ is the jump term, while $z \in \mathbb{C}$ is a perturbation parameter.
A detailed analysis of perturbations in a vicinity of a third order EP
can be found in \cite{Seyranian2003, Demange2011}. 

For the chain of right vectors $\{u_1, u_2, u_3\}$ one constructs the corresponding chain of vectors $\{v_1, v_2, v_3\}$, satisfying
\begin{align}
\nonumber  & v_1^Tu_1=0, \,\,\, v_1^Tu_2=0, \,\,\, v_1^Tu_3=1&\\
\label{eq:SelfOrthogonality} &v_3^Tu_2=0,\,\,\,v_3^Tu_3=0, \,\,\,  v_2^Tu_2=1. &
\end{align}
Note that the right vectors $\{u_1, u_2, u_3\}$ and the left 
vectors $\{v_1^T, v_2^T, v_3^T\}$ obey 
the self-orthogonality condition \ref{eq:SelfOrthogonality}
rather than the one of the type of 
Eq. (\ref{LRbasis}). 
Explicitly, we find
\begin{align}
    v_1=\begin{pmatrix}
        0 \\ 0 \\ 1
    \end{pmatrix}, \,\,\,  v_2=\begin{pmatrix}
        0 \\ 1 \\ 0
    \end{pmatrix}, \,\,\,  v_3=\begin{pmatrix}
        1 \\ 0 \\ 0
    \end{pmatrix}.
\end{align}
In general, the EP is lifted for the perturbed matrix $L_{EP}+z L_J$.
The perturbed eigenvalues $\lambda^{(m)}$ are given in terms of the Puiseux series. For a small perturbation $|z|\ll 1$, they can be approximated as
\begin{align}
 \label{EVpert}   \lambda^{(m)}=\lambda_0+\lambda_1^{(m)} r^{\frac{1}{3}}+o(r^{\frac{1}{3}}),
\end{align}
where $\lambda_0$ is the unperturbed degenerate eigenvalue at the exceptional point, $z=r e^{\textit{i}k}$ is the perturbation parameter of Eq. (\ref{L0J}), while $\lambda_1^{(m)}$ are the three complex roots of the equation  $(\lambda_1)^3=e^{\textit{i}k}v_1^TL_Ju_1$. Note that 
$e^{\textit{i}k}v_1^TL_Ju_1$ is real for $k=0$, thus one of the solutions has a positive real part, while two other have identical negative real parts. Thus even an infinitesimal positive $z$ immediately opens the Liouvillian gap, i.e., the longest living solution with the positive real part of $\lambda_1$ can no longer participate in any EP.  
Although, strictly speaking, this analysis is applied to $|z|\ll 1$, its interpolation up to $z=1$ is aligned with the results of \cite{Minganti2020} and restores the results of \cite{Minganti2019}. Interestingly, the structure of the perturbed eigenstates reverses at $k=\pi$, namely, the two longest lived eigenstates have the same real part and further form a second order EP.

Let us see how these general features appear in a particular example given by Eq. (\ref{ModelEff}).
Its effective Hamiltonian has an EP at $\gamma=2\omega$. Measuring all energies in units of $\omega$, we obtain the Jordan form of the EP Liouvillian $L_{EP}$ and the jump operator $L_J$:
\begin{align}
\label{Example1}    L_{EP}=\begin{pmatrix}
        -1 & 0 & 0 & 0 \\
        0 & -1 & 1 & 0 \\
        0 & 0 & -1 & 1 \\
        0 & 0 & 0 & -1
    \end{pmatrix}, \,\,\, L_J=\begin{pmatrix}
        0 & 0 & 0 & 0 \\
        0 & 0 & 0 & 0 \\
        0 & 0 & 0 & 0 \\
        0 & 1 & 0 & 0
    \end{pmatrix}.
\end{align}
The perturbed eigenvalues of Eq. (\ref{EVpert}) are constructed with
\begin{align}
    \lambda_0=-1, \,\,\, \lambda_1^{(n)}=e^{\textit{i}(k+2\pi n)/3}.
\end{align}
This gives the perturbed eigenvalues
\begin{align}
   & k=0: \, \lambda^{(3)}=-1+r^{\frac{1}{3}}, \, \lambda^{(1,2)}=-1+e^{\pm\textit{i}\frac{2\pi}{3}}r^{\frac{1}{3}};&\\
   & k=\pi: \, \lambda^{(3,2)}=-1+e^{\pm\textit{i}\frac{\pi}{3}}r^{\frac{1}{3}}, \, \lambda^{(1)}=-1-r^{\frac{1}{3}}. &
\end{align}
Due to simplicity of the perturbation term in Eq. (\ref{Example1}), these results turn out to be exact, even for $r=1$. The numerical results showing the effects of quantum jumps and the duality between $k=0$ and $\pi$ are shown in Fig. \ref{fig:pert}. As expected, one of the eigenvalues is not involved in the formation of the EP and is not affected by its perturbation.\\
Furthermore, the eigenvalues of $L_{EP}+zL_J$ in Eq. (\ref{ModelEff}) allow for analytical solutions for arbitrary parameters. After rescaling the eigenvalues in units of $\omega$, there is one $z$-independent eigenvalue, which does not participate in the EP and three others are given by roots of the third-order polynomial,
\begin{align}
    \lambda^{(1)}=\frac{1}{2}x_1, \, \lambda^{(2)}=\frac{1}{2}x_2, \, \lambda^{(3)}=\frac{1}{2}x_3, \, \lambda^{(4)}=-\frac{1}{2}\gamma,
\end{align}
where $x_{1,2,3}$ are roots of the polynomial
\begin{align}
    x^3+3\gamma x^2+(4+2\gamma^2)x+ 4\gamma(1-z) =0.
\end{align}
\setcounter{section}{5} 
\begin{figure*}[!t]
    \hspace{-0.5cm}
    \begin{minipage}[t]{.49\linewidth}
    \centering
    \includegraphics[width=1.1\linewidth]{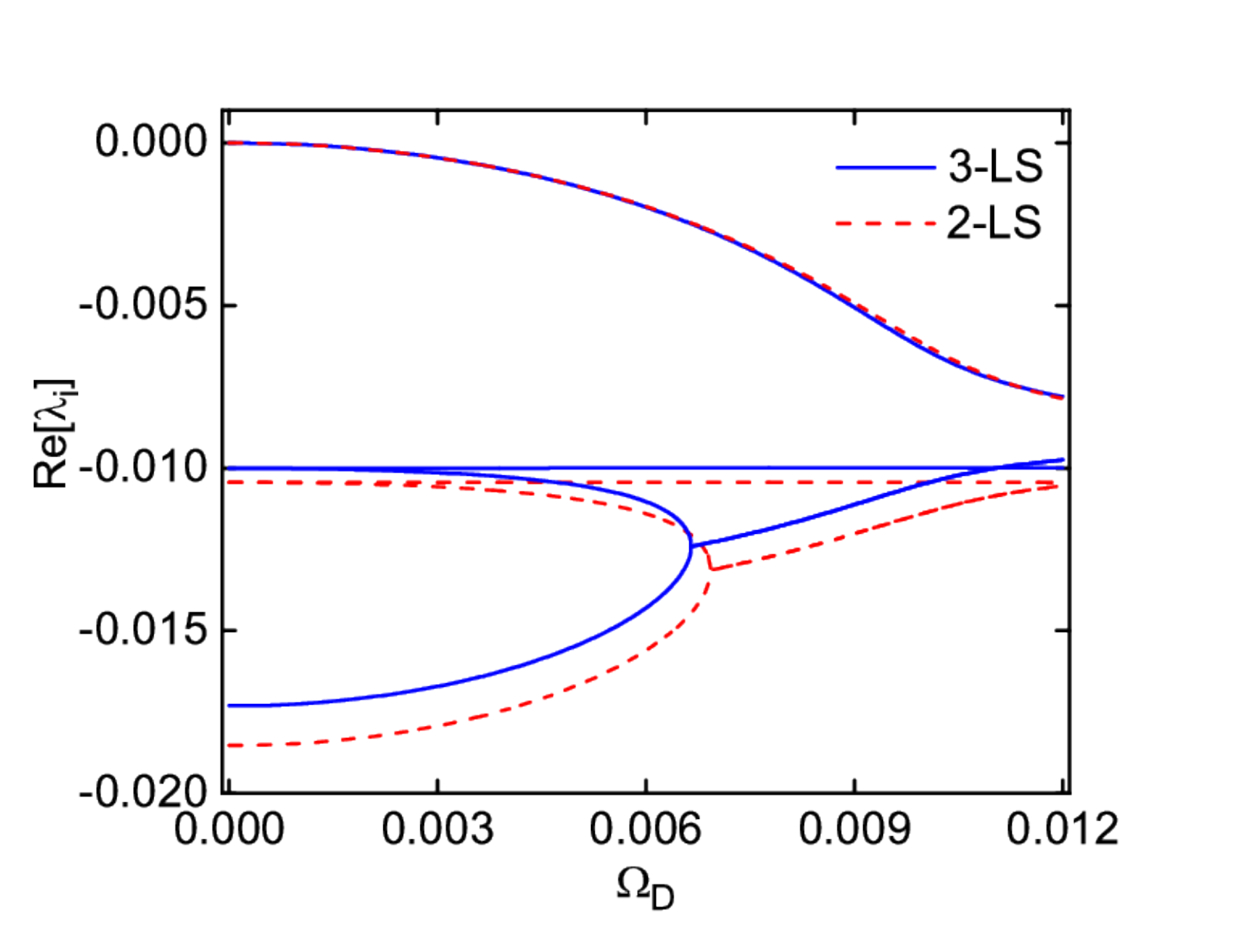}
    \end{minipage}
    \begin{minipage}[t]{.49\linewidth}
    \centering
    \includegraphics[width=1.1\linewidth]{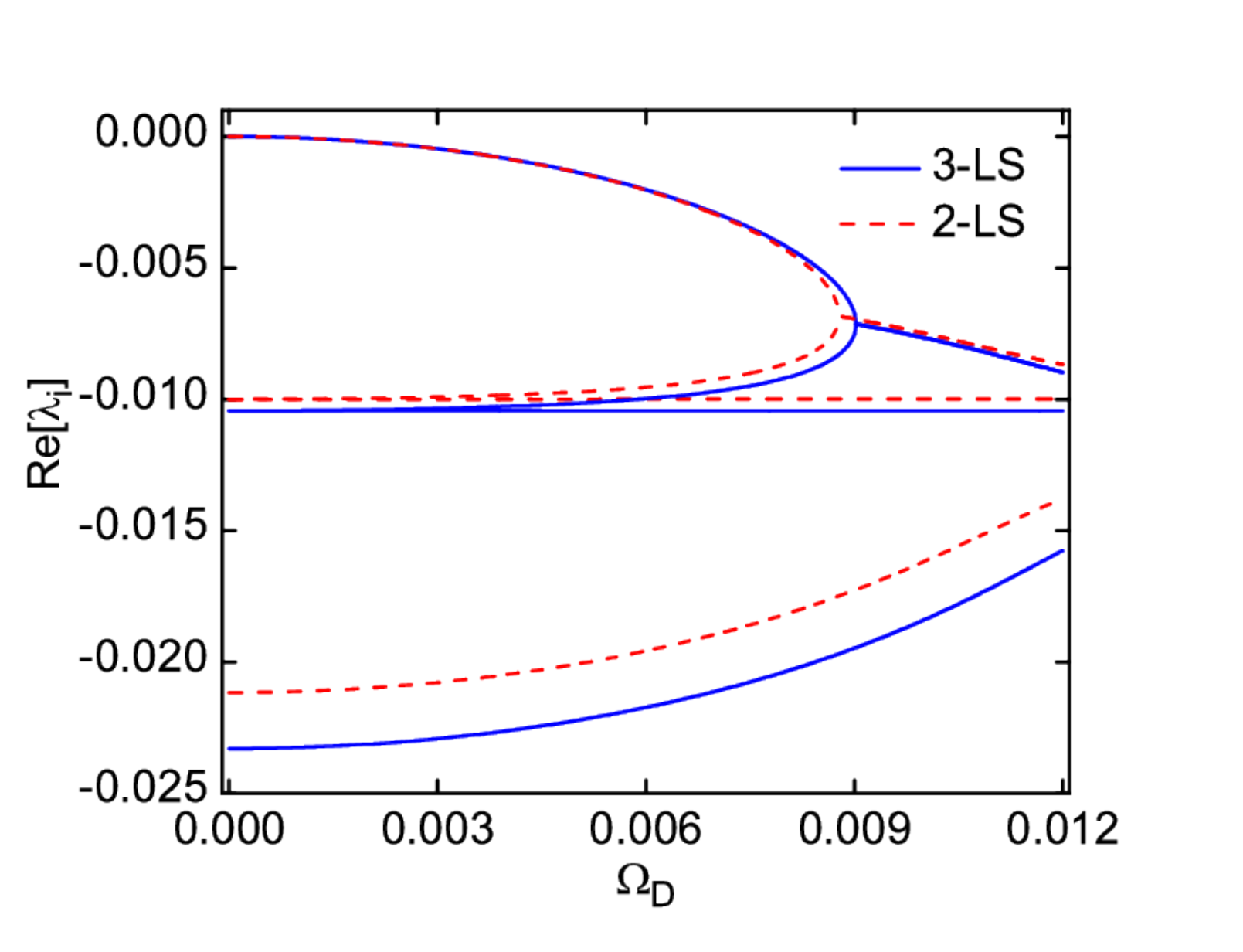}   
    \end{minipage}
    \caption{Eigenvalues (real parts) $\lambda_i$ of the three-level system Lindbladian Eq. (\ref{fullL}) with the jump term suppressed by $z=\pm 0.1$ (blue solid lines, 3-LS) and eigenvalues of the effective ground-dark two-level system's Lindbladian obtained via adiabatic elimination (red dashed line, 2-LS) as functions of the driving frequency $\Omega_D$. Left panel: $z=0.1$, right panel: $z=-0.1$. For the three-level system, only four relevant eigenvalues (long-living states forming an exceptional point) are plotted. The reduced system retains an exceptional point of the same structure in close proximity to the original exceptional point. }
    \label{fig:LReduced} 
\end{figure*}
\setcounter{section}{4}   
The real parts of the eigenvalues $\lambda^{(i)}$ for fixed $z$ are plotted in Fig. \ref{fig:pert} as functions of $\gamma$. Additionally, we plot real and imaginary parts of these exact solutions in Fig. \ref{fig:LzGamma} for various fixed $\gamma$ as functions of (real) $z$. It is evident from this plot that a change from $k=0$ to $k=\pi$ flips the structure of the eigenvalues simply due to the symmetry of the roots of a cubic equation. Moreover, the EP that appears without quantum jumps at some critical value of $\gamma$ shifts for nonzero $z$ but survives, though at different $\gamma$. For example, in Fig. \ref{fig:LzGamma} there are EPs at $\gamma=2\omega$ for $z=0$ (dotted-dashed blue line), at $\gamma=3\omega$ for $z=\pm 0.5$ (dashed red line), and at $\gamma=4$ for $z=\pm 1$ (solid green line).\\
It is worth mentioning that a special case  $v_1^TL_Ju_1=0$ is in principle possible. Then Eq. (\ref{EVpert}) is no longer applicable and the EP is perturbed by a linear term and a pair of square-root terms \cite{Seyranian2003, Demange2011}. This situation is unlikely for realistic quantum jump models in two-level-systems, but even in this case the phase $k=\pi$ flips the order of the perturbed eigenstates with respect to $k=0$, making a formation of an EP between two least decaying states possible.\\
\section{Adiabatic elimination}
\label{sec:AE}
When a system consists of two coupled subsystems with asymptotically stable hierarchy of characteristic times (one subsystem exhibits slow dynamics and the other has fast dynamics) there is a method for systematic elimination of the fast subsystem and for arriving at an effective description of the reduced slow subsystem. This method is known as adiabatic elimination. Most generally, it applies at the Lindbladian level such that 
the reduced density matrix of the subsystem
is governed by a new effective Lindbladian
\cite{Mirrahimi2009}. If the dynamics of the whole system can be described in terms of a non-Hermitian Hamiltonian, then an effective description of the reduced system has also the Hamiltonian form \cite{Brion2007}.\\  
In the following, we do not go into details on a formal and rigorous description of this procedure in terms of the projection operators and construction of the perturbation theory in terms of these operators (for details see \cite{Mirrahimi2009, Reiter2012, Azouit2017, Finkelstein2020}), but rather describe the exact (in the sense of perturbation theory) numerical approach to the adiabatic elimination, which can be easily implemented for small systems (\cite{Shnirman1998, Makhlin2000, Brion2007}). \\
Let us consider the Lindbladian dynamics of our three-level system $\{G,\,D,\,B\}$. We write its density matrix $\rho$ in the vector form, $\dot{\rho}(t)=\mathcal{L}\rho$, where $\mathcal{L}$ is the system's Lindbladian superoperator. Then we reorder the column representing the density matrix as well as the Lindbladian as follows
\begin{align}
    \rho=\begin{pmatrix}
        \rho_S \\ \rho_F
    \end{pmatrix}, \,\,\,
    \mathcal{L}=\begin{pmatrix}
        \mathbf{S} & \mathbf{C} \\ \mathbf{C}^T & \mathbf{F}
    \end{pmatrix}.
\end{align}
Here $\rho_S$ is a column of four elements comprising the slow system: populations of D and G states and coherences between them (namely, $\rho_{GG}$, $\rho_{GD}$, $\rho_{DG}$, $\rho_{DD}$), $\rho_{F}$ is a column of five elements with fast dynamics, which involve $B$ state and transitions to/from it. $\mathbf{S}$ and $\mathbf{F}$ are $4\times 4$ and $5\times 5$ matrices describing Lindbladian dynamics of the slow and fast sectors correspondingly. $\mathbf{C}$ and $\mathbf{C}^T$ are matrices describing transitions between these sectors. The essential idea behind the adiabatic elimination is that due to the large separation of time scales the fast system provides a constant background for the slow system's evolution. Hence, the $\rho_F$ elements can be considered constant in time. In our case, this assumption is justified by a clear hierarchy of transition rates $\gamma_B\gg \Gamma_B\gg \Gamma_D$, 
\begin{align}
    \begin{pmatrix}
        \dot{\rho}_S(t) \\ \dot{\rho}_F(t) 
    \end{pmatrix}\simeq
    \begin{pmatrix}
        \dot{\rho}_S(t) \\ 0
    \end{pmatrix}=
    \begin{pmatrix}
        \mathbf{S} & \mathbf{C} \\ \mathbf{C}^T & \mathbf{F}
    \end{pmatrix} 
    \begin{pmatrix}
        \rho_S(t) \\ \rho_F(t) 
    \end{pmatrix}.    
\end{align}
 This brings us to
 \begin{align}
     \dot{\rho_S}=\mathcal{L}_{eff}\rho_S=\left(\mathbf{S}-\mathbf{C}\mathbf{F}^{-1}\mathbf{C}^T\right)\rho_S.
 \end{align}
 This reduced effective Lindbladian describes the dynamics of the slow subsystem. For the numerical implementation of a similar procedure in an arbitrary large system see \cite{Spiecker2024} (with the difference that the reduction is done there to describe only populations of a full system, rather than populations and coherences of a subsystem). The same procedure can be straightforwardly implemented on the Schrödinger equation level in the absence of quantum jumps (e.g. see \cite{Minev2019} for the Hamiltonian adiabatic elimination in the three-level system). We compare the reduced effective two-level system Hamiltonian with the full three-level system Hamiltonian in Fig. \ref{fig:Hreduced}. Evidently, the two most important states are almost not affected by this procedure at all - they preserve the EP with only a minor shift in frequency $\Omega_D$. The same applies to the Lindbladian dynamics.
 As demonstrated in Fig. \ref{fig:LReduced}, for the Lindbladian with reduced weight of quantum jumps ($|z|=0.1$), the three longest living states preserve their structures both for $k=0$ and $k=\pi$. 
Additionally, we see that the adiabatic elimination is most accurate for the state with the smallest real part, regardless if it contains an EP. Further eigenstates are approximated less accurately due to increase of their decay, though they still preserve the EP structure.

\bibliography{EPref} 

\end{document}